\documentclass[12pt]{article}
\usepackage{amsfonts,amssymb,epsfig,amsmath,mathtools}
\usepackage{color}

\usepackage[enableskew]{youngtab}
\usepackage{mathrsfs}
\usepackage{graphicx}
\usepackage{subcaption}
\usepackage{hyperref}

\renewcommand{\baselinestretch}{1.2}

\newcommand{\be}{\begin{eqnarray}}
\newcommand{\ee}{\end{eqnarray}}

\newcommand{\bn}{\begin{enumerate}}
\newcommand{\en}{\end{enumerate}}

\begin{document}

\makeatletter \@addtoreset{equation}{section} \makeatother
\renewcommand{\theequation}{\thesection.\arabic{equation}}
\renewcommand{\thefootnote}{\alph{footnote}}

\begin{titlepage}

\begin{center}
\hfill {\tt SNUTP18-005}\\
\hfill{\tt KIAS-P18097}\\

\vspace{2cm}

{\Large\bf Large AdS black holes from QFT}

\vspace{2cm}

\renewcommand{\thefootnote}{\alph{footnote}}

{\large Sunjin Choi$^{1,2,3}$, Joonho Kim$^{2}$, Seok Kim$^1$ and June Nahmgoong$^{1}$}

\vspace{0.7cm}

\textit{$^1$Department of Physics and Astronomy \& Center for
Theoretical Physics,\\
Seoul National University, 1 Gwanak-ro, Gwanak-gu, Seoul 08826, Republic of Korea}\\

\vspace{0.2cm}

\textit{$^2$School of Physics, Korea Institute for Advanced Study,\\
85 Hoegi-ro, Dongdaemun-gu, Seoul 02455, Republic of Korea}\\

\vspace{0.2cm}

\textit{$^3$Kavli Institute for the Physics and Mathematics of the Universe (WPI),\\
The University of Tokyo Institutes for Advanced Study, The University of Tokyo,\\
Kashiwa, Chiba 277-8583, Japan}\\

\vspace{0.7cm}

E-mails: {\tt sunjin.choi@ipmu.jp, joonhokim@kias.re.kr, \\
seokkimseok@gmail.com, earendil25@snu.ac.kr}

\end{center}

\vspace{1cm}

\begin{abstract}

We study the index of $\mathcal{N}=4$ Yang-Mills theory on $S^3\times\mathbb{R}$ at
large angular momenta. A generalized Cardy limit exhibits macroscopic entropy
at large $N$. Our result is derived using free QFT analysis, and also a background field
method on $S^3$. The index sets a lower bound on the entropy.
It saturates the Bekenstein-Hawking entropy of known supersymmetric AdS$_5$ black holes,
thus accounting for their microstates. We further analyze
the so-called Macdonald index, exploring small black holes and possibly new black holes
reminiscent of hairy black holes. Finally, we study aspects of large supersymmetric
AdS$_7$ black holes, using background field method on $S^5$ and 't Hooft anomalies.

\end{abstract}

\end{titlepage}

\renewcommand{\thefootnote}{\arabic{footnote}}

\setcounter{footnote}{0}

\renewcommand{\baselinestretch}{1}

\tableofcontents

\renewcommand{\baselinestretch}{1.2}

\section{Introduction and summary}

Microscopic understanding of black holes is a major achievement of string theory.
Many successes are made using 2d QFT approaches, starting from the pioneering work of  \cite{Strominger:1996sh}. These are mostly related to the AdS$_3$/CFT$_2$ duality.
In fact, AdS/CFT \cite{Maldacena:1997re}
is an ideal setting to study black holes using quantum field theory.
In higher dimensional AdS$_d$, at $d>3$, there are interesting physics
of black holes to be better understood: see e.g. \cite{Witten:1998zw,Aharony:2003sx}
and references thereof. Especially, supersymmetric black holes
in $d>3$ suggest quantitative challenges to CFTs. (Magnetic/dyonic
black holes in AdS$_4$, with fluxes on $S^2$ boundary, were recently studied microscopically
from topologically twisted 3d QFTs \cite{Benini:2015eyy,Benini:2016rke}. Our interest in
this paper will be the electric black holes, whose microstates
consist of excitations from the unique vacuum of the radially quantized CFT.)

It has been believed that such BPS black holes in AdS$_5$ defied
quantitative understandings from indices of SCFTs on $S^{3}\times\mathbb{R}$
\cite{Kinney:2005ej,Romelsberger:2005eg}. There have been many speculations on why the
index fails to capture black holes. A possible reason is that bosonic/fermionic states
undergo big cancelation. For instance, the index cannot see the deconfinement phase
transition at an order $1$ temperature in the unit of AdS$_5$ radius \cite{Kinney:2005ej},
which is the QFT dual of the Hawking-Page transition of AdS black holes \cite{Hawking:1982dh}.
So the index cannot capture all the physics of generic supersymmetric
AdS$_5$ black holes. Direct studies of BPS operators at weak coupling did not discover
enough microstates for such black holes either  \cite{Berkooz:2006wc,Berkooz:2008gc,Grant:2008sk,Chang:2013fba}, at least
so far.

In this paper, we show that the index of 4d $\mathcal{N}=4$ Yang-Mills theory
\underline{\textit{does}} capture large supersymmetric AdS$_5$ black holes
\cite{Gutowski:2004ez,Chong:2005da,Cvetic:2005zi} in an asymptotic Cardy-like limit.
Our Cardy limit is more refined than \cite{DiPietro:2014bca},
in that the imaginary parts of chemical potentials are tuned to optimally obstruct
boson/fermion cancelations. The entropy of our asymptotic index is macroscopic,
meaning that it is proportional to $N^2$ when all the charges are at this order.
This sets a lower bound on the true microscopic entropy of BPS states, assuring
the existence of BPS black holes in AdS$_5\times S^5$.
In particular, when a charge relation is met, our asymptotic free energy agrees with
the Bekenstein-Hawking entropy of known supersymmetric AdS$_5$ black holes
\cite{Gutowski:2004ez}, thereby microscopically counting them. The asymptotic free energy
of our index is the recently suggested entropy function for supersymmetric AdS$_5$
black holes \cite{Hosseini:2017mds}, in our large black hole limit. At general values
of charges, perhaps our findings may have implications to possible supersymmetric
hairy black holes in AdS$_5\times S^5$ \cite{Markeviciute:2018cqs,Markeviciute:2018yal}.
The last suggestion is indirectly supported by studying the asymptotic free energy of the
so-called Macdonald index \cite{Gadde:2011uv}. Here, depending on charge regime, the
Cardy-like free energy differs from the entropy function of \cite{Hosseini:2017mds}, showing
properties reminiscent of hairy black holes in $AdS_5\times S^5$.

Our derivation is based on two methods. One is the free QFT.
Another is a background field method on $S^3$, in which the
Chern-Simons terms of these background fields yield the asymptotic free energy.
The relevant Chern-Simons terms are determined by 't Hooft anomalies. The
latter method can be useful for non-Lagrangian QFTs. We apply it to
the 6d $(2,0)$ theory and study aspects of large supersymmetric black holes in
AdS$_7\times S^4$ \cite{Chong:2004dy,Chow:2007ts}.

The rest of this paper is organized as follows. In section 2,
we derive the asymptotic free energy of the index of 4d $\mathcal{N}=4$ Yang-Mills
theory, in a generalized Cardy-like limit. This free energy counts
known supersymmetric AdS$_5$ black holes. In section 3, we study similar asymptotic
free energy of the index in the Macdonald limit, suggesting rich structures
such as small black holes and new saddle points reminiscent of hairy black holes.
In section 4, we apply the background field method to study supersymmetric AdS$_7$
black holes. In section 5, we summarize with comments on future directions.

There has been a lot of progress on this subject since this paper appeared: see \cite{Cabo-Bizet:2018ehj,Benini:2018ywd} and references thereof.

\section{Large supersymmetric AdS$_5$ black holes}

We study the the partition function of $\mathcal{N}=4$ Yang-Mills theory on
$S^3\times \mathbb{R}$, focussing on the index limit \cite{Kinney:2005ej}.
The partition function counts states carrying six charges.
The first one is the energy $E$, made dimensionless
by multiplying the $S^3$ radius. Three charges $Q_1,Q_2,Q_3$ are for the Cartans
of $SO(6)$ R-symmetry, defined to be the angular momenta on three orthogonal
2-planes on $\mathbb{R}^6$, being $\pm\frac{1}{2}$ for spinors. The final two are
the angular momenta $J_1,J_2$ on $S^3$, being
$\pm\frac{1}{2}$ for spinors. The BPS states of our interest saturate
the bound $E\geq Q_1+Q_2+Q_3+J_1+J_2$, but we shall impose the BPS limit at a
later stage to see more universal features.
Consider the general partition function:
\begin{equation}\label{partition-function}
  Z(\beta,\Delta_I,\omega_i)={\rm Tr}\left[e^{-\beta E}
  e^{-\sum_{I=1}^3\Delta_I Q_I}e^{-\sum_{i=1}^2\omega_i J_i}\right]\ .
\end{equation}
The complex chemical potentials $\Delta_I$, $\omega_i$ satisfy five periodicity
conditions $\Delta_I\sim\Delta_I+4\pi i$, $\omega_i\sim\omega_i+4\pi i$.
The $16$ supercharges are $\mathcal{Q}^{Q_1,Q_2,Q_3}_{J_1,J_2}$.
$16$ possible values of $Q_I,J_i$ carried by $\mathcal{Q}$ are
$\pm\frac{1}{2}$, where the product of all $5$ $\pm$ signs is
$+$. The conformal supercharges are $\mathcal{S}^{Q_1,Q_2,Q_3}_{J_1,J_2}$
with five charges being $\pm\frac{1}{2}$, where the product of signs is $-$.
Taking the trace without $(-1)^F$, the fermionic fields are anti-periodic along
temporal circle, twisted by $\Delta_I,\omega_i$. So the SUSY connecting periodic bosons
and anti-periodic fermions are generally broken. In a sense, the supercharges are
anti-periodic which has no zero modes on temporal $S^1$. However, if
\begin{equation}\label{BPS-hypersurface}
  \sum_{I=1}^3s_I\Delta_I-\sum_{i=1}^2t_i\omega_i=2\pi i\ (\textrm{mod }4\pi i)
  \ \ \ ,\ \ \ s_I,t_i=\pm 1\textrm{\ \ satisfying\ \ }s_1s_2s_3t_1t_2=+1\ ,
\end{equation}
(\ref{partition-function}) becomes an index if one takes $\beta\rightarrow 0^+$.
This is because
\begin{equation}
  e^{-\Delta\cdot Q-\omega\cdot J}\mathcal{Q}^{s_1,s_2,s_3}_{-t_1,-t_2}
  =e^{-\frac{s\cdot\Delta-t\cdot J}{2}}\mathcal{Q}^{s_1,s_2,s_3}_{-t_1,-t_2}
  e^{-\Delta\cdot Q-\omega\cdot J}=-\mathcal{Q}^{s_1,s_2,s_3}_{-t_1,-t_2}
  e^{-\Delta\cdot Q-\omega\cdot J}\ ,
\end{equation}
so that translating $\mathcal{Q}^{s_1,s_2,s_3}_{-t_1,-t_2}$ along the trace will
cause extra $-1$ sign, creating a zero mode of this supercharge.
So restricting $Z$ to this hypersurface of $\Delta_I,\omega_i$, it becomes an
index which counts $\frac{1}{16}$-BPS states annihilated by
$\mathcal{Q}\equiv\mathcal{Q}^{s_1,s_2,s_3}_{-t_1,-t_2}$ and
$\mathcal{S}\equiv \mathcal{S}^{-s_1,-s_2,-s_3}_{t_1,t_2}$. From the
algebra
\begin{equation}
  \{\mathcal{Q},\mathcal{S}\}=E-\sum_{I=1}^3s_IQ_I-\sum_{i=1}^2t_iJ_i\ ,
\end{equation}
one finds $E=s_IQ_I+t_iJ_i$.
Therefore, having in mind that we shall eventually live on one of the hyperspaces
(\ref{BPS-hypersurface}), we study $Z$ in the `formal high temperature limit'
$\beta\rightarrow 0^+$.

We shall analyze $\log Z$ in an asymptotic
Cardy-like limit $|\omega_i|\ll 1$. In our limit, $\Delta_I$ is kept complex,
$\mathcal{O}(1)$, and generic. The computation will be made
using two complementary approaches. One is the free QFT analysis, which is reliable
because $Z$ will be independent of the coupling constant at the hyperspace (\ref{BPS-hypersurface}). This will be presented in section 2.1.
Another method, explored in section 2.2, is a background field approach
on $S^3$. To understand this, note that $Z$ has a path integral
representation on $S^3\times S^1$, where the size of the temporal circle is given by
vanishing $\beta$. $\beta,\Delta_I$, $\omega_i$ are realized as
background fields on this space. At small $\beta$, we reduce the system on small circle, integrating out the KK modes on $S^1$. The partition function will then acquire contribution
from dynamical zero modes on $S^3$, while the KK modes will yield
an effective action of 3d background fields. Although the KK reduction is a bit
subtle than it may naively sound, we shall argue that the derivative expansion of
the effective action is a series expansion in small $\beta,\omega_1,\omega_2$.
The leading $\log Z$ is given by the Chern-Simons terms of background fields,
determined either from a weakly-coupled 4d QFT or using 't Hooft
anomalies \cite{DiPietro:2014bca}. Evaluating these Chern-Simons terms, we obtain
same asymptotic $\log Z$. Then in section 2.3, we study the physics of the derived
$\log Z$ and discuss the dual black holes.

\subsection{Free QFT analysis}

{\allowdisplaybreaks
The partition function (\ref{partition-function}) of weakly-coupled $\mathcal{N}=4$
Yang-Mills theory is given by \cite{Aharony:2003sx}
\begin{eqnarray}\label{index-weak}
  \hspace*{-1cm}Z&=&\frac{1}{N!}\oint\prod_{a=1}^N\frac{d\alpha_a}{2\pi}\cdot
  \prod_{a<b}\left(2\sin\frac{\alpha_{ab}}{2}\right)^2
  \exp\left[\!\!\frac{}{}\right.\sum_{a,b=1}^N
  \sum_{n=1}^\infty\frac{1}{n}\left(\!\!\frac{}{}\right.
  f^{\rm v}_B(n\beta,n\omega_i)+(-1)^{n-1}f^{\rm v}_F(n\beta,n\omega_i)\\
  \hspace*{-1cm}&&+\chi_{\bf 3}(n\Delta_I)(f^{\rm c}_B(n\beta,n\omega_i)
  \!+\!(-1)^{n-1}f^{\rm c}_F(n\beta,n\omega_i))+\chi_{\bar{\bf 3}}(n\Delta_I)
  (f^{\rm a}_B(n\beta,n\omega_i)\!+\!(-1)^{n-1}f^{\rm a}_F(n\beta,n\omega_i))
  \!\left.\frac{}{}\!\!\right)e^{in\alpha_{ab}}\left.\frac{}{}\!\!\right]\nonumber
\end{eqnarray}
where $\alpha_{ab}\equiv\alpha_a-\alpha_b$,
$\chi_{\bf 3}=\sum_{I=1}^3e^{\Delta_I}$,
$\chi_{\bar{\bf 3}}=\sum_{I=1}^3e^{-\Delta_I}$, and
\begin{eqnarray}
  f^{\rm v}_B&=&\frac{e^{-\beta}(1-e^{-2\beta})(e^{\omega_1}+e^{\omega_2}
  +e^{-\omega_1}+e^{-\omega_2})-1+e^{-4\beta}}
  {(1-e^{-\beta+\omega_1})(1-e^{-\beta+\omega_2})
  (1-e^{-\beta-\omega_1})(1-e^{-\beta-\omega_2})}+1\\
  f^{\rm v}_F&=&\frac{e^{-\frac{3}{2}\beta}(e^{\Delta}-e^{-\Delta}e^{-\beta})
  (e^{\omega_+}+e^{-\omega_+})+e^{-\frac{3}{2}\beta}
  (e^{-\Delta}-e^{\Delta}e^{-\beta})(e^{\omega_-}+e^{-\omega_-})}
  {(1-e^{-\beta+\omega_1})(1-e^{-\beta+\omega_2})
  (1-e^{-\beta-\omega_1})(1-e^{-\beta-\omega_2})}\nonumber\\
  f^{\rm c}_B=f^{\rm a}_B&=&\frac{e^{-\beta}(1-e^{-2\beta})}
  {(1-e^{-\beta+\omega_1})(1-e^{-\beta+\omega_2})
  (1-e^{-\beta-\omega_1})(1-e^{-\beta-\omega_2})}\nonumber\\
  f^{\rm c}_F&=&\frac{e^{-\frac{3}{2}\beta-\Delta}
  \left((e^{\omega_+}+e^{-\omega_+})-e^{-\beta}((e^{\omega_-}+e^{-\omega_-})\right)}
  {(1-e^{-\beta+\omega_1})(1-e^{-\beta+\omega_2})
  (1-e^{-\beta-\omega_1})(1-e^{-\beta-\omega_2})}\nonumber\\
  f^{\rm a}_F&=&\frac{e^{-\frac{3}{2}\beta+\Delta}
  \left((e^{\omega_-}+e^{-\omega_-})-e^{-\beta}((e^{\omega_+}+e^{-\omega_+})\right)}
  {(1-e^{-\beta+\omega_1})(1-e^{-\beta+\omega_2})
  (1-e^{-\beta-\omega_1})(1-e^{-\beta-\omega_2})}\ ,\nonumber
\end{eqnarray}
with $\Delta\equiv\frac{\Delta_1+\Delta_2+\Delta_3}{2}$,
$\omega_\pm\equiv\frac{\omega_1\pm\omega_2}{2}$. The superscripts v, c, a refers to
$\mathcal{N}=1$ vector, chiral, anti-chiral multiplets, respectively, with
the chiral supercharges
$\mathcal{Q}_\alpha\equiv\mathcal{Q}^{+,+,+}_\alpha$ (at $(t_1,t_2)=(+,+),(-,-)$).

With the understanding that one of the BPS index conditions (\ref{BPS-hypersurface})
will be taken, we study the $\beta\rightarrow 0^+$ limit of this partition function.
One might worry that, before reaching $\beta\rightarrow 0$,
the factors $1-e^{-\beta+\omega_{1,2}}$ in the denominators will hit zeros or
make the sum divergent
if ${\rm Re}(\omega_{1,2})>0$ (for BPS states with $t_1=t_2=+1$).
These are divergences caused by two non-BPS derivatives, losing fugacity
factors smaller than $1$. In general partition function, going beyond this point
will probably have no meaning, analogous to going beyond infinite temperature.
However, having in mind imposing (\ref{BPS-hypersurface}) at $t_1=t_2=1$, these
poles are canceled between bosons/fermions, so
that one can reduce $\beta$ below $\omega_{1,2}$.
Anyway, later in this subsection, we shall present a complementary derivation
manifestly within the index. (However, we think the analysis presented now has
a conceptual advantage.) In this limit, one finds $f^{\rm c,a}_B\rightarrow 0$ due
to the vanishing of the equation of motion factor $1-e^{-2\beta}\rightarrow 0$ on
the numerators. Also, one finds $f^{\rm v}_B\rightarrow 1$ for the same reason.
The fermionic letter partition functions reduce to
\begin{eqnarray}\label{fermion-letter-limit}
  f^{\rm v}_F&\rightarrow&\frac{(e^{\Delta}-e^{-\Delta})
  (e^{\omega_+}+e^{-\omega_+}-e^{\omega_-}-e^{-\omega_-})}
  {(1-e^{\omega_1})(1-e^{\omega_2})(1-e^{-\omega_1})(1-e^{-\omega_2})}
  =\frac{e^{\Delta}-e^{-\Delta}}{2\sinh\frac{\omega_1}{2}\cdot
  2\sinh\frac{\omega_2}{2}}\nonumber\\
  f^{\rm c,a}_F&\rightarrow&\pm\frac{e^{\mp\Delta}}
  {2\sinh\frac{\omega_1}{2}\cdot 2\sinh\frac{\omega_2}{2}}\ .
\end{eqnarray}
$Z$ then becomes
\begin{equation}\label{Z-beta-0}
  Z\rightarrow\frac{1}{N!}\oint\prod_{a=1}^N\frac{d\alpha_a}{2\pi}\cdot
  \prod_{a<b}\left(2\sin\frac{\alpha_{ab}}{2}\right)^2
  \exp\left[\sum_{a,b=1}^N\sum_{n=1}^\infty\frac{1}{n}\left(1+\!\!\!
  \sum_{s_1,s_2,s_3=\pm 1}\frac{s_1s_2s_3(-1)^{n-1}e^{\frac{ns_I\Delta_I}{2}}}
  {2\sinh\frac{n\omega_1}{2}\cdot 2\sinh\frac{n\omega_2}{2}}\right)e^{in\alpha_{ab}}\right]\ .
\end{equation}
Note that the sum over $n$ in the exponent is convergent with nonzero real parts of
$\omega_{1,2}$. For instance, let us have in mind imposing
$\sum_I\Delta_I=\sum_i\omega_i+2\pi i$
for an index, with all chemical potentials having positive real part.
For the terms with given $s_1,s_2,s_3$, the sum over $n$ is separately convergent
if $(s_1,s_2,s_3)\neq (+,+,+)$. This is because, for large $n$, one finds
\begin{equation}\label{convergence1}
  \sim\sum_n\frac{(-1)^{n-1}}{n}e^{-\frac{n}{2}(\omega_1+\omega_2)}e^{in\alpha_{ab}}
  e^{\frac{ns\cdot \Delta}{2}}=-\sum_n\frac{1}{n}e^{-\frac{n}{2}\sum_I(1-s_I)\Delta_I}
  e^{in\alpha_{ab}}\ .
\end{equation}
If some $s_I$ is $-1$, this sum is convergent at large $n$, due to an exponential damping.
On the other hand, the remaining terms in the exponent are the first term `$1$' and the
term with $(s_1,s_2,s_3)=(+,+,+)$. The sum over each term over $n$ may be divergent,
for instance at $\alpha_{ab}=0$. For $a\neq b$,
divergence at $\alpha_{ab}=0$ is fine because there
is a suppression factor given by the Haar measure $\left(2\sinh\frac{\alpha_{ab}}{2}\right)^2$.
For the Cartans, $a=b$, one has to study the possible convergence
of the sum of these two terms without resorting to the phase factor $e^{i\alpha_{ab}}$
or the Haar measure. The sum of these two terms at large $n$ behaves as
\begin{eqnarray}\label{convergence2}
  \sum_n\frac{1}{n}\left(1+\frac{(-1)^{n-1}e^{\frac{\Delta_1+\Delta_2+\Delta_3}{2}}}
  {2\sinh\frac{n\omega_1}{2}\cdot 2\sinh\frac{n\omega_2}{2}}\right)e^{i\alpha_{ab}}
  &=&\sum_n\frac{1}{n}\left(1+\frac{(-1)^{n-1}e^{\frac{\omega_1+\omega_2+2\pi i}{2}}}
  {2\sinh\frac{n\omega_1}{2}\cdot 2\sinh\frac{n\omega_2}{2}}\right)e^{i\alpha_{ab}}
  \nonumber\\
  &\sim&\sum_n\frac{1}{n}\left[1-\left(1-\mathcal{O}(e^{-n\omega})\right)\right]
  e^{i\alpha_{ab}}\ .
\end{eqnarray}
So even at $\alpha_{ab}=0$, or $a=b$, the sum over $n$ converges.

Having realized that the sum converges at $\sum_I\Delta_I=\omega_1+\omega_2+2\pi i$,
we also note here that it will be useful later to consider this sum slightly away
from this surface. Namely, we shall consider the approximation of the index in the `Cardy limit'
$|\omega_i|\ll 1$. Imposing the relation $\sum_I\Delta_I=\omega_1+\omega_2+2\pi i$,
$\Delta_I$'s will share the $\mathcal{O}(1)$ imaginary part $2\pi i$, and furthermore will
have small real parts to match ${\rm Re}(\omega_1+\omega_2)$.
However, for convenient intermediate manipulations, we shall take $\Delta_I$'s slightly
away from this surface by temporarily demanding them to be of order $1$ and purely imaginary.
This parameter deformation clearly does not affect the convergence analysis of
(\ref{convergence1}) for $(s_I)\neq (+,+,+)$. So as for this part,
the function is well defined even after slight deformations.
However, for (\ref{convergence2}), the convergence issue becomes tricky
after the deformation. Just working with the left hand side of
(\ref{convergence2}) with $\Delta_1+\Delta_2+\Delta_3$ being imaginary,
the second term containing $\Delta_I,\omega_i$ will be convergent by itself, for
any $a,b$, while the first term `$1$' will remain divergent at $a=b$.
Therefore, in the analysis below, we shall separate the Cartan parts at $a=b$ and
the off-diagonal parts at $a\neq b$. The former has an exponent proportional to $N$,
and it can be taken out of the $\alpha_a$ integral. The latter part has
$N^2-N$ terms, and only for these terms we shall make a deformation to purely
imaginary $\Delta_I$'s. Ignoring the former contribution to the free energy
$\sim\mathcal{O}(N^1)$ will be justified if one obtains a free energy and entropy of
order $N^2$ from the latter part only. So with this understanding,
we shall often ignore the exponents at $a=b$ in the discussions
below. Note also that, for the off-diagonal parts, the term `$1$' in the exponent
completely cancels the Haar measure part, so we can ignore this term together with
the Haar measure.\footnote{Probably, using asymptotic properties of special functions in the
integrand carefully, one can do the approximation below without using our small
deformations of $\Delta_I$. We just regard it as a short-cut derivation,
similar to familiar `$i\epsilon$' prescriptions which often makes many calculus
more straightforward.}

Now we consider the Cardy limit $|\omega_i|\ll 1$, keeping $\Delta_I$ order $1$
and purely imaginary. The sum over $n$ can be divided into two parts: the `dominant part'
till $n\ll|\omega_i|$, and the `suppressed part' from $n\gtrsim |\omega_i|$.
As for the `dominant' part, we can approximate $2\sinh\frac{n\omega_i}{2}\approx n\omega_i$.
The terms in the exponent of (\ref{Z-beta-0}) from these $n$'s is given by
\begin{equation}\label{dominant-1}
  \frac{s_1s_2s_3}{\omega_1\omega_2}\sum_{n<n_{\rm 0}}\frac{(-1)^{n-1}}{n^3}
  e^{n\left(\frac{s\cdot\Delta}{2}+i\alpha_{ab}\right)}
\end{equation}
where $n_0\ll |\omega|^{-1}$ is a `cut-off' which defines the `dominant part.'
(Note again that we considser the terms at $a\neq b$ only, and we ignored the term $1$
which cancels the Haar measure.) The summation over $n$ is now
independent of the cut-off value $n_0$, as the summand is independent of
$\omega_i$ and converges when $e^{\frac{s\cdot \Delta}{2}+i\alpha_{ab}}$ is a pure
phase. So one obtains the dominant part given by
\begin{equation}\label{dominant-2}
  \frac{s_1s_2s_3}{\omega_1\omega_2}\sum_{n<n_{\rm 0}}\frac{(-1)^{n-1}}{n^3}
  e^{n\left(\frac{s\cdot\Delta}{2}+i\alpha_{ab}\right)}\
  \stackrel{n_0\rightarrow\infty}{\longrightarrow}\
  -\frac{s_1s_2s_3}{\omega_1\omega_2}{\rm Li}_3
  \left(-e^{\frac{s\cdot \Delta}{2}+i\alpha_{ab}}\right)\ .
\end{equation}
Before proceeding, we note that if one wishes, one can take the cut-off $n_0$
to be as big as $|\omega|^{-1}$. This is because at $n\sim|\omega|^{-1}$,
both summands $\frac{e^{n(\frac{s\cdot\Delta}{2}+i\alpha_{ab})}}{n^3\omega_1\omega_2}$
and $\frac{1}{n}\frac{e^{n\left(\frac{s\cdot\Delta}{2}+i\alpha_{ab}\right)}}
{2\sinh\frac{n\omega_1}{2}\cdot 2\sinh\frac{n\omega_2}{2}}$ are very small,
much smaller than the final asserted result (\ref{dominant-2}) which
is $\frac{\mathcal{O}(1)}{\omega_1\omega_2}$. So we proceed with
assuming $n_0\sim|\omega|^{-1}$ below.
Now we discuss the `suppressed' part. It is easy to see that it is indeed suppressed
at $|\omega_i|\ll 1$. This is because
\begin{equation}
  \left|\sum_{n\gtrsim |\omega|^{-1}}\frac{s_1s_2s_3}{n}
  \frac{(-1)^{n-1}e^{n\left(\frac{s\cdot \Delta}{2}+i\alpha_{ab}\right)}}
  {2\sinh\frac{n\omega_1}{2}\cdot 2\sinh\frac{n\omega_2}{2}}\right|\lesssim
  \sum_{n\gtrsim|\omega|^{-1}}\frac{1}{n\cdot 2\sinh\frac{n\omega_1}{2}
  \cdot 2\sinh\frac{n\omega_2}{2}}\lesssim|\omega|
  \sum_{n\gtrsim|\omega|^{-1}}
  \left(2\sinh\frac{n\omega}{2}\right)^{-2}
\end{equation}
which is indeed much smaller than $\frac{1}{\omega^2}$.
With these approximations, one then obtains
\begin{equation}
  Z\sim\frac{1}{N!}\oint\prod_{a=1}^N\frac{d\alpha_a}{2\pi}
  \exp\left[-\frac{1}{\omega_1\omega_2}
  \sum_{a\neq b}\sum_{s_1,s_2,s_3=\pm 1}s_1s_2s_3
  {\rm Li}_3\left(-e^{\frac{s_I\Delta_I}{2}}e^{i\alpha_{ab}}\right)\right]
\end{equation}
where we used the series definition ${\rm Li}_3(x)=\sum_{n=1}^\infty\frac{x^n}{n^3}$
when $|x|\leq 1$. The summations over $a\neq b$, $(s_1,s_2,s_3)$ can be arranged so
that
\begin{equation}
  \hspace*{-0.5cm}Z\sim\frac{1}{N!}\oint\prod_{a=1}^N\frac{d\alpha_a}{2\pi}\cdot
  \exp\left[-\frac{1}{\omega_1\omega_2}\!
  \sum_{a\neq b}\sum_{s_1s_2s_3=+1}\!\!\left(
  {\rm Li}_3\left(-e^{\frac{s\cdot\Delta}{2}+i\alpha_{ab}}\right)
  -{\rm Li}_3\left(-e^{-\frac{s\cdot\Delta}{2}-i\alpha_{ab}}\right)
  \right)\right]\ .
\end{equation}
Here, we note an identity
\begin{equation}\label{Li3-identity}
  {\rm Li}_3(-e^x)-{\rm Li}_3(-e^{-x})=-\frac{x^3}{6}-\frac{\pi^2 x}{6}\ ,
\end{equation}
valid for $-\pi<{\rm Im}(x)<\pi$, taking
$-e^x=e^{x+\pi i}$, $-e^{-x}=e^{-(x+\pi i)}$, respectively.
When $(2p-1)\pi<{\rm Im}(x)<(2p+1)\pi$ for $p\in\mathbb{Z}$, similar identity holds
with $x\rightarrow x-2\pi ip$ on the right hand side.
This identity can be continued to include either positive or negative real parts of $x$.

Now we treat the integrals over
$\alpha_a$'s by a saddle point approximation at $|\omega_1\omega_2|\ll 1$.
Considering a pair of terms ${\rm Li}_3(-e^{\frac{s_I\Delta_I}{2}+i\alpha_{ab}})
+{\rm Li}_3(-e^{\frac{s_I\Delta_I}{2}-i\alpha_{ab}})$ at given $s_I$, one finds that $\alpha_a$
derivative of these are all zero at $\alpha_1=\alpha_2=\cdots=\alpha_N$.
We assume the dominance of this $U(N)$ saddle point in our generalized Cardy
limit. The dominance of such a saddle point was assumed in the Cardy limit of
\cite{DiPietro:2014bca}. But it may fail to be dominant in certain models, e.g.
for other gauge groups than $U(N)$, with fields in certain representations
\cite{Ardehali:2015bla}. Here and later, we shall basically assume the dominance of
our saddle point. In particular, it will reproduce the physics of known large black holes.
As a very basic check, we confirmed at $N=2$ that
$\alpha_1=\alpha_2$ is the global maximum of $\log Z$, making its real part maximal
and imaginary part stationary, along the line of \cite{Ardehali:2015bla}. However, since
our free energy will depend on various complex parameters $\Delta_I,\omega_i$, we have
tested it self-consistently at the extremal values of $\Delta_I,\omega_i$ found in
section 2.3, only at $Q_1=Q_2=Q_3$, $J_1=J_2$. More conceptually, \cite{Ardehali:2015bla}
discussed the relation between other possible saddle points and the behaviors of
the $S^3$ partition function of 4d QFT reduced on small $S^1$. Depending on how bad
the IR divergence of this partition function is \cite{Ardehali:2015bla,DiPietro:2014bca},
one may either expect more nontrivial saddle points to be dominant, or otherwise
zero modes like $\alpha_a$ to cause subleading $N^1\log \omega$ corrections. As we shall
discuss further in section 2.2, our reduced QFT on $S^3$ is maximal SYM, belonging to
the latter class \cite{Ardehali:2015bla,DiPietro:2014bca}. The expected log correction
at $N^1$ order should come from the Cartan terms that we have ignored.
So mostly in this paper, we shall proceed assuming that the above `maximally deconfining'
saddle point is dominant. (In only section 3, we discuss a different saddle point
in a non-Cardy scaling limit.)

Perhaps as a related issue, one may worry from the Haar measure factor $\sim \left(2\sin\frac{\alpha_a-\alpha_b}{2}\right)^2$ that there is a net factor of $0$ when
all $\alpha_a$ are the same, making this saddle point suppressed.
Indeed, in the usual large $N$ saddle point analysis (see e.g. \cite{Aharony:2003sx}),
the Haar measure provides relative repulsions between pairs of $\alpha_a$'s, forbidding
them to be on top of each other. However, in our Cardy saddle point,
log of Haar measure is sub-dominant $\mathcal{O}(\omega^0)$.
So $\alpha_1=\cdots=\alpha_N$ should make sense only as the asymptotic
Cardy saddle point at $\omega\ll 1$.

So assuming this saddle point, one finds
\begin{equation}
  \log Z\sim-\frac{N^2}{\omega_1\omega_2}\sum_{s_1s_2s_3=+1}\left[
  {\rm Li}_3\left(-e^{\frac{s_I\Delta_I}{2}}\right)
  -{\rm Li}_3\left(-e^{-\frac{s_I\Delta_I}{2}}\right)\right]
\end{equation}
where we used $N^2-N\sim N^2$.
Now using the identity (\ref{Li3-identity}), one obtains
\begin{equation}\label{free-energy-general}
  \log Z\sim\frac{N^2}{6\omega_1\omega_2}\sum_{s_1s_2s_3=+1}
  \left[\left(\frac{s\cdot\Delta}{2}-2\pi p_s\right)^3
  +\pi^2\left(\frac{s\cdot\Delta}{2}-2\pi p_s\right)\right]
\end{equation}
in the chamber defined by
\begin{equation}
  (2p_s-1)\pi<\sum_{I=1}^3{\rm Im}\left(\frac{s_I\Delta_I}{2}\right)<(2p_s+1)\pi\ \ ,\ \
  p_s\in\mathbb{Z}\ \ ,\ \ s_1s_2s_3=+1\ .
\end{equation}
Let us consider the `canonical chamber,' with all four $p_s=0$. This
chamber is an octahedron in the space of ${\rm Im}(\Delta_I)$. In this chamber, summing
over $4$ values of $s$, one obtains
\begin{equation}\label{free-energy-final}
  \log Z\sim\frac{N^2\Delta_1\Delta_2\Delta_3}{2\omega_1\omega_2}\ .
\end{equation}
This is the final form of our free energy in the generalized Cardy-like limit.
Now we can continue $\Delta_I$'s to have (small) real parts, to go back to one of the
surfaces (\ref{BPS-hypersurface}).
This formula is reliable at strong coupling on any hypersurface
(\ref{BPS-hypersurface}). Note that in our notation, it appears that
\cite{DiPietro:2014bca} restricted
their interest to $\omega_1=\omega_2\equiv\omega\ll 1$, one of $\Delta_I$'s
$2\pi i+\mathcal{O}(\omega)$, and the remaining two of $\Delta_I$'s at $\mathcal{O}(\omega)$.
The partition function is trivial in this setting. However, as we shall explain in section 2.3,
complex $\Delta_I\sim\mathcal{O}(1)$ are required for all $I=1,2,3$
to see the black hole saddle points, with minimally
obstructed boson/fermion cancelation by the phases of fugacities.

We discussed the asymptotic free energy
in the octahedral `canonical chamber,' defined by
\begin{eqnarray}
  &&-2\pi<{\rm Im}(\Delta_1+\Delta_2+\Delta_3)<2\pi\ ,\ \
  -2\pi<{\rm Im}(\Delta_1-\Delta_2-\Delta_3)<2\pi\ ,\\
  &&-2\pi<{\rm Im}(-\Delta_1+\Delta_2-\Delta_3)<2\pi\ ,\ \
  -2\pi<{\rm Im}(-\Delta_1-\Delta_2+\Delta_3)<2\pi\ .\nonumber
\end{eqnarray}
Here, note that we should seek for an expression on one of the surfaces
(\ref{BPS-hypersurface}). For instance, let us consider
$\Delta_1+\Delta_2+\Delta_3-\omega_1-\omega_2=2\pi i$. Since $\omega_{1,2}$ are very small
in our scaling limit, our hypersurface is very close to the right boundary of
the first inequality, ${\rm Im}(\Delta_1+\Delta_2+\Delta_3)=2\pi$. Whether one is within
the octahedral chamber or not will depend on the small imaginary parts of $\omega_i$'s.
So one may wonder if the expression (\ref{free-energy-final}) can be used or not.
This issue does not matter, as (\ref{free-energy-general}) is continuous
across $\Delta_1+\Delta_2+\Delta_3=2\pi i$. To see this, note that one uses
\begin{equation}\label{Li3-identity-new}
  {\rm Li}_3(-e^x)-{\rm Li}_3(-e^{-x})=-\frac{(x-2\pi i)^3}{6}-\frac{\pi^2 (x-2\pi i)}{6}\ ,
\end{equation}
outside the boundary, instead of (\ref{Li3-identity}),
where $x=\frac{\Delta_1+\Delta_2+\Delta_3}{2}$.
However, the differences between the right hand sides of (\ref{Li3-identity-new}) and
(\ref{Li3-identity}) is $\pi i(x-\pi i)^2$, being continuous and differentiable at
$x=\pi i$. We shall therefore use (\ref{free-energy-final}) at the surface
(\ref{BPS-hypersurface}).

Note that we used large $N$ limit very trivially so far, just to ignore the Cartans.
We basically relied on $|\omega_i|\ll 1$ to approximate the calculations.
This is similar to the Cardy limit
of 2d QFT's describing black holes or strings. There, central charge $c$ is kept fixed while
the chemical potential $\tau$ conjugate to the left Hamiltonian is taken small.
However, the entropy in the Cardy limit is useful to study black holes with large $c$
\cite{Strominger:1996sh}, sometimes beyond the Cardy regime.
To derive the true large $N$ free energy in the non-Cardy regime,
one should consider the large $N$ saddle point approximation of $\alpha_a$ integrals,
at finite $\Delta_I,\omega_i$. As we explained above, we expect a more complicated saddle
point. Also, we are not sure how the graviton phase will get
converted to the black hole phase as we change chemical potentials.
In section 3, in the Macdonald limit, we find that (\ref{free-energy-final}) may
\underline{\textit{not}} be true in general. However, still there might be other
regime in which (\ref{free-energy-final}) is true, which we shall partly probe
in the Macdonald limit. With this in mind, in section 2.3, we shall also explore the
`thermodynamics phenomenology' of (\ref{free-energy-final}) beyond Cardy limit,
especially pointing out the existence of a Hawking-Page transition of this free energy.

So far, we took the limit $\beta\rightarrow 0$ first, having in mind imposing the index
condition (\ref{BPS-hypersurface}) later. We think this is completely fine, but some
people might think that this way of thinking is dangerous.
Appreciating possible worries, we start from the index given by \cite{Kinney:2005ej}
and rederive (\ref{free-energy-final}) at (\ref{BPS-hypersurface}).
A direct consideration of the index will also give interesting lessons beyond
the Cardy limit, in the Macdonald limit \cite{Gadde:2011uv}.
Let us insert the following shifted values to the chemical potentials in
(\ref{partition-function}),
\begin{equation}
  \Delta_I\rightarrow \Delta_I-\beta\ ,\ \ \omega_i\rightarrow\omega_i-\beta\ ,
\end{equation}
after which the partition function is given by
\begin{equation}\label{index-trace}
  Z(\beta,\Delta_I,\omega_i)={\rm Tr}\left[
  e^{-\beta(E-\sum_IQ_I-\sum_iJ_i)}e^{-\Delta_I Q_I-\omega_i J_i}\right]\ .
\end{equation}
Now imposing the condition $\Delta_1+\Delta_2+\Delta_3-\omega_1-\omega_2=2\pi i$,
the measure in the trace commutes with the supercharge $\mathcal{Q}^{+++}_{--}$,
$\mathcal{S}^{---}_{++}$, at any value of $\beta$. We take $\beta\rightarrow\infty$
to suppress the contributions from all non-BPS letters. Let us redefine one of the
chemical potentials, say $\Delta_1-2\pi i$ as the new $\Delta_1$, so that
the index condition becomes
\begin{equation}\label{4d-chemical-relation}
  \Delta_1+\Delta_2+\Delta_3=\omega_1+\omega_2\ .
\end{equation}
Then, the shift by $2\pi i$ generates extra $e^{-2\pi iQ_1}=(-1)^F$ in the trace
formula (\ref{index-trace}), making it a manifest index.
(This redefinition can be made with any one of the five chemical potentials.)
After this redefinition, and taking $\beta\rightarrow\infty$ in (\ref{index-weak}),
one obtains \cite{Kinney:2005ej}
\begin{equation}
  Z=\frac{1}{N!}\int\prod_{a=1}^N\frac{d\alpha_a}{2\pi}
  \prod_{a<b}\left(2\sin\frac{\alpha_{ab}}{2}\right)^2
  PE\left[\left(1-\frac{\prod_{I=1}^3(1-t^2v_I)}{(1-t^3y)(1-t^3/y)}\right)
  \sum_{a,b=1}^Ne^{in\alpha_{ab}}\right]\ ,
\end{equation}
where $v_i$'s satisfying $v_1v_2v_3=1$ are the fugacities for $SU(3)\subset SO(6)$ part of R-symmetry. The parameters $t,v_i,y$ are related to our parameters in (\ref{index-weak})
by $(e^{-\omega_1},e^{-\omega_2},e^{-\Delta_I})=(t^3y,t^3/y,t^2v_I)$, manifestly
satisfying (\ref{4d-chemical-relation}). This is rewritten as
\begin{equation}\label{index-canonical}
  Z=\frac{1}{N!}\int\prod_{a=1}^N\frac{d\alpha_a}{2\pi}
  \prod_{a<b}\left(2\sin\frac{\alpha_{ab}}{2}\right)^2
  \exp\left[\sum_{n=1}^\infty\frac{1}{n} \left(1-
  \frac{\prod_{I=1}^3 2\sinh\frac{n\Delta_I}{2}}
  {2\sinh\frac{n\omega_1}{2}\cdot 2\sinh\frac{n\omega_2}{2}} \right)
  \sum_{a,b=1}^Ne^{in\alpha_{ab}}\right]\ .
\end{equation}

We again take $|\omega_i|\ll 1$, keeping them
complex with ${\rm Re}(\omega_{1,2})>0$. Had we been taking this limit with real positive
$\Delta_i$'s, which is the canonical range for the chemical potentials, $\Delta_i$'s should
also vanish at order $\mathcal{O}(\omega_{1,2})$ due to the relation (\ref{4d-chemical-relation}). This will make the free energy to be small, $\sim\frac{\Delta^3}{\omega^2}\ll 1$, making
the index uninteresting. However, we keep finite
imaginary parts of $\Delta_i$'s while taking the limit $\omega_{1,2}\rightarrow 0$.
Physically, we take advantage of the possibility of tuning the phases of bosonic/fermionic
terms to maximally obstruct their cancelations. The asymptotic limit of
(\ref{4d-chemical-relation}) is $\Delta_1+\Delta_2+\Delta_3\approx 0$, so we take
all $\Delta_I$'s to be purely imaginary whose sum is zero, and
continue back to complex numbers later. The details of the approximation
is the same as we presented above. Following very similar procedures, againg taking out
the Cartan parts and ignoring them, one obtains
\begin{equation}\label{Li3-index}
  Z\sim\frac{1}{N!}\int\prod_{a=1}^N\frac{d\alpha_a}{2\pi}
  \exp\Bigg[-\frac{1}{\omega_1\omega_2}\sum_{s_1s_2s_3=+1}
  \sum_{a\neq b}\bigg(
  {\rm Li}_3\left(e^{\frac{s\cdot\Delta}{2}
  +i\alpha_{ab}}\right)-{\rm Li}_3\left(e^{-\frac{s\cdot\Delta}{2}-i\alpha_{ab}}
  \right)\bigg)\Bigg]\ .
\end{equation}
Here, note that ${\rm Li}_3(e^x)-{\rm Li}_3(e^{-x})=-\frac{(2\pi i)^3}{6}B_3(\frac{x}{2\pi i})$ for ${\rm Re}(x)\geq 0$ and $0<{\rm Im}(x)<2\pi$, with $B_3(x)=x^3-\frac{3}{2}x^2+\frac{1}{2}x$.
For $2\pi p<{\rm Im}(x)<2\pi(p+1)$ with an integer $p$, one finds
\begin{equation}
  {\rm Li}_3(e^x)-{\rm Li}_3(e^{-x})=-\frac{(2\pi i)^3}{6}
  B_3\left(\frac{x}{2\pi i}-p\right)
  =-\frac{(x\!-\!2\pi ip)^3}{6}+\frac{\pi i(x\!-\!2\pi ip)^2}{2}
  +\frac{\pi^2 (x\!-\!2\pi ip)}{3}\ .
\end{equation}
When the arguments are pure phase, one finds
\begin{equation}
  {\rm Li}_3(e^{ix})-{\rm Li}_3(e^{-ix})
  =i\left[\frac{(x\!-\!2\pi p)^3}{6}-\frac{\pi(x\!-\!2\pi p)^2}{2}
  +\frac{\pi^2 (x\!-\!2\pi p)}{3}\right]\equiv if(x)\ .
\end{equation}
in the interval $x\in(2\pi p,2\pi(p+1))$. $f$, defined piecewise as above, is
an odd function. Inserting $\Delta_I=i\, T_I$ with real $T_I$'s, one obtains
\begin{equation}\label{SYM-SCI}
  Z=\frac{1}{N!}\int\prod_{a=1}^N\frac{d\alpha_a}{2\pi}
  \exp\left[-\frac{i}{\omega_1\omega_2}\sum_{s_1s_2s_3=+1}\sum_{a\neq b}f\left(
  \frac{s\cdot T}{2}+\alpha_{ab}\right)\right]\ .
\end{equation}
Using $\sum_IT_I=0$, one can easily notice that
\begin{equation}
\mathcal{F}(\alpha_a)\equiv\sum_{s_1s_2s_3=+1} \sum_{a\neq b}f
\left(\frac{s\cdot T}{2} +\alpha_{ab}\right) = \sum_{a\neq b}^N
\left(f(\alpha_{ab})+\sum_{I=1}^3 f(T_I+\alpha_{ab})\right)
=\sum_{a\neq b} \sum_{I=1}^3 f(T_I+\alpha_{ab}).
\end{equation}
At the last step, we used the fact that $f$ is an odd function to set $f(\alpha_{ab})+f(\alpha_{ba})=0$. Since $\mathcal{F}(\alpha_a)$
is an even function in all $(\alpha_{ab})$'s, its derivatives with respect to all
$\alpha_a$'s vanish at the `completely deconfining configuration'
$\alpha_1=\alpha_2=\cdots=\alpha_N$ as before. 
Again assuming the dominance of this  saddle point,
one can asymptotically evaluate the integral \eqref{SYM-SCI} by the saddle point
method as
\begin{equation}
\log Z\sim - \frac{i}{\omega_1 \omega_2} \mathcal{F}(\alpha_1=\alpha_2=\cdots=\alpha_N)
=- \frac{iN^2}{\omega_1 \omega_2} \sum_{I=1}^3 f(T_I)\ .
\end{equation}

Without loss of generality, we now assume that
\begin{equation}
2\pi p_I<T_I<2\pi (p_I+1), \quad p_I \in \mathbb{Z}.
\end{equation}
Since $\sum_IT_I=0$ for the index, we find a constraint on $\sum_I p_I$ as
\begin{equation}
\begin{aligned}
2\pi \sum_{I=1}^3 p_I < \sum_{I=1}^3 T_I=0 < 2\pi (\sum_{I=1}^3 p_I +3)
\; \Rightarrow \; -3< \sum_{I=1}^3 p_I <0
\; \Rightarrow \; \sum_{I=1}^3 p_I = -1, -2.
\end{aligned}
\end{equation}
When $\sum_I p_I = -1$, one can show that
\begin{equation}
\sum_{I=1}^3 f(T_I)=\sum_{I=1}^3 f(T_I-2\pi p_i)
= \frac{1}{2}(T_1-2\pi p_1)(T_2-2\pi p_2)(T_3-2\pi p_3).
\end{equation}
On the other hand, when $\sum_I p_I = -2$, one will find that
\begin{equation}
\sum_{I=1}^3 f(T_I)=\sum_{I=1}^3 f(T_I-2\pi p_I) = \frac{1}{2}\big(T_1-2\pi (p_1+1)\big)\big(T_2-2\pi (p_2+1)\big)\big(T_3-2\pi (p_3+1)\big).
\end{equation}
These two results can be all expressed as the following equation,
\begin{equation}
\sum_{I=1}^3 f(T_I)=\frac{1}{2} \big(T_1+2\pi (1+p_2+p_3)\big)\big(T_2+2\pi (1+p_3+p_1)\big)\big(T_3+2\pi (1+p_1+p_2)\big).
\end{equation}
Therefore, one obtains
\begin{equation}
\log Z\sim - \frac{iN^2}{2\omega_1 \omega_2} \big(T_1+2\pi (1+p_2+p_3)\big)\big(T_2+2\pi (1+p_3+p_1)\big)\big(T_3+2\pi (1+p_1+p_2)\big)\ .
\end{equation}
Converting back to $\Delta_I=iT_I$, one obtains
\begin{equation}
  \log Z\sim\frac{N^2}{2\omega_1\omega_2}\prod_{I=1}^3(\Delta_I+2\pi i n_I)
\end{equation}
where $n_1\equiv 1+p_2+p_3$, $n_2\equiv 1+p_3+p_1$, $n_3\equiv 1+p_1+p_2$,
satisfying $\sum_{I=1}^3 n_I=\pm 1$. This agrees with the previous analysis,
supposing that $\Delta_1+\Delta_2+\Delta_3$ there and here are related by
a shift of $2\pi i$ (mod $4\pi i$).

\subsection{Background field analysis on $S^3$}

We consider an alternative approach to compute the asymptotic free energy of the index.
The chemical potentials $\beta$, $\omega_i$ are reflected in the background metric of
$S^3\times S^1$ as
\begin{equation}\label{4d-metric}
  ds^2=r^2\left[d\theta^2+\sum_{i=1}^2n_i^2\left(d\phi_i-\frac{i\omega_i}{\beta}d\tau
  \right)^2\right]+d\tau^2\ ,
\end{equation}
where $(n_1,n_2)=(\cos\theta,\sin\theta)$, $0\leq \theta\leq\frac{\pi}{2}$.
The Euclidean time $\tau$ has period $\tau\sim\tau+\beta$,
and we restored the radius $r$ of $S^3$.
$\Delta_I$ are encoded in the background
$U(1)^3\subset SO(6)$ gauge fields
\begin{equation}
  A^I=-\frac{i\Delta_I}{\beta}d\tau\ .
\end{equation}
The partition function is given by a path integral over the $\mathcal{N}=4$ Yang-Mills
fields at coupling constant $g_{\rm YM}$,
coupled to these background fields in a canonical manner.
Again having in mind imposing (\ref{BPS-hypersurface}) to get the index,
we take $\beta\rightarrow 0^+$. Very naively, one might think that
a Kaluza-Klein reduction to $S^3$ would be possible, integrating out heavy KK fields,
because the circle size $\beta$ is small. If one can integrate out the heavy fields,
they will contribute to an effective action of the
background fields, arranged in the derivative expansion which is a series in
small $\beta$. This will turn out to be a much subtler issue, because $\beta^{-1}$ appears
in other background fields. Indeed, naively doing the KK reduction, one would see shortly
that the 3d metric, dilaton and $U(1)^3$ fields all see inverse powers of $\beta$.
Still, when $\omega_{1,2}\ll 1$, we will show that the KK fields can be integrated
out, whose effect will be arranged in a derivative expansion. The expansion will be a series
in small $\beta,\omega_{1,2}$, whose leading terms will be given by Chern-Simons terms.
The effect of 3d zero modes is also expected to be subleading in our model. The analysis
is similar to \cite{DiPietro:2014bca}, except that our setting is subtler with
new aspects.

Having these in mind, we arrange the 4d background fields as 3d background fields.
To this end, we rewrite (\ref{4d-metric}) in terms of 3d metric,
gravi-photon $a$, and the dilaton $\Phi$ as
\begin{eqnarray}\label{background-3d}
  ds_4^2&=&r^2\left[d\theta^2+\sum_in_i^2d\phi_i^2
  +\frac{r^2(\sum_i\omega_in_i^2d\phi_i)^2}
  {\beta^2(1-r^2\sum_i\frac{n_i^2\omega_i^2}{\beta^2})}\right]
  +e^{-2\Phi}\left(d\tau+a\right)^2\equiv ds_3^2+e^{-2\Phi}(d\tau+a)^2\nonumber\\
  e^{-2\Phi}&=& 1-r^2\sum_i\frac{n_i^2\omega_i^2}{\beta^2}
  \ \ ,\ \ \ a=-i\frac{r^2\sum_i\omega_i n_i^2d\phi_i}
  {\beta(1-r^2\sum_i\frac{n_i^2\omega_i^2}{\beta^2})}\ .
\end{eqnarray}
The 4d $U(1)^3$ background fields $A^I$ are arranged to 3d gauge field $\mathcal{A}^I$
and the scalar $A_4^I$ as $A^I=A_4^I(d\tau+a)+\mathcal{A}^I$, where
\begin{equation}\label{background-R}
  A_4^I=-\frac{i\Delta^I}{\beta}\equiv\frac{\alpha^I}{\beta}\ ,\ \
  \mathcal{A}^I=-A_4^Ia\ .
\end{equation}
We take $\beta$ to be the smallest variable, eventually intending to take the limit
$\beta\rightarrow 0^+$. $\omega_i\ll 1$ are also small, but still satisfying
$\frac{\beta}{r\omega_i}\ll 1$. One might worry that some background fields may behave
badly due to the factor $1-r^2\sum_i\frac{n_i^2\omega_i^2}{\beta^2}$ in denominators.
We temporarily circumvent this issue by taking $\omega_i$ to be complex and generic,
evading the poles. Physically, this has to do with the fact that non-BPS derivatives'
effect is present before imposing (\ref{BPS-hypersurface}).

We first consider the limiting behaviors of the 3d background fields
for $\frac{\beta}{r}\ll|\omega_i|\ll 1$:
\begin{eqnarray}
  ds_3^2&\sim&r^2\left[ds^2(S^3_{\rm round})-
  \frac{(\sum_i\omega_in_i^2d\phi_i)^2}{\sum_in_i^2\omega_i^2}+\cdots\right]
  \nonumber\\
  \beta^2e^{-2\Phi}&\sim&-r^2\sum_in_i^2\omega_i^2+\cdots\ ,\ \
  \frac{a}{\beta}\sim\frac{i(\sum_i\omega_in_i^2d\phi_i)}{2\sum_in_i^2\omega_i^2}+\cdots\ .
\end{eqnarray}
The omitted terms $\cdots$ are suppressed by positive powers of
$\frac{\beta}{r\omega_i}\ll 1$. Note that in the 3d metric, one has a
canonical round sphere metric, accompanied by the second term
which is an $\mathcal{O}(1)$ negative length element along one direction.
For instance, if $\omega_1=\omega_2\equiv\omega$, this direction is the Hopf fiber
of $S^3$. Along this direction, leading $\mathcal{O}(1)$ length elements cancel
and its [length]$^2$ becomes smaller, at a positive power in $\frac{\beta}{\omega}$.
This is one reason why a naive KK reduction
becomes subtle in our case. The dilaton field $\beta^2 e^{-2\Phi}$ for the
[circumference]$^2$ of temporal circle is suppressed to be small $|\omega_i|\ll 1$,
which is an intuitive reason why we should also keep $\omega_i$ small
to trust the derivative expansion. The 3d background fields are
highly singular (e.g. $\omega,\beta$ dependence), presumably having short wavelength
components on $S^3$, so that one might wonder if the whole spirit of using derivative
expansion is relevant or not. In general, using these fields will be highly problematic in the
general effective field theory. For instance, if one wishes to make variation of this
effective action in background fields to generate correlation functions, this probably
might be tricky. However, our strategy here is very practical, having
in mind using this EFT just for our particular background. In other
words, we use it just as a way of expressing the series expansion of a particular
observable $\log Z$ in $\beta,\omega_1,\omega_2$. So no matter how singular
the fields may look, we just care about whether the actual values of terms
after spatial integrals are sequentially suppressed as an infinite series.
We will show (more precisely, strongly illustrate) that this is indeed true.

In this background, we consider the path integral of 4d $\mathcal{N}=4$ Yang-Mills
theory. We formally decompose the 4d dynamical fields into 3d `zero modes' and `KK fields,'
depending on the momentum mode on $S^1$. We schematically call the
zero modes $\Phi_L$ and KK modes $\Phi_H$, where $L$/$H$ stand for `light/heavy.'
$\Phi_H$ couples to the background field $a$, while $\Phi_L$ does not.
The path integral is done by integrating over $\Phi_H$ at fixed $\Phi_L$,
and then integrating over $\Phi_L$.

We discuss the structure of the path integral over $\Phi_H$,
at fixed $\Phi_L$. In our scaling limit of small $S^1$ radius, the path integral
over $\Phi_H$ gives an effective action that depends only on the
3d background fields, but not on $\Phi_L$ which are held fixed for the moment.
To see this, consider the schematic structure of the 3d action for $\Phi_H$.
It takes the form of
\begin{equation}
  \mathcal{L}\sim \Phi_H(\partial^2+M_{\rm KK}^2)\Phi_H
  +g_{\rm 3d}^2V(\Phi_L,\Phi_H)
\end{equation}
where $V$ denotes a potential quartic in $\Phi_H,\Phi_L$, with order $1$ coefficients.
Here we consider the case in which $\Phi_H$, $\Phi_L$ are bosonic, for simplicity.
Both $M_{\rm KK}$ and
$g_{\rm 3d}^2$ have dimension of mass, proportional to the inverse-radius of
the temporal circle $\sim\frac{1}{r\omega}$ (where $\omega\sim\omega_{1,2}$.)
The solution to $\Phi_H$ at given $\Phi_L$ is schematically
given by $\Phi_H\sim\frac{g_{\rm 3d}^2}{\partial^2+M_{\rm KK}^2}\partial_{\Phi_H}V$.
The propagator factor scales like $\frac{g_{\rm 3d}^2}{\partial^2+M_{\rm KK}^2}\lesssim
r\omega$, which suppresses the $\Phi_H$ tadpole and fluctuations depending on
$\Phi_L$.\footnote{We expect a caveat when $\Phi_L$ has zero
modes held at large value without a potential cost, making
$\partial_{\Phi_H}V$ large. There are two types of such modes,
again depending on the IR divergent behaviors of $Z_{S^3}$ for $\Phi_L$
\cite{Ardehali:2015bla}. In our 4d $U(N)$ theory, or 6d $(2,0)$ theory for $N$
M5-branes, we assume the absence of
such dangerous modes. See the next two paragraphs for more discussions.}
$\Phi_H$'s path integral is effectively
Gaussian, depending on background fields only. So after integrating out $\Phi_H$,
$Z$ consists of two factors: one given by the 3d background fields, and
another given by the path integral of `zero modes' $\Phi_L$ canonically coupled to
3d background fields, obtained by classical dimensional reduction of 4d $\mathcal{N}=4$
Yang-Mills theory. In the latter sector, the dilaton appears as the 3d coupling constant
(which may depend on spatial coordinate if $\omega_1\neq\omega_2$), while the gravi-photon
$\beta^{-1}a$ does not couple to the classical 3d Yang-Mills.

We first consider the factor coming from the path integral over $\Phi_L$. It consists
of the fields of 3d maximal super-Yang-Mills, whose action is deformed to be less
supersymmetric by various parameters. Here, we simply discuss how its contribution
to $\log Z$ will depend on various parameters. The 3d effective coupling is given by
$g_{\rm 3d}^2\sim\frac{1}{r\omega}$. The 3d metric consists of 2d base whose
length scale is $r$, and a fiber whose length scale is $\frac{\beta}{\omega}\ll r$.
As we shall see below from background effective actions (which is also obvious from
BPS kinematics), the leading free energy will be of order
$\sim\frac{\beta^0}{\omega^2}$ at $\frac{\beta}{r}\ll \omega\ll 1$. We can argue that
the path integral of $\Phi_L$ will yield much smaller terms than this. Suppose otherwise,
and the $\Phi_L$'s path integral contributes a term at this order.
Then, the divergent $\omega^{-2}$ part would come either from positive power in
the 3d gauge coupling $g_{\rm 3d}^2\sim \frac{g_{\rm YM}^2}{r\omega}$, or positive power in the
Hopf fiber radius $\sim(\beta\omega^{-1})^{\#}$. But acquiring this factor from the
Hopf fiber radius is accompanied by a positive power in $\beta$, which is subleading.
So $\beta^0\omega^{-2}$ dependence would come from the divergent
3d coupling, $g_{\rm 3d}^2\sim\omega^{-1}$.

However, it is also hard to imagine (probably
inconsistent) that a 3d QFT partition function diverges as the coupling grows, as the
3d QFT seems to be perfectly well defined. The only way in which we can imagine
a divergent dependence on large $g_{\rm 3d}$ is when the observable suffers from
infrared divergence, since $g_{\rm 3d}\rightarrow\infty$ is a sort of IR limit in 3d.
More concretely, the partition function of 3d maximal SYM on $S^3$ is well known to
have an IR divergence \cite{Kapustin:2010xq}. As studied in
\cite{DiPietro:2014bca,Ardehali:2015bla}, this is due to the $N$ gauge holonomies
of $U(N)$ on $S^1$ being non-compact in the small circle limit. At small but finite
circle radius, $\sim r\omega$, the holonomies have period given by $\sim\frac{1}{r\omega}$,
thus providing an IR cutoff. This would yield a factor of $\sim\omega^{-N}$ to $Z$,
contributing at a subleading order $\sim N\log\omega$ to the free energy.
Thus, we expect the divergent leading part $\propto\beta^0\omega^{-2}$ of the net
free energy to be unaffected by the 3d dynamical fields.

So it suffices to consider the effect of integrating out the `KK fields' $\Phi_H$,
yielding an effective action of $g_{\mu\nu}$, $a_\mu$, $\Phi$, $\mathcal{A}_\mu^I$,
$A^I_4$. There are infinitely many terms in this effective action, arranged in a
derivative expansion, whose coefficients are mostly unknown.
At generic points of
the background fields, before imposing the BPS index constraint (\ref{BPS-hypersurface}),
all fermions of the 4d theory will go to $\Phi_H$, due to the anti-periodic boundary
conditions. At (\ref{BPS-hypersurface}), some fermion modes may be massless.
Across this surface, as we shall see, these transiently
massless fermions at (\ref{BPS-hypersurface}) will simply change some Chern-Simons
coefficients, without further effects on the effective action. Below, we will show that:
(1) the derivative expansion is arranged in a series of $\beta,\omega_1,\omega_2$;
(2) the leading terms are at order $\frac{\beta^0}{\omega_1\omega_2}$, completely
coming from the Chern-Simons terms; (3) the Chern-Simons coefficients
can be determined either from the free 4d QFT, or by an anomaly consideration.
We shall discuss these issues in the order of $(3)\rightarrow(2)\rightarrow(1)$.

We first discuss possible Chern-Simons terms of $\mathcal{A}^I$, $a$. (One might also
think of the gravitational Chern-Simons term $\sim \omega \wedge R$. We think its
coefficient is zero, but anyway it will be subleading in our scaling limit, as illustrated
below.) There can be standard gauge-invariant Chern-Simons terms of the forms
\cite{Banerjee:2012iz,DiPietro:2014bca}
\begin{equation}\label{gauge-inv-CS}
  \beta^{-2}\int a\wedge da\ ,\ \
  \beta^{-1}\int\mathcal{A}^I\wedge da\ ,\ \
  \int\mathcal{A}^I\wedge d\mathcal{A}^J\ ,
\end{equation}
whose coefficients are dimensionless and quantized.
There can also be gauge non-invariant Chern-Simons terms which are needed
for anomaly matching \cite{Banerjee:2012iz,DiPietro:2014bca}. Since their coefficients
are all quantized, either from gauge invariance or anomaly matching,
one can determine them by integrating out KK fermions of the 4d QFT at weak coupling.

We follow \cite{DiPietro:2014bca} to compute these coefficients for
$U(1)^3\subset SO(6)$ times the gravi-photon $U(1)$.
There are four Weyl fermions $\Psi^{Q_1,Q_2,Q_3}_{\alpha}$,
where $\alpha=\pm\frac{1}{2}$, and
with $(Q_1,Q_2,Q_3)=(-,+,+)$, $(+,-,+)$, $(+,+,-)$, $(-,-,-)$. $\pm$'s for
$Q_I$'s denote $\pm\frac{1}{2}$.
The fermions with anti-periodic boundary conditions are labeled by
the Kaluza-Klein level $n\in\mathbb{Z}+\frac{1}{2}$.
The contributions to the Chern-Simons terms from the $n$'th KK modes
are given by \cite{DiPietro:2014bca}\footnote{The overall sign is chosen
to be consistent with our chirality/parity convention.}
\begin{equation}
  S_{\rm CS}^{(n)}=\frac{iN^2}{8\pi}\!\!
  \sum_{(Q_1,Q_2,Q_3)}\!\!\!{\rm sgn}\left(n\!-\!\frac{\beta}{2\pi}A_4^IQ_I\right)
  \int_{S^3}\left(\!Q_IQ_J\mathcal{A}^I\wedge d\mathcal{A}^J
  +2Q_I\frac{2\pi n}{\beta}\mathcal{A}^I\wedge da+\frac{(2\pi n)^2}{\beta^2}
  a\wedge da\!\right)\ .
\end{equation}
There are infinitely many contributions from the tower of KK modes, which
should be regularized. Following \cite{DiPietro:2014bca}, we sum over all
$n\in\mathbb{Z}+\frac{1}{2}$
using the zeta function regularization.\footnote{There are various proposals
for regularizing $Z[S^3\times S^1]$
\cite{DiPietro:2014bca,Assel:2014paa,Ardehali:2015hya,Assel:2015nca},
concerning the supersymmetric Casimir energy
\cite{Kim:2012ava,Bobev:2015kza,Cassani:2014zwa}. Employing the
regularization of \cite{DiPietro:2014bca}, we obtain a free energy unspoiled by
the formal Casimir energy factor of \cite{Bobev:2015kza}. Although we have no clear
reasoning for this, note that Casimir energy is very sensitive to regularization,
while the integral spectrum part should be more robust. Especially, our setup
respects all the periodicities of holonomies, which is a property of the spectral
part of $\log Z$ but not of the Casimir energy \cite{Bobev:2015kza}. So our
regularization appears
to disallow a room for vacuum energy factor like \cite{Bobev:2015kza}.}
To start with, when $-\frac{1}{2}<\mu\equiv\frac{\beta}{2\pi}Q_IA^I_4<\frac{1}{2}$
for a fermion mode with given $Q_I$, one obtains \cite{DiPietro:2014bca}
\begin{equation}
  \sum_{n}{\rm sgn}(n-\mu)\sim 2\mu\ ,\ \
  \sum_n{\rm sgn}(n-\mu)n\sim\mu^2+\frac{1}{12}\ ,\ \
  \sum_n{\rm sgn}(n-\mu)n^2\sim\frac{2}{3}\mu^3\ .
\end{equation}
If $A_4^I$'s are chosen so that $\frac{\beta}{2\pi}Q_IA^I_4$ is in the
range $(-\frac{1}{2},\frac{1}{2})$ for all possible $Q_I$'s, one obtains
\begin{eqnarray}\label{CS-non-invariant}
  S_{\rm CS}\!&\!=\!&\!\frac{iN^2}{4\pi}\!\sum_{(Q_1,Q_2,Q_3)}\int
  \left[\frac{\beta}{2\pi}Q_IQ_JQ_KA^I_4\mathcal{A}^J\wedge d\mathcal{A}^K
  +\frac{2\pi}{\beta}Q_I\!\left(Q_JQ_K\frac{\beta^2}{(2\pi)^2}A_4^JA_4^K+\frac{1}{12}\right)
  \mathcal{A}^I\wedge da\right.\nonumber\\
  &&\hspace{3.3cm}\left.+\frac{\beta}{3\cdot 2\pi}Q_IQ_JQ_KA_4^IA_4^JA_4^K
  a\wedge da\right]\ .
\end{eqnarray}
Here, note that
\begin{equation}
  \sum_{(Q_1,Q_2,Q_3)}Q_IQ_JQ_K=-\frac{1}{2}C_{IJK}\ ,\ \
  \sum_{(Q_1,Q_2,Q_3)}Q_I=0\ ,
\end{equation}
where $C_{IJK}$ is symmetric in $I,J,K$, $C_{123}=1$, and $C_{IJK}=0$ if any two
of $I,J,K$ are same. (These are the anomaly coefficients of $U(1)^3$.) Using
these facts, one obtains
\begin{equation}\label{CS-canonical}
  S_{\rm CS}=-\frac{iN^2}{8\pi}\cdot\frac{\beta}{2\pi}\int_{S^3}
  C_{IJK}\left(A_4^I\mathcal{A}^J\wedge d\mathcal{A}^K+A_4^IA_4^J
  \mathcal{A}^K\wedge da+\frac{1}{3}A_4^IA_4^JA_4^Ka\wedge da\right)\ .
\end{equation}
Note that the gauge invariant Chern-Simons terms (\ref{gauge-inv-CS}) are all
zero in this chamber, with
$-\frac{1}{2}\leq\frac{\beta}{4\pi}(\pm A_4^1\pm A_4^2\pm A_4^3)\leq\frac{1}{2}$
for all four possible sign choices satisfying $\pm\cdot\pm\cdot\pm=-1$.

In general chambers of $A_4^I$, one takes
\begin{equation}
  -\frac{1}{2}+p_Q\leq\frac{\beta}{2\pi}Q_IA_4^I\leq\frac{1}{2}+p_Q\ ,
\end{equation}
where $Q$ runs over $4$ possible cases, with integral $p_Q$'s.
In this chamber, the regularized sums are now given by
\begin{eqnarray}
  \sum_{n}{\rm sgn}(n-\mu)\!&\!=\!&\!\sum_{n^\prime}{\rm sgn}(n^\prime-\mu^\prime)
  \sim 2(\mu-p)\\
  \sum_n{\rm sgn}(n-\mu)n\!&\!=\!&\!\sum_{n^\prime}{\rm sgn}(n^\prime-\mu^\prime)
  (n^\prime+p)
  \sim(\mu-p)^2+\frac{1}{12}+2p(\mu-p)\nonumber\\
  \sum_n{\rm sgn}(n-\mu)n^2\!&\!=\!&\!
  \sum_{n^\prime}{\rm sgn}(n^\prime-\mu^\prime)((n^\prime)^2+2pn^\prime+p^2)
  \sim\frac{2}{3}(\mu-p)^3+2p(\mu-p)^2+\frac{p}{6}+
  2p^2(\mu-p)\ ,\nonumber
\end{eqnarray}
where $n^\prime=n-p$, $\mu^\prime=\mu-p$.
In this chamber, one obtains
\pagebreak
\begin{eqnarray}\label{CS-general}
  \hspace*{-1.5cm}&&S_{\rm CS}=\frac{iN^2}{4\pi}\cdot\frac{\beta}{2\pi}\sum_{(Q_1,Q_2,Q_3)}\int
  \left[\left(Q_IA_4^I-\frac{2\pi p_Q}{\beta}\right)
  Q_JQ_K\mathcal{A}^J\wedge d\mathcal{A}^K\right.\\
  \hspace*{-1.5cm}&&\left.+Q_I\left(\!\left(Q\cdot A_4-\frac{2\pi p_Q}{\beta}\right)^2
  +\frac{1}{12}\cdot\frac{(2\pi)^2}{\beta^2}+2p_Q\cdot\frac{2\pi}{\beta}
  \left(Q\cdot A_4-\frac{2\pi p_Q}{\beta}\right)\!\right)
  \mathcal{A}^I\wedge da\right.\nonumber\\
  \hspace*{-1.5cm}&&\left.+\left(\frac{1}{3}\left(Q\cdot A_4-\frac{2\pi p_Q}{\beta}\right)^3
  +\frac{2\pi p_Q}{\beta}\left(Q\cdot A_4-\frac{2\pi p_Q}{\beta}\right)^2
  +\frac{(2\pi p_Q)^2}{\beta^2}\left(Q\cdot A_4-\frac{2\pi p_Q}{\beta}\right)
  +\frac{p_Q}{12}\cdot\frac{(2\pi)^3}{\beta^3}\right)
  a\wedge da
  \right].\nonumber
\end{eqnarray}
We shall mostly work with the result (\ref{CS-canonical}) in the canonical chamber.

One can also determine (\ref{CS-canonical}) by just knowing 't Hooft anomalies
and discrete symmetries. Firstly, the gauge non-invariant terms (\ref{CS-canonical})
are completely fixed in \cite{Banerjee:2012iz,DiPietro:2014bca},
by demanding that its gauge variation yields the expected 't Hooft anomaly of the
4d $U(1)^3\subset SO(6)_R$ symmetry. (More precisely, (\ref{CS-canonical}) matches the
covariant anomalies.) To complete the argument, we discuss why gauge invariant CS terms
(\ref{gauge-inv-CS}) should vanish. Firstly, $a\wedge da$ is forbidden
by the 3d parity after $S^1$ reduction, which is a symmetry of the mother
4d theory if an object is blind to $SO(6)_R$, such as $a\wedge da$.
Similarly, $\mathcal{A}^I\wedge d\mathcal{A}^I$ with a given $I$ is forbidden
since the mother 4d $\mathcal{N}=4$ theory is invariant under parity with
sign flip of odd number of $\mathcal{A}^I$ fields. The latter flip is charge
conjugation, flipping ${\bf 4}\leftrightarrow\overline{\bf 4}$.
The remaining gauge invariant CS terms are forbidden simply from
the Weyl symmetry of $SO(6)$. We consider the Weyl reflections which reflects
two of the three $\mathcal{A}^I$'s, leaving one invariant. This reflection also
acts on $A^I_4$. But they cannot affect the gauge invariant CS terms, so in the canonical
chamber which is left invariant under these reflections, the gauge invariant CS terms
should respect this symmetry. For $\mathcal{A}^I\wedge da$ with any given $I$,
a reflection which flips $I$ and another $J(\neq I)$ flips sign of this term, forbidding
its generation. Similarly, for $\mathcal{A}^I\wedge d\mathcal{A}^J$ at given
pair $I\neq J$, reflection of $I$ and $K(\neq I,J)$ forbids its generation.
This completes a symmetry-based argument for (\ref{CS-canonical}).
Such an approach may be useful for some non-Lagrangian theories, if there
are enough discrete symmetries. In section 4, we shall make similar studies with
6d $(2,0)$ theory, although it appears that such intrinsic arguments are
less predictive there.

We now evaluate these CS terms for our backgrond fields, in
the canonical chamber. We first consider the background R-symmetry fields
(\ref{background-R}) with real $\alpha^I=-i\Delta^I$, and
later continue to complex $\Delta^I$. Also, we keep $\epsilon_i\equiv -i\omega_i$
real for a moment, and
later continue to complex $\omega_i$. (\ref{CS-canonical}) is given by
\begin{equation}
  S_{\rm CS}=-\frac{iN^2}{48\pi^2 \beta^2}C_{IJK}\alpha^I\alpha^J\alpha^K
  \int_{S^3}a\wedge da\ .
\end{equation}
Inserting $a$ in (\ref{background-3d}), one finds
\begin{eqnarray}
  \int a\wedge da&=&\frac{r^4}{\beta^2}\int\frac{\epsilon_in_i^2d\phi_i\wedge
  \epsilon_jd(n_j^2)\wedge d\phi_j}
  {\left(1+\frac{r^2n_i^2\epsilon_i^2}{\beta^2}\right)^2}=
  \frac{(2\pi)^2r^4\epsilon_1\epsilon_2}{\beta^2}\int\frac{ydx-xdy}
  {\left(1+\frac{r^2(\epsilon_1^2x+\epsilon_2^2y)}{\beta^2}\right)^2}\\
  &=&\frac{(2\pi)^2r^4\epsilon_1\epsilon_2}{\beta^2}\int_0^1\frac{dx}
  {\left(1+\frac{r^2}{\beta^2}(\epsilon_2^2+(\epsilon_1^2-\epsilon_2^2)x)\right)^2}
  =\frac{(2\pi)^2r^4\epsilon_1\epsilon_2}
  {\beta^2(1+\frac{r^2\epsilon_1^2}{\beta^2})(1+\frac{r^2\epsilon_2^2}{\beta^2})}
  \ ,\nonumber
\end{eqnarray}
where $x\equiv n_1^2$, $y\equiv n_2^2=1-x$. So one finds
\begin{equation}\label{CS-evaluation}
  S_{\rm CS}=-\frac{iN^2r^4\epsilon_1\epsilon_2}{12\beta^4
  (1+\frac{r^2\epsilon_1^2}{\beta^2})(1+\frac{r^2\epsilon_2^2}{\beta^2})}
  C_{IJK}\alpha^I\alpha^J\alpha^K
\end{equation}
in the canonical chamber. Inserting $\alpha^I=-i\Delta^I$,
$\epsilon_i=-i\omega_i$ and taking $\beta\rightarrow 0^+$,
one obtains
\begin{equation}\label{CS-final}
  S_{\rm CS}\rightarrow-\frac{N^2C_{IJK}\Delta^I\Delta^I\Delta^K}{12\omega_1\omega_2}
  =-\frac{N^2\Delta_1\Delta_2\Delta_3}{2\omega_1\omega_2}
\end{equation}
in the canonical chamber. If $S_{CS}$ is the dominant term in the effective action
(which we will show shortly), this yields the asymptotic free energy by
the relation $Z\sim e^{-S_{\rm CS}}$. So $\log Z\sim -S_{\rm CS}$ completely
agrees with the free QFT analysis in the previous subsection. The extension of this
result to different chambers also agrees with the result from free QFT.

Now to complete the analysis of the free energy, we show that all the other
terms in the effective action are subleading in our scaling limit, suppressed by
small $\beta,\omega_{1,2}$. The background fields are the 3d metric $g_{\mu\nu}$,
dilaton $\Phi$, graviphoton $a_{\mu}$, gauge boson $\mathcal{A}_\mu^I$, and scalar
$A_{4}^I$. Greek indices run over the coordinates $\{\phi_1, \phi_2, \theta\}$,
and small Latin indices used below will run over the locally flat coordinates
$\{1,2,3\}$. There are rich possibilities in constructing the effective action. However, many possible terms are eliminated by taking into account the actual background value
(\ref{background-3d}) and (\ref{background-R}). First, the Riemann curvature $\mathcal{R}_{\mu\nu\rho\sigma}$ has non-zero components only at $\{\mu,\nu\}=\{\rho,\sigma \}$ or $\{\mu,\nu\} \cap \{\rho,\sigma\} = \{ \theta \}$. Second, the background value (\ref{background-3d}) and (\ref{background-R}) depends only on the $\theta$ coordinate,
so that the field strengths $\mathcal{F}^0_{\mu\nu} \equiv \frac{1}{2\beta} (\partial_{\mu} a_\nu - \partial_{\nu} a_\mu)$ and $\mathcal{F}_{\mu\nu}^I \equiv \frac{1}{2} (\partial_{\mu} \mathcal{A}^I_\nu - \partial_{\nu} \mathcal{A}^I_\mu)$ of the graviphoton $a_{\mu}$ and gauge field $\mathcal{A}_\mu^I$ have non-zero components only at $\{\mu, \nu\} \supset \{\theta\}$. For the same reason, the derivative of any scalar function of the background fields $\partial_\mu f(\omega_{\rho}{}^{ab}, \Phi, a_{\rho}, \mathcal{A}_\rho^I, A_{4}^I)$ can have non-zero components only at $\{\mu\} = \{\theta\}$.
Third, the graviphoton $a_{\mu}$ and gauge field $\mathcal{A}_\mu^I$ have non-zero components only at $\{\mu \} \not\supset \{\theta\}$.
We will further assume that $\omega_1 = \omega_2 \equiv \omega$ for simplification, so that the dilaton $\Phi$ becomes a constant.

Let us first examine the possible terms that involve the volume integral $\int d^3x \sqrt{g}$ of gauge-invariant Lagrangian densities, formed by contracting tensors without $\epsilon^{\mu\nu\rho}$. When we consider the scalar contraction between the curvature $\mathcal{R}_{\mu\nu\rho\sigma}$ and the field strength $\mathcal{F}^0_{\mu\nu}$ or $\mathcal{F}_{\mu\nu}^I$, only an \emph{even} number of $\mathcal{F}^0_{\mu\nu} $ or $\mathcal{F}_{\mu\nu}^I$ can appear in the non-vanishing Lagrangian densities. It can be shown as follows: the scalar contraction of $\mathcal{R}_{\mu\nu\rho\sigma}$, $\mathcal{F}^0_{\mu\nu} $, $\mathcal{F}_{\mu\nu}^I$ can be encoded in the circular sequence of antisymmetric pairs of tensor indices $[\alpha\beta][\gamma\delta]\cdots[\zeta\alpha]$, where adjacent indices in adjoining pairs are contracted to each other. We distinguish the curvature tensor indices by using capital letters. Then the contraction to a Lorentz scalar can be generally written as
\begin{align*}
	[\alpha_{1,1}\beta_{1,1}]\cdots [\alpha_{1,n_{1}}\beta_{1,n_{1}}] [A_1 B_1][\alpha_{2,1}\beta_{2,1}]\cdots [\alpha_{2,n_{2}}\beta_{2,n_{2}}] [A_2 B_2] \cdots [A_{2j} B_{2j}] \quad \text{ with }\quad \textstyle\sum_{i=1}^{2j}n_i  \in 2\mathbb{Z}+1.
\end{align*}
The set of the field strength indices $\{\alpha_{k,1}, \beta_{k,n_k}\}$ in $[\alpha_{k,1}\beta_{k,1}]\cdots [\alpha_{k,n_{k}}\beta_{k,n_{k}}]$ can only be either
\begin{align}
	\{a_{k,1}, b_{k,n_k}\} = \begin{dcases}
 \{\phi_1,\theta\} \text{ or } \{\phi_2, \theta\} & \text{if } n_k \in 2\mathbb{Z}+1  \\
  \{\theta\} \text{ or } \{\phi_1, \phi_2\} \text{ or } \{\phi_1\} \text{ or } \{\phi_2\} & \text{if } n_k \in 2\mathbb{Z}.
 \end{dcases}
\end{align}
Collecting the sets of the curvature indices $\{A_{k},B_k\}$ for $k=1,\cdots,\,2j$, there are always an odd number of $\{\phi_1,\theta\}$ or $\{\phi_2, \theta\}$ and an odd number of $\{\phi_1,\phi_2\}$. Any complete pairings in this collection have at least one pair between $\{\phi_1,\phi_2\}$ and  $\{\phi_{1,2},\theta\}$, so each term in the contraction refers to $\mathcal{R}_{\phi_1\phi_2\phi_{i}\theta} = 0$. This exhausts many possible terms in the effective action. Here we evaluate and list all non-vanishing terms which involve up to 4 derivatives:  (Below we assume $I,J,K,L$ run over $0,1,2,3$, and $\Delta^0 \equiv -i$.)
{\allowdisplaybreaks
\begin{align}
\label{eq:s3action}
   \frac{1}{(2\pi)^2}\int\beta^{-3} e^{3\Phi} \sqrt{g} & = \frac{\beta r^3 }{2 (\beta ^2-r^2 \omega ^2)^2} =  \frac{\beta }{2 r \omega ^4}+ \mathcal{O}\left(\frac{\beta^3}{r^3 \omega^6}\right)\\
   \frac{1}{(2\pi)^2}\int\beta^{-1} e^{\Phi} \sqrt{g} \,\mathcal{R}^{ab}{}_{ab} & =\frac{r (3 \beta ^3-4 \beta  r^2 \omega ^2)}{(\beta ^2-r^2 \omega ^2)^2} =  -\frac{4 \beta }{r \omega ^2}+ \mathcal{O}\left(\frac{\beta^3}{r^3 \omega^4}\right) \nonumber \\
   \frac{1}{(2\pi)^2} \int\beta e^{-\Phi} \sqrt{g} \,\mathcal{F}_{ab}^I \mathcal{F}_{ab}^J &= \frac{\beta  \Delta ^I \Delta ^J r^3 \omega ^2}{(\beta ^2-r^2 \omega ^2)^2} = \frac{\beta \Delta^I \Delta^J }{r \omega^2}+ \mathcal{O}\left(\frac{\beta^3}{r^3 \omega^4}\right) \nonumber\\
 \frac{1}{(2\pi)^2}\int\beta^3 e^{-3\Phi} \sqrt{g} \,(\nabla_c \mathcal{F}^I_{ab})(\nabla^c\mathcal{F}^{Jab}) &=  \frac{2 \beta ^3 r \omega ^2 \Delta ^{I}\Delta ^{J}}{(\beta ^2-r^2 \omega ^2)^2} =\frac{2 \beta ^3  \Delta ^{I}\Delta ^{J}}{r^3 \omega ^2} +\mathcal{O}\left(\frac{\beta^5}{r^5 \omega^4}\right)\nonumber\\
  \frac{1}{(2\pi)^2}\int\beta^3 e^{-3\Phi} \sqrt{g} \,(\nabla_c \mathcal{F}^I_{ab})(\nabla^a \mathcal{F}^{Jcb}) &=  \frac{\beta ^3 r \omega ^2 \Delta ^{I}\Delta ^{J}}{(\beta ^2-r^2 \omega ^2)^2} =\frac{\beta ^3  \Delta ^{I}\Delta ^{J}}{r^3 \omega ^2} +\mathcal{O}\left(\frac{\beta^5}{r^5 \omega^4}\right)\nonumber\\
\frac{1}{(2\pi)^2}\int\beta e^{-\Phi} \sqrt{g} \,\mathcal{R}^{ab}{}_{a}{}^c \mathcal{R}_b{}^d{}_{cd}&=\frac{2 \beta   (8 r^4 \omega ^4-8 \beta ^2 r^2 \omega ^2 +3 \beta ^4 )}{r (\beta ^2 -r^2  \omega ^2)^2} = \frac{16 \beta  }{r}+\mathcal{O}\left( \frac{\beta^3}{r^3 \omega^2}\right)\nonumber\\
\frac{1}{(2\pi)^2}\int\beta e^{-\Phi} \sqrt{g} \,\mathcal{R}_{abcd}\mathcal{R}^{abcd}& = \frac{32 \beta  r^4 \omega ^4 -16 \beta ^3 r^2 \omega ^2 +6 \beta ^5 }{r (\beta ^2 -r^2  \omega ^2)^2} = \frac{32 \beta  }{r}+\mathcal{O}\left( \frac{\beta^3}{r^3 \omega^2}\right)\nonumber\\
\frac{1}{(2\pi)^2}\int\beta^3 e^{-3\Phi} \sqrt{g} \,\mathcal{F}^{Iab} \mathcal{F}_{a}^J{}^{c} \mathcal{R}_b{}^d{}_{cd} &=\frac{2 \beta  \Delta^I\Delta^J r \omega ^2 (\beta ^2-2 r^2 \omega ^2)}{(\beta ^2-r^2 \omega ^2)^2}
    = -\frac{4 \beta  \Delta^I \Delta^J }{r} + \mathcal{O}\left( \frac{\beta^3}{r^3 \omega^2}\right)\nonumber\\
\frac{1}{(2\pi)^2}\int\beta^3 e^{-3\Phi} \sqrt{g} \,\mathcal{F}^I_{ab}\mathcal{F}^J_{cd}\mathcal{R}^{abcd} &= \frac{2 \beta  \Delta ^I \Delta ^J r \omega ^2 (\beta ^2-4 r^2 \omega ^2)}{(\beta ^2-r^2 \omega ^2)^2} = -\frac{8 \beta  \Delta^I \Delta^J }{r} + \mathcal{O}\left( \frac{\beta^3}{r^3 \omega^2}\right)\nonumber\\
\frac{1}{(2\pi)^2}\int\beta^5 e^{-5\Phi} \sqrt{g} \,\mathcal{F}^{Iab} \mathcal{F}^J_{a}{}^c \mathcal{F}^K_b{}^d\mathcal{F}^L_{cd} &= \frac{\beta  \Delta^I \Delta^J \Delta^K \Delta^L r^3 \omega ^4 }{(\beta ^2 -r^2  \omega ^2)^2} = \frac{\beta  \Delta^I \Delta^J \Delta^K \Delta^L }{r} + \mathcal{O}\left( \frac{\beta^3}{r^3 \omega^2}\right) \nonumber\\
\frac{1}{(2\pi)^2}\int\beta^5 e^{-5\Phi} \sqrt{g} \,\mathcal{F}^{Iab} \mathcal{F}^J_{ab} \mathcal{F}^{Kcd} \mathcal{F}^L_{cd} &= \frac{2 \beta  \Delta^I \Delta^J \Delta^K \Delta^L r^3 \omega ^4 }{(\beta ^2 -r^2  \omega ^2)^2} = \frac{2 \beta \Delta^I \Delta^J \Delta^K \Delta^L }{r} + \mathcal{O}\left( \frac{\beta^3}{r^3 \omega^2}\right)\ . \nonumber
\end{align}
These terms are all much smaller than (\ref{CS-evaluation}) in the scaling limit
${\beta}/{r} \ll \omega \ll 1$.} Extrapolating a pattern from the above terms, an action made of $n_1$ curvature tensors, $n_2$ graviphoton field strengths, $n_3$ background $U(1) \subset SO(6)$ field strengths, and $n_4$ derivatives should behave as
\begin{align}
\label{eq:scale1}
\frac{\beta^{1+ n_4} \Delta^{n_3}}{r^{1+ n_4}\omega^{4-2n_1 -n_2- n_3}}  + \mathcal{O}\left( \frac{\beta^{3 + n_4}}{r^{3 + n_4} \omega^{6-2n_1 -n_2- n_3}}\right)
\end{align}
which would be suppressed in the limit ${\beta}/{r} \ll  \omega \ll 1$.

As a next step, we turn to the effective action that contains a totally antisymmetric tensor
$\epsilon^{\mu\nu\rho}$. This consists of Chern-Simons terms and those terms associated with
a gauge invariant Lagrangian density. We can further distinguish the gauge non-invariant
Chern-Simons terms from the gauge invariant ones. The gauge non-invariant Chern-Simons terms
are entirely dictated by the chiral anomaly, so that no other terms than
(\ref{CS-non-invariant}) can arise \cite{Banerjee:2012iz,DiPietro:2014bca}. And also, the gauge invariant Chern-Simons terms displayed in (\ref{gauge-inv-CS}) are already shown to be absent
in the canonical chamber.
The gravitational Chern-Simons term $\text{tr}\,(\omega \wedge \mathcal{R} + \frac{2}{3} \omega\wedge\omega\wedge\omega)$, even if present, makes only a sub-dominant contribution in the limit  ${\beta}/{r} \ll \omega \ll 1$:
\begin{align}
	&\frac{1}{3!} \frac{1}{(2\pi)^2}\int  \epsilon^{\mu\nu\rho} \left(\omega_{\mu}{}^{ab} \mathcal{R}_{\nu\rho}{}^{ab} + \frac{2}{3} \omega_{\mu}{}^{ab}\omega_{\nu}{}^{bc}\omega_{\rho}{}^{ca } \right)  = -\frac{4 \beta ^2}{r^2 \omega^2} + \mathcal{O}\left(\frac{\beta^4}{r^4 \omega^4}\right).
\end{align}
Other gauge invariant Lagrangian densities containing $\epsilon^{\mu\nu\rho}$ are constrained by the symmetry-based argument, which was used to argue the gauge invariant CS terms \eqref{gauge-inv-CS} are absent. Each allowed term should have odd numbers of three different $U(1) \subset SO(6)$ field strengths $\mathcal{F}^{1,2,3}_{\mu\nu}$. So even a minimal term of this sort has 3 $U(1)^3 \subset SO(6)$ field strengths coupled to one another. Some non-vanishing sample terms are evaluated and displayed below:
\begin{align}
	\int \frac{\beta^6e^{-6\Phi}}{3!\,(2\pi)^2}\ \epsilon^{\mu\nu\rho} \mathcal{F}^1_{\mu\nu} (\nabla_{\alpha}\mathcal{F}_{\rho\beta}^2) \mathcal{F}^{3\alpha\sigma}\mathcal{F}^0_{\sigma}{}^\beta &= -\frac{i \beta ^2 r^2 \omega ^4 \Delta_1 \Delta_2\Delta_3}{3(\beta ^2-r^2 \omega ^2)^{2}} = -\frac{ i\beta ^2 \Delta_1 \Delta_2\Delta_3}{3r^2 }+\mathcal{O}\left(\frac{\beta ^{4}}{r^{4} \omega^2}\right)\nonumber\\
	\int \frac{\beta^{10} e^{-10\Phi}}{3!\,(2\pi)^2} \  \epsilon^{\mu\nu\rho} \mathcal{F}^1_{\mu\alpha} \mathcal{F}^1_{\nu\beta}\mathcal{F}^{1\alpha\beta} (\nabla_{\lambda}\mathcal{F}_{\rho\sigma}^2) \mathcal{F}^{3\lambda\delta}\mathcal{F}^0_{\delta}{}^\sigma &=-\frac{i \beta ^{2} r^2 \omega ^6 \Delta_1^3 \Delta_2\Delta_3}{3(\beta ^2-r^2 \omega ^2)^{2}}
	= -\frac{ i\beta ^2 \omega^2\Delta_1 \Delta_2\Delta_3 }{3r^2 }  +\mathcal{O}\left(\frac{\beta ^{4} }{r^{4}} \right)
\end{align}
Notice that these leading corrections exhibit the same scaling behavior as \eqref{eq:scale1}.
In any case, all these terms become sub-dominant in the limit ${\beta}/{r} \ll \omega \ll 1$. One can probably make a systematic proof of this statement, but we content ourselves here by illustrating the suppressions. This establishes our claimed result (\ref{CS-final}), rederived from an effective action approach.

\subsection{AdS$_5$ black holes}

In this subsection, we make a Legendre transformation of the free energy
(\ref{free-energy-final}) to the microcanonical ensemble, as the macroscopic saddle point
approximation of the inverse Laplace
transformation. One should extremize the following entropy function
\begin{equation}\label{entropy-function}
  S(\Delta_I,\omega_i;Q_I,J_i)=
  \frac{N^2}{2}\frac{\Delta_1\Delta_2\Delta_3}{\omega_1\omega_2}
  +\sum_{I=1}^3Q_I\Delta_I+\sum_{i=1}^2J_i\omega_i\ .
\end{equation}
Since this free energy is reliable only at one of the surfaces (\ref{BPS-hypersurface}),
we make variation with $4$ independent variables, which couples to four combinations of
$5$ charges. This is our ignorance due to restricting considerations to the
index. We consider the surface
\begin{equation}\label{index-constraint}
  \Delta_1+\Delta_2+\Delta_3-\omega_1-\omega_2=2\pi i\ ,
\end{equation}
for BPS states saturating $E\geq Q_1+Q_2+Q_3+J_1+J_2$. Extremization of \eqref{entropy-function} under \eqref{index-constraint} was first proposed in \cite{Hosseini:2017mds}.

Let us first make a basic consideration on what this extremization does. Although
the entropy function (\ref{entropy-function}) has real coefficients only, it should
have complex solutions for $\Delta_I,\omega_i$ due to the constraint
(\ref{index-constraint}). During the extremization, we will be led to distribute
$2\pi i$ on the right hand side suitably to the $5$ chemical potentials.
We assert that one should pay attention to nontrivial distribution of this phase
to the fugacities.  Allowing nontrivial imaginary parts of
$\Delta_I,\omega_i$ (mod $2\pi i$) satisfying (\ref{index-constraint}),
one can hope to reduce unnecessary boson/fermion cancelations in
the index. Namely, we insert $(-1)^F$ in the index
because we want pairs of bosonic/fermionic states related by $\mathcal{Q},\mathcal{S}$
to cancel. If the index does not acquire contributions from such states, it can be
computed at any coupling constant. However, inserting $-1$ factor to all fermions,
it may cause unnecessary cancelations between bosonic/fermionic states which
are not superpartners of each other. So as long as it is allowed by
(\ref{index-constraint}), we attempt to insert extra phase factor $e^{-i\varphi}$
for each state, defined by $e^{-i{\rm Im}(\Delta_I Q_I+\omega_i J_i)}\equiv
(-1)^Fe^{-i\varphi}$, trying to maximally
obstruct cancelations. Converting to microscopic ensemble at definite charges,
the `entropy' is counted with such phase factor inserted for each state:
\begin{equation}\label{phase-sum}
  e^{S(Q_I,J_i)}\sim\sum_{B}e^{-i\varphi_B}-\sum_Fe^{-i\varphi_F}=
  \sum_{B}e^{-i\varphi_B}+\sum_Fe^{-i(\varphi_F+\pi)}\ .
\end{equation}
Morally, the real parts of chemical potentials are extremized to tune the system
to definite charges in the microscopic ensemble, while imaginary parts are tuned
to make (\ref{phase-sum}) maximally unobstructed. However, the two extremizations
are intertwined, so that both real and imaginary parts participate in both processes.
If one is lucky so that all phases $\varphi_B,\varphi_F+\pi$ at a saddle point
are same (mod $2\pi$) for all microstates, then ${\rm Re}(S)$ of the index would be
the true BPS entropy.
In the unlucky case that one cannot make all these phases collinear, ${\rm Re}(S)$ would be smaller than the entropy.
In any case, ${\rm Re}(S)$ computed from our index sets a lower bound on the true
entropy, and there is no a priori way of knowing when this bound saturates the true
entropy. In particular, there seems to be no a priori reason to care about ${\rm Im}(S)$,
as the saturation may happen or not irrespective of whether ${\rm Im}(S)$ assumes
a specific value.
With this in mind, we consider the extremization of (\ref{entropy-function}).

This extremization problem was considered in \cite{Hosseini:2017mds}.
Below, we shall be essentially reviewing the calculations of
\cite{Hosseini:2017mds}, however employing our viewpoints stated above,
and hopefully making some calculus more explicit and transparent.

One first solves the constraint $\sum_I \Delta_I-\sum_i\omega_i=2\pi i$ by
the following parametrization:
\begin{equation}\label{solve-chemical-constraint}
  \Delta_I=\frac{2\pi iz_I}{1+z_1+z_2+z_3+z_4}\ ,\ \
  \omega_1=-\frac{2\pi i z_4}{1+z_1+z_2+z_3+z_4}\ ,\ \
  \omega_2=-\frac{2\pi i}{1+z_1+z_2+z_3+z_4}\ .
\end{equation}
Now $z_{1,2,3,4}$ are unconstrained variables.
Extremization in $z_1$ yields
\begin{equation}\label{z123}
  \frac{N^2}{2}\frac{\Delta_1\Delta_2\Delta_3}{\omega_1\omega_2}
  +Q_I\Delta_I+J_i\omega_i=
  \pi i N^2\frac{\Delta_2\Delta_3}{\omega_1\omega_2}+2\pi i Q_1\ ,
\end{equation}
while one obtains similar equations with cyclic permutations of three $(Q_I,\Delta_I)$,
to get the extremization conditions for $z_2,z_3$.
Extremization in $z_4$ yields
\begin{equation}\label{z4}
  \frac{N^2}{2}\frac{\Delta_1\Delta_2\Delta_3}{\omega_1\omega_2}
  +Q_I\Delta_I+J_i\omega_i=
  \pi i N^2\frac{\Delta_1\Delta_2\Delta_3}{\omega_1^2\omega_2}-2\pi i J_1\ .
\end{equation}
Inserting $\omega_1=\sum_I\Delta_I-\omega_2-2\pi i$ to the term $J_1\omega_1$
on the left hand side of (\ref{z4}), one obtains
\begin{equation}
 (Q_I+J_1)\Delta_I+(J_2-J_1)\omega_2=
 \frac{N^2}{2}\frac{\Delta_1\Delta_2\Delta_3}{\omega_1\omega_2}
 \left(\frac{2\pi i}{\omega_1}-1\right)\ ,
\end{equation}
which can be rewritten as
\begin{equation}\label{eqn4}
 (Q_I+J_1)z_I-(J_2-J_1)=
 -\frac{N^2}{2}\frac{z_1z_2z_3}{z_4}
 \left(\frac{1+z_1+z_2+z_3+z_4}{z_4}+1\right)\ .
\end{equation}
On the other hand, subtracting (\ref{z123}) and (\ref{z4}), and doing similar
subtractions for $I=2,3$, one obtains
\begin{equation}\label{eqn123}
  N^2\frac{\Delta_1\Delta_2\Delta_3}{\omega_1\omega_2}
  \left(\frac{1}{\Delta_I}-\frac{1}{\omega_1}\right)=-2(Q_I+J_1)
  \rightarrow N^2\frac{z_1z_2z_3}{z_4}
  \left(\frac{1}{z_4}+\frac{1}{z_I}\right)=-2(Q_I+J_1)\ .
\end{equation}
Viewing these four equations (\ref{eqn4}), (\ref{eqn123})
as equations for $Q_I+J_1$, $J_2-J_1$, one can `solve' them for
these charges and obtain
\begin{equation}\label{charge-chemical}
  Q_I+J_1=-\frac{N^2}{2}\frac{z_1z_2z_3}{z_4}\left(\frac{1}{z_I}+\frac{1}{z_4}\right)
  \ ,\ \ J_2-J_1=\frac{N^2}{2}\frac{z_1z_2z_3}{z_4}\left(\frac{1}{z_4}-1\right)\ .
\end{equation}
To get further useful arrangements, we view this equation as those for
$\frac{1}{z_{1,2,3,4}}$ with given overall $\frac{z_1z_2z_3}{z_4}$ factor.
Namely, with $f\equiv \frac{N^2}{2}\frac{z_1z_2z_3}{z_4}$,
one obtains
\begin{equation}
  \frac{1}{z_4}=\frac{J_2-J_1}{f}+1\ ,\ \
  \frac{1}{z_I}=-\frac{Q_I+J_2}{f}-1\ .
\end{equation}
From the definition of $f$, one obtains the following equation for $f$:
\begin{equation}
  f=-\frac{N^2}{2}\frac{f^2(J_2-J_1+f)}{(Q_1+J_2+f)(Q_2+J_2+f)(Q_3+J_2+f)}\ .
\end{equation}
This is a cubic equation of $f$,
\begin{equation}\label{cubic-f}
  (f+Q_1+J_2)(f+Q_2+J_2)(f+Q_3+J_2)+\frac{N^2}{2}f(f+J_2-J_1)=0\ .
\end{equation}
After suitably applying the saddle point equations, the entropy $S$ can be expressed as
\begin{equation}
  S=-2\pi i(f+J_2)\ .
\end{equation}
Then (\ref{cubic-f}) yields the following a cubic equation for $S$,
\begin{equation}\label{S-cubic}
  (S-2\pi iQ_1)(S-2\pi iQ_2)(S-2\pi iQ_3)-\pi i N^2
  (S+2\pi iJ_1)(S+2\pi iJ_2)=0\ .
\end{equation}
If we solve this equation, we will get three solutions. At general charges, all solutions
for $S$ will be complex. If ${\rm Re}(S)\leq 0$, that solution will not represent
black holes. Furthermore, if the solution for any of $\Delta_I$, $\omega_i$ has negative
real part, that solution will not describe a good saddle point.
Finally, if the constructed solution has negative ${\rm Re}(\log Z)$, that saddle point
will lose against thermal gravitons whose free energy is of subleading order $N^0$,
meaning that the system is below the Hawking-Page
transition \cite{Hawking:1982dh}. If there are multiple physical solutions, one should
compare their ${\rm Re}(\log Z)$ to see which one is thermodynamically dominant.
All these issues will be gradually addressed, below in this subsection and also in
section 3.

Before analyzing the solutions of (\ref{S-cubic}) in detail, we first seek for a special
situation, in which case ${\rm Im}(S)=0$. As emphasized, we see no a priori physical reason
to expect this to be a special locus, but it will just turn out that the saddle points for
known black hole solutions stay there. Demanding a real solution for $S$ will force the $5$
charges $Q_I,J_i$ to stay on a codimension $1$ locus, to be determined below.
In this setting, one can demand that the real and imaginary parts of
(\ref{S-cubic}) are separately zero, which are of the form $S^3+ \alpha S=0$,
$\beta S^2+\gamma=0$ with real $\alpha,\beta,\gamma$. These equations determine
$S$ twice, leading to
\begin{eqnarray}\label{entropy-two}
  S\ (=\sqrt{-\alpha})&=&2\pi\sqrt{Q_1Q_2+Q_2Q_3+Q_3Q_1-\frac{N^2}{2}(J_1+J_2)}\nonumber\\
  S\ (=\sqrt{-\gamma/\beta})&=&2\pi\sqrt{
  \frac{Q_1Q_2Q_3+\frac{N^2}{2}J_1J_2}{\frac{N^2}{2}+Q_1+Q_2+Q_3}}\ .
\end{eqnarray}
Compatibility of the two expressions yields the charge relation for ${\rm Im}(S)=0$.

Here note that, all known BPS black hole solutions
of \cite{Gutowski:2004ez,Cvetic:2005zi} satisfy a charge relation (whose physical
reason is unclear, at least to us, if any).
The $4$ parameter solutions in the last reference of \cite{Gutowski:2004ez}
have the following charges:
\begin{eqnarray}\label{BH-charges}
  Q_1&=&\frac{N^2}{2\ell^2}\left[\mu_1+\frac{1}{2\ell^2}(\mu_1\mu_2+\mu_1\mu_3-\mu_2\mu_3)\right]\\
  Q_2,Q_3&=&(\textrm{obtained from $Q_1$ by cyclic permutations of }\mu_1,\mu_2,\mu_3)
  \frac{}{}\nonumber\\
  J_1&=&\frac{N^2}{2\ell^4}\left[\frac{1}{2}(\mu_1\mu_2+\mu_2\mu_3+\mu_3\mu_1)
  +\frac{\mu_1\mu_2\mu_3}{\ell^2}+\ell^4\left(\sqrt{\Xi_b/\Xi_a}-1\right)\mathcal{J}\right]
  \nonumber\\
  J_2&=&(\textrm{obtained from }J_1\textrm{ by }a\leftrightarrow b)\frac{}{}\ .\nonumber
\end{eqnarray}
$\ell=g^{-1}$ is the radius of AdS$_5$ (and also $S^5$), $N^2=\frac{\pi\ell^3}{2G}$ where
$G$ is the 5d Newton constant on AdS$_5$, and
\begin{equation}
  \Xi_a=1-\frac{a^2}{\ell^2}\ ,\ \ \Xi_b=1-\frac{b^2}{\ell^2}\ ,\ \
  \mathcal{J}=\prod_{I=1}^3\left(1+\frac{\mu_I}{\ell^2}\right)\ .
\end{equation}
(We multiplied $\ell$ to
the expressions of $Q_I$'s presented in \cite{Gutowski:2004ez} to get charges in our
convention.) The four independent parmaters are $\mu_I$'s and $a,b$ constrained by
\begin{equation}\label{parameter-relation}
  \mu_1+\mu_2+\mu_3=\frac{1}{\sqrt{\Xi_a\Xi_b}}\left[2\ell(a+b)+2ab
  +\frac{3}{\ell^2}(1-\sqrt{\Xi_a\Xi_b})\right]\ .
\end{equation}
Inserting these expressions, One can show
\begin{equation}\label{charge-relation}
  Q_1Q_2Q_3+\frac{N^2}{2}J_1J_2=
  \left(\frac{N^2}{2}+Q_1+Q_2+Q_3\right)\left(Q_1Q_2+Q_2Q_3+Q_3Q_1-\frac{N^2}{2}(J_1+J_2)
  \right)\ .
\end{equation}
The charge relation of these known solutions is precisely the equation obtained
by equating the two right hand sides of (\ref{entropy-two}). This means that, somehow,
the technically chosen surface ${\rm Im}(S)=0$ is where the known BPS black holes sit.
Furthermore, assuming this charge relation, the first expression in
(\ref{entropy-two}) was shown to be equal to the Bekenstein-Hawking entropy of
these black holes \cite{Kim:2006he}. Since the lower bound of entropy given by our
index saturates the black hole entropy, we have microscopically accounted for all
their microstates. We do not have a good understanding on why/whether the locus
${\rm Im}(S(Q,J))=0$ is physically special.

So far, the analysis was general, without assuming the Cardy limit, as first
discovered in \cite{Hosseini:2017mds}. So even though we managed to derive it only in
our Cardy limit, the free energy (\ref{entropy-function}) could be the correct one
describing the known black holes. However, beyond the Cardy regime $|\omega_i|\ll 1$,
it is not guaranteed that there are no more black hole saddle points, so that the true
free energy of large $N$ $\mathcal{N}=4$ Yang-Mills may be more complicated.
Indeed, in section 3, we find that the true free energy may be more complicated
than (\ref{entropy-function}), by studying another special limit.
Now focussing on our Cardy limit, it demands $z_4$ to be order $1$ while
$z_{1,2,3}$ to be much larger than $1$. From (\ref{charge-chemical}), this implies
that the four combinations of charges $Q_I+J_1$, $J_2-J_1$ are much larger than
$N^2$ (unless $z_4=1$, so one considers microstates with equal rotations $J_1=J_2$).
This is all one can say intrinsically from the index. However, we can discuss
the implication of the Cardy limit on the known black hole solutions that we have
just counted, on the surface (\ref{charge-relation}). From the expressions
(\ref{BH-charges}), $Q_I+J_1$, $J_2-J_1$ can be taken to be much larger than
$N^2$ by taking $\mu_I\gg \ell^2$,assuming that $a,b$ are further tune to meet (\ref{parameter-relation}). So generically,
$Q_I\propto N^2\mu^2$, $J_i\propto N^2\mu^3$. One can then approximate
the right hand side of (\ref{charge-relation}) by dropping $\frac{N^2}{2}$ term on
the first factor, and $-\frac{N^2}{2}(J_1+J_2)$ term on the second factor, yielding
the asymptotic relation
\begin{equation}
  (Q_1+Q_2+Q_3)(Q_1Q_2+Q_2Q_3+Q_3Q_1)-Q_1Q_2Q_3\approx\frac{N^2}{2}J_1J_2\ .
\end{equation}
When all charges are equal, $\equiv Q$, and also when all angular
momenta are equal $\equiv J$, it becomes $(J/N^2)^2\approx 16(Q/N^2)^3$.
So our Cardy limit on known solutions demands $J/N^2\gg Q/N^2\gg 1$.

To finalize our discussion on the saddle points on the surface ${\rm Im}(S)=0$,
we should confirm that all ${\rm Re}(\Delta_I)$, ${\rm Re}(\omega_i)$
agree with the BPS chemical potentials
of the black hole solutions. $z_{1,2,3,4}$, can be
determined from $S$ that we just expressed in terms of $Q_I,J_i$.
From (\ref{charge-chemical}), and from the relation $S=-2\pi i(f+J_2)$ with
$f\equiv\frac{N^2}{2}\frac{z_1z_2z_3}{z_4}$, one obtains
\begin{equation}
  Q_I+J_1=\left(\frac{S}{2\pi i}+J_2\right)\left(\frac{1}{z_I}+\frac{1}{z_4}\right)
  \ ,\ \ J_1-J_2=\left(\frac{S}{2\pi i}+J_2\right)\left(\frac{1}{z_4}-1\right)\ .
\end{equation}
This determines $z_{1,2,3,4}$ as
\begin{equation}
  z_I=-\frac{S+2\pi iJ_2}{S-2\pi iQ_I}\ ,\ \
  z_4=\frac{S+2\pi iJ_2}{S+2\pi iJ_1}\ .
\end{equation}
Inserting these into (\ref{solve-chemical-constraint}), one obtains $\Delta_I$, $\omega_i$,
whose real parts ${\rm Re}(\Delta_I)\equiv\xi_I$, ${\rm Re}(\omega_i)\equiv\zeta_i$
are the chemical potentials coupling to $Q_I,J_i$.
We have shown that $\xi_I$ and $\zeta_i$ agree with those of the dual black holes.
Since this involved computerized calculations of complicated functions, we simply
outline the procedures.

To compute these chemical potentials from gravity, one has to start from the non-BPS
solutions and take the zero temperature BPS limit to find $\xi_I,\zeta_i$. The general
non-BPS solutions with unequal $Q_I,J_i$ is known in \cite{Wu:2011gq}.
Here, we study the solutions of \cite{Cvetic:2005zi} with independent charges $Q_I$,
but only at equal angular momenta. So we made comparisons only at $J\equiv J_1=J_2$, $
\zeta\equiv \zeta_1=\zeta_2$.
The non-BPS black holes of \cite{Cvetic:2005zi} at \(J_{1,2}=J\) has $5$
parameters \(m, \ a\) and \(\delta_{1,2,3}\). Its energy, charges and entropy
are given by\footnote{Compared to (3.10), (3.11) of \cite{Cvetic:2005zi},
we multiplied factors containing $g$ and/or $G$, to meet our convention.}
 \begin{align}
 E&={1\over gG}\cdot{1\over 4}m\pi \Bigr(3+a^2 g^2+2s_1^2+2s_2^2+2s_3^2 \Bigr)
 \ \ ,\ \
 Q_I={1\over g G} \cdot {1\over 2}m\pi s_I c_I
 \nonumber\\
 J&={1\over G} \cdot {1\over 2}ma \pi \Bigr( c_1 c_2 c_3-s_1 s_2 s_3\Bigr)
 \ \ ,\ \
 S={1\over G}  \cdot {\pi^2\over 2} \sqrt{f_1(r_+)}\ ,
 \label{153}
 \end{align}
 where \(G\) is the Newton's constant and
 \begin{align}
 s_I&=\sinh{\delta_I}, \quad c_I=\cosh{\delta_I}, \quad
 H_I=1+{2m\over r^2} s_I^2
 \nonumber \\
 f_1&=r^6 H_1 H_2 H_3 +2ma^2 r^2+4ma^2 \Bigr(2(c_1 c_2 c_3-s_1 s_2 s_3)s_1 s_2 s_3-s_1^2 s_2^2-s_2^2 s_3^2-s_3^2 s_1^2 \Bigr)
 \nonumber \\
 f_2&=2ma(c_1 c_2 c_3 - s_1 s_2 s_3)r^2 +4m^2 a s_1 s_2 s_3
 \nonumber \\
 f_3&=2ma^2 (1+g^2 r^2)+4g^2 m^2 a^2 \Bigr(2(c_1 c_2 c_3-s_1 s_2 s_3)s_1 s_2 s_3-s_1^2  s_2^2 -s_2^2 s_3^2 -s_3^2 s_1^2 \Bigr)
 \nonumber \\
 Y&=f_3+g^2 r^6 H_1 H_2 H_3+r^4-2mr^2\ .
 \end{align}
\(r=r_+\) is the largest positive root of \(Y(r)=0\).
The BPS condition and smooth horizon condition yield
 \begin{align}
 a={1\over g}e^{-\delta_1 -\delta_2-\delta_3}, \quad
 m={4e^{2\delta_1+2\delta_2+2\delta_3}\over g^2 (e^{2\delta_1+2\delta_2}-1)(e^{2\delta_2+2\delta_3}-1)(e^{2\delta_3+2\delta_1}-1) }\ .
 \label{171}
 \end{align}
 For BPS black holes, the outer horizon is located
at\footnote{Here, a typo of (3.75) in \cite{Cvetic:2005zi} is corrected. \((r_+^2)_\text{here}=-(r_0^2)_\text{there}\)}
\begin{align}
r_+=\sqrt{4 e^{2\delta_1+2\delta_2+2\delta_3}-2e^{2\delta_1+2\delta_2}-2e^{2\delta_2+2\delta_3}
-2e^{2\delta_3+2\delta_1}+2
\over g^2(e^{2\delta_1+2\delta_2}-1)(e^{2\delta_2+2\delta_3}-1)(e^{2\delta_3+2\delta_1}-1) }\ .
\label{177}
\end{align}
Inserting (\ref{171}) and (\ref{177}) to  (\ref{153}), one can check \(E=2J+Q_1+Q_2+Q_3\)
and (\ref{entropy-two}). Chemical potentials are given
by\footnote{We changed normalization by multiplying $\frac{1}{g}$ to $T$, and
$\frac{1}{2g}$ to $\Omega$. We also corrected a typo in (3.10) of \cite{Cvetic:2005zi}:
the $+$ sign in front of the second term of $\Phi_I$ in our (\ref{AdS5-chemical})
was $-$ there.}
\begin{align}\label{AdS5-chemical}
T={1\over g}\cdot{1\over 4\pi r \sqrt{f_1}}{\partial Y\over \partial r}, \quad
\Omega={1\over 2g}\cdot{2f_2\over f_1}, \quad
\Phi_I={2m\over r^2 H_I} \Bigr( s_I c_I
+ {2a f_2\over f_1}( c_I s_J s_K-s_I c_J c_K) \Bigr)
\end{align}
where functions are evaluated at \(r=r_+\). They satisfy the
following first law of thermodynamics:
\begin{align}
dE=TdS+2\Omega dJ+\sum_{I=1}^3 \Phi_I dQ_I\ .
\end{align}
The free energy $F$ in the canonical ensemble for all $\Omega,\Phi_I$ is given by
\begin{equation}
  F=E-TS-2\Omega J - \Phi_I Q_I\ .
\end{equation}
Defining $\Delta E=E-\sum_I Q_I-2J$, the energy beyond BPS bound, one finds
\begin{equation}
  \frac{F}{T}=\frac{\Delta E}{T}-S+\sum_{I}\frac{1-\Phi_I}{T}Q_I+2\frac{1-\Omega}{T}J\ .
\end{equation}
Taking the BPS limit (\ref{171}), the black hole chemical potentials
approach $\Phi_I\rightarrow 1$, $\Omega\rightarrow 1$ in our normalization.
Since $T\rightarrow 0$ is associated with taking the BPS limit, one finds that
\begin{equation}\label{BPS-chemical}
\xi_I\equiv\lim_{T\to 0}{1-\Phi_I\over T}, \quad
\zeta\equiv\lim_{T\to 0}{1-\Omega\over T}
\end{equation}
is finite. Since $S$ is finite in the limit, one finds that the BPS limit
of $\frac{F-\Delta E}{T}\equiv F_{\rm BPS}(\mu_I,\nu)$ should exist.
Therefore, one finds
\begin{equation}\label{BPS-free}
  -F_{\rm BPS}=S-\sum_I\xi_IQ_I-2\zeta J\ .
\end{equation}
$-F_{\rm BPS}$ is nothing but $\log Z$, where $Z$ is the partition function in the BPS
limit. $\xi_I,\zeta$ defined by (\ref{BPS-chemical}) are functions of $Q_I,J$ subject to
a charge relation. This can be directly compared to our result from the entropy function.
We find that the two results agree.

Having found that both entropy and chemical potentials derived from
(\ref{entropy-function}) agree those of known BPS black holes, even away from the
Cardy limit $|\omega_i|\ll 1$, we can understand
$-{\rm Re}(\log Z)=\-{\rm Re}\left(\frac{N^2\Delta_1\Delta_2\Delta_3}{2\omega_1\omega_2}\right)$
more generally as the free energy of the known black holes. For instance,
$F_{\rm BPS}$ defined by the right hand side of (\ref{BPS-free}) will
automatically be the same as the extremal value of it. Of course
an unclear part is whether this saddle point is thermodynamically dominant
or not, against other possible black holes. However, there is a point in discussing
(\ref{free-energy-final}) beyond the Cardy limit, as a tool to study
known black holes better. We shall sometimes assume this attitude below.

Now we work more intrinsically within the index without imposing any charge relation
by hand (e.g. ${\rm Im}(S)=0$), and study ${\rm Re}(S)$. This will depend only
on $4$ combinations of the $5$ charges $Q_I,J_i$. For simplicity, let us consider the case
with equal electric charges, $Q_1=Q_2=Q_3\equiv Q$, and equal angular momenta
$J_1=J_2\equiv J$. Then we also set the corresponding chemical potentials to be equal,
$\Delta_1=\Delta_2=\Delta_3\equiv\Delta$, $\omega_1=\omega_2\equiv\omega$.
The constraint on chemical potentials is $3\Delta=2\pi i+2\omega$. Inserting this,
the entropy function is given by
\begin{align}
S={N^2\over 2}{\Bigr({2\pi i+2\omega \over 3} \Bigr)^3\over \omega^2}+2\omega(J+Q)+2\pi i Q\ .
\end{align}
We ignore the last constant term $2\pi iQ$, as this will not contribute to
${\rm Re}(S)$. (In fact, $e^{2\pi i Q}=\pm 1$ from charge quatization.)
The saddle point equation \({\partial S\over \partial \omega}=0\) yields
\begin{align}
J+Q={N^2\over 54}\Bigr({(2\pi i+2\omega)^3\over \omega^3}-3{(2\pi i+2\omega)^2\over \omega^2} \Bigr)\ .
\label{261}
\end{align}
As mentioned in the previous paragraph, we allow general $\omega$, not necessarily small.
$\omega$ will be complex, but since the left hand side of (\ref{261}) is real,
it is helpful to write \(\omega=\omega_R + i\omega_I\) with real \(\omega_{I,R}\).
Then (\ref{261}) can be separated to real and imaginary parts.
Setting the imaginary part to zero, one obtains
three solutions for \(\omega_R\) at given \(\omega_I\):
\begin{equation}
  \omega_R=\left\{
  \begin{array}{ll}
    0&\textrm{for }\omega_I \in (-\infty,\infty)\\
    \pm \omega_I\sqrt{3\pi +3\omega_I \over \pi-3\omega_I}&\textrm{for }
  \omega_I \in (-\pi, {\pi\over 3})
  \end{array}\right.\ .
\label{278}
\end{equation}
If one inserts (\ref{278}) to (\ref{261}), the real part of this equation becomes
\begin{equation}
  J+Q=\left\{
  \begin{array}{ll}
    {2N^2\over 27} {(2\pi -\omega_I) (\pi+\omega_I)^2 \over \omega_I^3 }&
    \quad \textrm{if }\ \omega_R=0\\
    -{N^2\over 54} {(\pi -2\omega_I)^2 (\pi+\omega_I) \over \omega_I^3 }&
    \quad \textrm{if }\ \omega_R=\pm \omega_I\sqrt{3\pi +3\omega_I \over \pi-3\omega_I}
    \end{array}\right.\ .
\label{287}
\end{equation}
Also, the `free energy' (\(\log Z={N^2\over 2}{\Delta^3\over \omega^2}\)) becomes
\begin{equation}
  \log Z=\left\{
  \begin{array}{ll}
  i{4N^2\over 27}{ (\pi+ \omega_I)^3\over \omega_I^2} &\text{if }\omega_R=0\\
  \mp{N^2\over 18} {\pi^3 -9 \pi \omega_I^2 -8 \omega_I^3 \over \omega_I^2 } \sqrt{\pi+\omega_I\over 3\pi-9\omega_I}
-i{N^2\over 54}{(\pi- 8\omega_I)(\pi+\omega_I)^2\over \omega_I^2 }
 &\text{if }\omega_R=\pm \omega_I\sqrt{3\pi +3\omega_I \over \pi-3\omega_I}
 \end{array}\right.\ .
\label{296}
\end{equation}
The solution with $\omega_R=0$ will yield imaginary $\log Z$ and therefore
${\rm Re}(S)=0$, making it an irrelevant solution. In the remaining two
solution, the free parameter $\omega_I$ is related to the unique charge combination
$J+Q$ captured by the index, which can be used to express $\log Z$ and $S$.

We further discuss which of the remaining solutions corresponds to black holes.
Since $\omega_R$ should be positive, one should choose the upper sign
for \(0<\omega_I < {\pi\over 3}\), and lower sign for \(-\pi<\omega_I < 0\).
Also, since $J+Q$ has to be positive, one obtains \(\omega_I<0\)
from the second line of (\ref{287}). Therefore the physical solution
is \(\omega_R=-\omega_I \sqrt{3\pi+3\omega_I\over \pi-3\omega_I}\) for
\(-\pi<\omega_I < 0\). Various quantities labeled by $\omega_I$ are summarized as
\begin{align}
\omega&=-\omega_I\sqrt{3\pi+3\omega_I\over \pi-3\omega_I}+i\omega_I,
\quad -\pi < \omega_I <0
\nonumber \\
J+Q&=-{N^2\over 54} {(\pi -2\omega_I)^2 (\pi+\omega_I) \over \omega_I^3 }
\nonumber \\
\log Z&={N^2\over 18} {\pi^3 -9 \pi \omega_I^2 -8 \omega_I^3 \over \omega_I^2 } \sqrt{\pi+\omega_I\over 3\pi-9\omega_I}
-i{N^2\over 54}{(\pi- 8\omega_I)(\pi+\omega_I)^2\over \omega_I^2 }\ .
\label{305}
\end{align}
In the Cardy limit we derived, $\omega_I$ should be a small negative number.

\begin{figure}[t!]
\begin{center}
	\includegraphics[width=8cm]{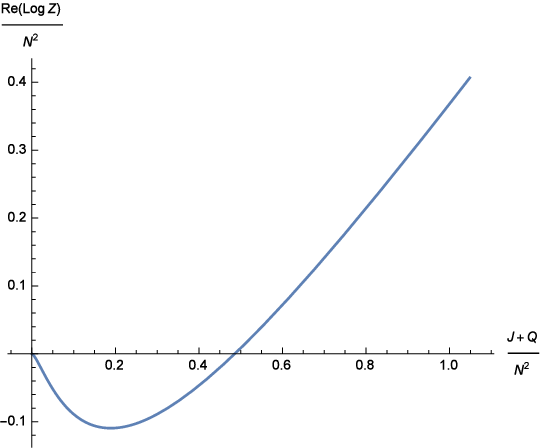}
\caption{$\log Z$ as a function of $Q+J$, extrapolating (\ref{free-energy-final})
beyond the Cardy limit}
	\label{hp}
\end{center}
\end{figure}

As emphasized, this is the free energy of known black hole saddle points. Our microscopic
analysis assures that this is the dominant one for large black holes in the Cardy limit.
But for not-so-large or small black holes, the situation is unclear.
In particular, numerical studies are made recently on hairy BPS black holes
\cite{Markeviciute:2018cqs},
predicting more general black holes as one approaches the zero temperature BPS limit.
In particular, as far as we see from the reported charge regimes in \cite{Markeviciute:2018cqs},
evidences for new black holes are found for small angular momenta,
at around $\frac{J}{N^2}\lesssim 0.05$. If we take these results seriously, the true
free energy may deviate from (\ref{free-energy-final}) for small black holes. Of course,
there could be a possibility that the intrinsic prediction from the index has its own
ambiguity, in that the physical charges $Q$, $J$ cannot be separately be specified.
In any case, we find it worthwhile to investigate the `phenomenology' of the thermodynamics
shown by this entropy function, all the way from large to small black holes.
For instance, one may ask if this saddle point is more dominant or not compared
to the thermal graviton phase in AdS$_5$.
This can be answered by comparing the free energy of
black holes and thermal gravitons. At given temperature, gravitons do not see $N$ so that
their free energy is $\mathcal{O}(N^0)$. This is much smaller than the free energy of our
entropy function. Therefore, we should compare the extremal value of
$-F_{\rm BPS}={\rm Re}(\log Z)$ in (\ref{305}) with $0$. The plot of the free energy
as a function of $Q+J$ is shown in Fig. \ref{hp}. One finds that $\log Z=0$  at
\(\omega_I=-\pi\) or \(\omega_I=-{1+\sqrt{33}\over 16}\pi\), for
\(J+Q=0\) and \(J+Q={3+\sqrt{33}\over 18}N^2 \simeq 0.486 N^2\), respectively.
So taking this entropy function (\ref{free-energy-final}) down to order $1$ values of
$\frac{Q+J}{N^2}$, one finds that the Hawking-Page transition would happen at
\begin{equation}
  \omega_R=\frac{\pi}{16} \sqrt{414-66\sqrt{33}}\simeq 1.16\ ,\ \
  J+Q={3+\sqrt{33}\over 18}N^2 \simeq 0.486 N^2\ ,
\end{equation}
if the phase structure does not get interrupted by other factors, like
yet unknown saddle points.

\section{The $\frac{1}{8}$-BPS Macdonald sector}

In this section, we investigate the Cardy-like and non-Cardy-like free energy of the
index in the so-called Macdonald limit \cite{Gadde:2011uv}. We first explain the
Macdonald index in the context of $\mathcal{N}=4$ Yang-Mills theory.
Consider the index
\begin{equation}
  Z={\rm Tr}\left[(-1)^Fe^{-\Delta_IQ_I-\omega_iJ_i}\right]
\end{equation}
at $\Delta_1+\Delta_2+\Delta_3=\omega_1+\omega_2$, which is obtained from
(\ref{index-trace}) by shifting a chemical potential by $2\pi i$, and by sending
$\beta\rightarrow\infty$. This is an index counting $\frac{1}{16}$-BPS states
preserving $\mathcal{Q}^{+++}_{--}$ and $\mathcal{S}^{---}_{++}$.
Eliminating $\omega_2=\Delta_1+\Delta_2+\Delta_3-\omega_1$, one obtains
\begin{equation}
  Z={\rm Tr}\left[(-1)^Fe^{-\Delta_1(Q_1+J_2)-\Delta_2(Q_2+J_2)-\Delta_3(Q_3+J_2)
  -\omega_1(J_1-J_2)}\right]
\end{equation}
in terms of four independent variables $\Delta_I,\omega_i$ with positive real parts.
Now we take the limit $\Delta_3\rightarrow\infty$, projecting to states satisfying
$Q_3+J_2=0$. One can show that this projection demands the BPS states to be
annihilated by an extra pair of supercharges, $\mathcal{Q}^{++-}_{-+}$, $\mathcal{S}^{--+}_{+-}$.
A quick way to see this is that the new pair demands the BPS energy relation
$E=Q_1+Q_2-Q_3+J_1-J_2$, which is satisfied by imposing the original BPS bound
$E=Q_1+Q_2+Q_3+J_1+J_2$ and the new projection condition $Q_3+J_2=0$.
This is a limit which takes $\Delta_3,\omega_2\rightarrow\infty$, with
$\frac{\Delta_3}{\omega_2}\rightarrow 1$. One also has to keep
$\Delta_3-\omega_2$ ($=\omega_1-\Delta_1-\Delta_2$) finite. This way, one
obtains the Macdonald index
for $\frac{1}{8}$-BPS states depending on $\Delta_1,\Delta_2,\omega_1$.

In the weakly interacting theory, $\frac{1}{16}$-BPS operators are made of:
$3$ anti-chiral scalars $\overline{\Phi}^{Q_I}$ with $(Q_I)=(1,0,0)$,
$(0,1,0)$, $(0,0,1)$; three chiralinos $\Psi^{Q_I}_{+\frac{1}{2},+\frac{1}{2}}$
with $(Q_I)=(-\frac{1}{2},\frac{1}{2},\frac{1}{2})$,
$(\frac{1}{2},-\frac{1}{2},\frac{1}{2})$,
$(\frac{1}{2},\frac{1}{2},-\frac{1}{2})$; two gauginos
$\Psi^{\frac{1}{2},\frac{1}{2},\frac{1}{2}}_{\pm\frac{1}{2},\mp\frac{1}{2}}$;
one self-dual component of field strength $f_{+1,+1}$;
two covariant derivatives $D_{1,0}$, $D_{0,1}$. In the Macdonald limit,
$\frac{1}{8}$-BPS operators are made of: two complex scalars
$\overline{\Phi}^{1,0,0}$, $\overline{\Phi}^{0,1,0}$;
two fermions $\Psi^{+\frac{1}{2},+\frac{1}{2},\pm\frac{1}{2}}_{+\frac{1}{2},\mp\frac{1}{2}}$;
one derivative $D_{1,0}$.
Despite preserving enhanced SUSY, the full spectrum of
this sector is not completely solved yet even at weak coupling, to the best of our knowledge.
This is in contrast to other $\frac{1}{8}$-BPS sectors of $\mathcal{N}=4$ Yang-Mills theory.
There are two more inequivalent $\frac{1}{8}$-BPS subsectors of the above canonical
$\frac{1}{16}$-BPS sector,
specified by either $J_1+J_2=0$ or $Q_1+Q_2=0$. The former is the well-known chiral
ring sector, completely solved in, e.g. \cite{Kinney:2005ej}. The solution in the second sector
can be found, e.g. in \cite{Grant:2008sk}. It might be surprising that the last
$\frac{1}{8}$-BPS sector given by the Macdonald limit is still unsolved. As we shall
see below, perhaps the reason is that this sector is too rich to admit a
simple exact solution.\footnote{However, \cite{Bourdier:2015wda} solved the
Schur index problem, which is an unrefined version of the Macdonald index. The Schur limit
of the general $\frac{1}{16}$-BPS index (\ref{index-canonical}) is defined as
$\Delta_3=\omega_2$. In the Macdonald index, to be studied shortly, one further
unrefines as $\Delta_1+\Delta_2=\omega_1$ to get the Schur index.}

We shall study a new Cardy-like limit and a non-Cardy-like limit of the Macdonald index
at $|\omega_1|\ll 1$. Although we also call the former a Cardy limit, it is different from
the one in section $2$ in that $\omega_2$ is sent large. In a way, the previous one is a
4d Cardy limit, acquiring large contributions from two BPS derivatives. Here, it is more
like a 2d Cardy limit.

In the Macdonald limit $\Delta_3,\omega_2\rightarrow\infty$,
$\Delta_3/\omega_2\rightarrow 1$, the index (\ref{index-canonical}) reduces to
\begin{equation}\label{index-macdonald}
  Z=\frac{1}{N!}\int\prod_{a=1}^N\frac{d\alpha_a}{2\pi}
  \prod_{a<b}\left(2\sin\frac{\alpha_{ab}}{2}\right)^2
  \exp\left[\sum_{n=1}^\infty\frac{1}{n} \left(1-\frac{(1-e^{-n\Delta_1})
  (1-e^{-n\Delta_2})}{1-e^{-n\omega_1}}\right)
  \sum_{a,b=1}^Ne^{in\alpha_{ab}}\right]
  \ .
\end{equation}
As before, we ignore the exponents for the Cartans, $a=b$, which will give
$\mathcal{O}(N^1)$ contribution to the free energy. Then, for $a\neq b$, the term `$1$'
in the exponent will cancel with the Haar measure.
Taking $\omega_1\ll 1$ with the remaining non-Abelian terms,
with $\Delta_{1,2}$ kept fixed,
and again assuming the maximally deconfining saddle point
$\alpha_1\approx\cdots\approx \alpha_N$, one obtains
\begin{equation}\label{macdonald-free}
  \log Z\sim-\frac{N^2}{\omega_1}\left[{\rm Li}_2(1)-{\rm Li}_2(e^{-\Delta_1})
  -{\rm Li}_2(e^{-\Delta_2})+{\rm Li}_2(e^{-\Delta_1-\Delta_2})\right]
\end{equation}
with unconstrained $\Delta_1,\Delta_2,\omega_1$. This is the Macdonald-Cardy
limit of the index.

On the other hand, had (\ref{free-energy-final}) or the result of \cite{Hosseini:2017mds}
been exact for general $\omega_{1,2}$, one would have obtained a very different result from
(\ref{macdonald-free}). Namely, taking the Macdonald limit of
(\ref{free-energy-final}) assuming its validity at general $\omega_{1,2}$,
$\Delta_3,\omega_2\rightarrow+\infty$ with $\Delta_3/\omega_2\rightarrow 1$,
one would have obtained
\begin{equation}\label{macdonald-hypothetic}
  \log Z\sim\frac{N^2\Delta_1\Delta_2}{2\omega_1}
\end{equation}
without any constraint on $\Delta_1,\Delta_2,\omega_1$. But keeping
$\omega_1\ll 1$ and $\Delta_{1,2}$ finite, we derive (\ref{macdonald-free}) instead
of (\ref{macdonald-hypothetic}) (assuming maximally deconfining saddle points).
So the true phase structure of black holes may be
richer than simply the known black holes, or \cite{Hosseini:2017mds}, even in
the $\frac{1}{8}$-BPS Macdonald sector.

However, before proceeding, we explain that there appears to be
a scaling limit of the Macdonald
index which yields (\ref{macdonald-hypothetic}). To see this, let us scale
$\omega_1\ll 1$, but also take $\Delta_1,\Delta_2\ll 1$ keeping
$\frac{\Delta_1\Delta_2}{\omega_1}$ finite. In this case, we take large $N$
and disregard the integrand factors for the Cartans, $a=b$, assuming that this
$\mathcal{O}(N^1)$ term will not affect our scaling free energy at $\mathcal{O}(N^2)$.
In fact, as we shall see later, the last assumption will fail, with an interesting
implication: however, let us proceed for now to derive (\ref{macdonald-hypothetic})
first. With the summation in the exponent restricted to $a\neq b$,
(\ref{index-macdonald}) can be written as
\begin{equation}
  Z\sim\frac{1}{N!}\int\prod_{a=1}^N\frac{d\alpha_a}{2\pi}
  \exp\left[-\frac{\Delta_1\Delta_2}{\omega_1}\sum_{n=1}^\infty
  \sum_{a\neq b}e^{in\alpha_{ab}}\right]\ .
\end{equation}
Since
\begin{equation}
  \sum_{n=1}^\infty\sum_{a\neq b}e^{in\alpha_{ab}}=\sum_{n\neq 0}\sum_{a<b}e^{in\alpha_{ab}}
  =\sum_{a<b}\left(2\pi\delta(\alpha_a-\alpha_b)-1\right)\ ,
\end{equation}
one obtains
\begin{equation}
  Z\sim\frac{1}{N!}\int\prod_{a=1}^N\frac{d\alpha_a}{2\pi}
  \exp\left[-\sum_{a<b}V_{\rm eff}(\alpha_a-\alpha_b)\right]\ \ ,\ \ \
  V_{\rm eff}(\theta)\equiv\frac{\Delta_1\Delta_2}{\omega_1}
  \left[2\pi\delta(\theta)-1\right]\ ,
\end{equation}
where $\delta(\theta)$ is the delta function on a circle, with $\theta\sim\theta+2\pi$.
Therefore, by keeping ${\rm Re}(\frac{\Delta_1\Delta_2}{\omega_1})>0$, one finds
an effective potential with very small repulsive core. Whether this is satisfied or not
will be controversial at the end, for a reason to be explained shortly. In any case,
let us assume this and proceed. In this case, if $\alpha_a$'s are not equal,
the potential is at its flat minimum, with constant negative energy. Since the repulsive
core is scaling to zero size in our scaling limit, one can take
$V_{\rm eff}=-\frac{\Delta_1\Delta_2}{\omega_1}$ for most values
of $\alpha_a$. It makes real part of $\log Z$ maximal,
and imaginary part stationary. Therefore, one approximates
\begin{equation}
  \log Z\sim\frac{(N^2-N)\Delta_1\Delta_2}{2\omega_1}\approx
  \frac{N^2\Delta_1\Delta_2}{2\omega_1}\ .
\end{equation}
In fact, as we will show below, the assumption that $\mathcal{O}(N^1)$ terms are ignorable
will fail, by the free energy (\ref{macdonald-hypothetic}) failing to have nontrivial
large $N$ saddle point with $\log Z\sim N^2$. But we shall use this free energy as a
probe of small black holes.

We shall now discuss the thermodynamic aspects of two free energies
(\ref{macdonald-free}) and (\ref{macdonald-hypothetic}).

It is first illustrative to see what is the consequence of (\ref{macdonald-hypothetic}).
As we emphasized in section 2.3, we can regard (\ref{free-energy-final}) as
describing known black holes, even beyond the Cardy limit.
Firstly, from the known
black hole solutions, one can show that the horizon area vanishes as one takes limit
$Q_3+J_2\rightarrow 0^+$. To see this, we start from the charge relation
(\ref{charge-relation}). Plugging in $J_2=-Q_3$ on both sides, and rearranging, one
obtains
\begin{equation}\label{charge-relation-mac}
  0=\left(Q_1+Q_2+\frac{N^2}{2}\right)\left(Q_1Q_2+Q_2Q_3+Q_3Q_1-\frac{N^2}{2}(J_1+J_2)
  +Q_3^2\right)\ ,
\end{equation}
where we suitably inserted back $Q_3\rightarrow -J_2$ on the second factor.
The first factor is positive since $Q_1+Q_2\geq 0$ in the BPS sector.
On the second factor, $Q_3^2\geq 0$ for the last term. The remaining terms in the second factor
are simply square of the black hole entropy $\left(\frac{S}{2\pi}\right)^2$, from the
first line of (\ref{entropy-two}). So the solution becomes meaningless if this is negative.
So from the vanishing of (\ref{charge-relation-mac}) on the solutions
without naked singularities, one finds
\begin{equation}
  Q_3\rightarrow 0\ ,\ \
  Q_1Q_2+Q_2Q_3+Q_3Q_1-\frac{N^2}{2}(J_1+J_2)=\left(\frac{S}{2\pi}\right)^2\rightarrow 0\ .
\end{equation}
We conclude that the known black solutions become `small black holes' in
the Macdonald limit. Here `small' and `large' is an entropic notion,
different from those used in the other part of this paper: the above configuration
has small entropy at large charges.
Collecting all the conditions, the charges carried by these small black holes satisfy
\begin{equation}\label{bh-charge-relation-mac}
  Q_1Q_2=\frac{N^2}{2}J_1\ ,\ \ Q_3=J_2=0\ ,
\end{equation}
where the first relation is the vanishing condition of the horizon area when $Q_3=J_2=0$.

Similar conclusion can be obtained from (\ref{macdonald-hypothetic}),
in a rather curious manner. Note that
\begin{equation}\label{entropy-function-mac}
  \frac{N^2\Delta_1\Delta_2}{2\omega_1}+(Q_1+J_2)\Delta_1+(Q_2+J_2)\Delta_2
  +(J_1-J_2)\omega_1
\end{equation}
is homogeneous degree $1$ in three independenet
$\Delta_1,\Delta_2,\omega_1$. Therefore, the overall scaling mode of them plays the
role of Lagrange multiplier, making the extremized entropy to vanish. Since the
remaining two ratios of the chemical potentials determine three charges
$Q_1+J_2$, $Q_2+J_2$, $J_1-J_2$, the charges satisfy a relation.
The relation is
\begin{equation}
  (Q_1+J_2)(Q_2+J_2)=\frac{N^2}{2}(J_1-J_2)\ .
\end{equation}
We find it as closest as one can get to (\ref{bh-charge-relation-mac}) from the index,
without extra input on the charges that the index cannot see (such as `$Q_3=J_2=0$').
However, we emphasize that both approaches predict small black holes
$S\rightarrow 0$ in the $\frac{1}{8}$-BPS Macdonald limit. And coming back to the
derivation of (\ref{macdonald-hypothetic}) ignoring $\mathcal{O}(N^1)$ terms,
we simply arrive at the conclusion that we may have to include them to obtain the
leading entropy. In any case, both known black hole solutions and the QFT analysis
in the non-Cardy scaling limit predicts small black holes. As an additional comment,
we cannot determine in this framework whether
${\rm Re}\left(\frac{\Delta_1\Delta_2}{\omega_1}\right)$ is positive or not,
because an overall scaling mode is a Lagrange multiplier which cannot be determined.
The sign of this quantity was important above, when we want to regard
(\ref{entropy-function-mac}) as derived from the Macdonald index in a scaling limit.
Perhaps it is related to the degenerate nature of this saddle point,
which one may resolve clearly by going slightly beyond the Macdonald limit and doing
a more careful calculation. We leave a more detailed study to the future.

Now we study the free energy (\ref{macdonald-free}).
We study the associated entropy function:
\begin{equation}
\begin{aligned}
S&=\log Z +  Q_1 \Delta_1 +  Q_2 \Delta_2 + Q_3 \Delta_3  + J_1 \omega_1 + J_2 \omega_2 \\
&= -\frac{N^2}{\omega_1} \left[\textrm{Li}_2 \left(1 \right) - \textrm{Li}_2 \left(e^{-\Delta_1 }\right) - \textrm{Li}_2 \left(e^{-\Delta_2 }\right) +\textrm{Li}_2 \left(e^{-\Delta_1-\Delta_2}\right) \right] \\
&\qquad \qquad +  (Q_1+J_2) \Delta_1 +  (Q_2+J_2) \Delta_2 + (J_1-J_2) \omega_1\ .
\end{aligned}
\end{equation}
Extremizing, one obtains
\begin{equation}\label{MDCh}
\begin{aligned}
Q_1+J_2=&\frac{N^2}{\omega_1} \left[  -\log \left(1-e^{-\Delta_1 }\right) +\log \left(1-e^{-\Delta_1-\Delta_2}\right)  \right], \\
Q_2+J_2=&\frac{N^2}{\omega_1} \left[  -\log \left(1-e^{-\Delta_2 }\right) +\log \left(1-e^{-\Delta_1-\Delta_2}\right)  \right], \\
J_1-J_2=&-\frac{N^2}{\omega_1^2} \left[\textrm{Li}_2 \left(1 \right) - \textrm{Li}_2 \left(e^{-\Delta_1 }\right) - \textrm{Li}_2 \left(e^{-\Delta_2 }\right) +\textrm{Li}_2 \left(e^{-\Delta_1-\Delta_2}\right)\right].
\end{aligned}
\end{equation}
From now on, we shall use some identities of ${\rm Li}_2$ to make a semi-analytic
study. However, all solutions below are cross-checked numerically against (\ref{MDCh}).

Using the following identity (\textrm{W. Schaeffer, 1846})
\begin{equation}\label{five-Li2}
\textrm{Li}_2 \left(x y\right) - \textrm{Li}_2 \left(x\right) - \textrm{Li}_2 \left(y\right) +\textrm{Li}_2 \left(1 \right)   = \textrm{Li}_2 \left(\frac{1-x}{1-x y} \right) -\textrm{Li}_2 \left(y\frac{1-x}{1-x y} \right) +  \log (x) \log\left (\frac{1-x}{1-x y}\right),
\end{equation}
the extremized entropy becomes
\begin{equation}\label{comment}
\begin{aligned}
S=\frac{N^2}{\omega_1}\Bigg[-\textrm{Li}_2 \left(\frac{1-e^{-\Delta_1}}{1-e^{-\Delta_1-\Delta_2}}\right)&-\textrm{Li}_2 \left(\frac{1-e^{-\Delta_2}}{1-e^{-\Delta_1-\Delta_2}}\right)\\
&+\textrm{Li}_2 \left(e^{-\Delta_2}\frac{1-e^{-\Delta_1}}{1-e^{-\Delta_1-\Delta_2}}\right)+\textrm{Li}_2 \left(e^{-\Delta_1}\frac{1-e^{-\Delta_2}}{1-e^{-\Delta_1-\Delta_2}}\right)\Bigg].
\end{aligned}
\end{equation}
From this formula, one finds $S<0$ if $\Delta_1, \Delta_2, \omega_1$ are strictly real and
positive. This is because ${\rm Li}_2(x)$ is an increasing function of $x>0$, so that
first plus third terms are negative, and second plus fourth terms are also negative.
Hence, in order to get black holes with $\textrm{Re} (S)>0$ at positive chemical
potential, we should turn on the imaginary part
of chemical potentials. Physically, this again implies that one should turn on phases of
fugacities to obstruct boson/femrion cancelation in the index to see black holes.

\begin{figure}[!t]
\centering
\begin{subfigure} [b]{0.55\textwidth}
\includegraphics[width=\textwidth]{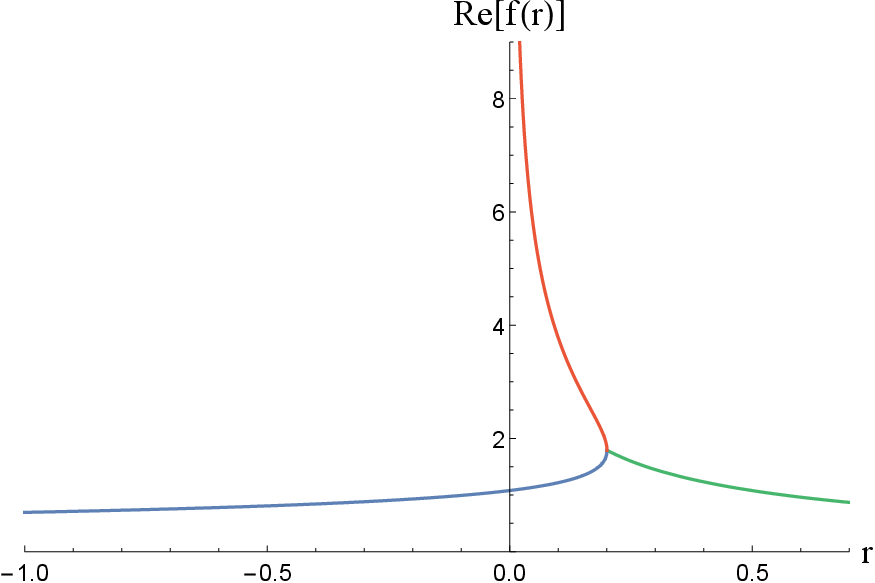}
\subcaption{$\textrm{Re}[f(r)]$: Green line denotes real parts
of both yellow and green lines in Fig. 3(b).}
\end{subfigure}
\quad
\begin{subfigure} [b]{0.4\textwidth}
\includegraphics[width=\textwidth]{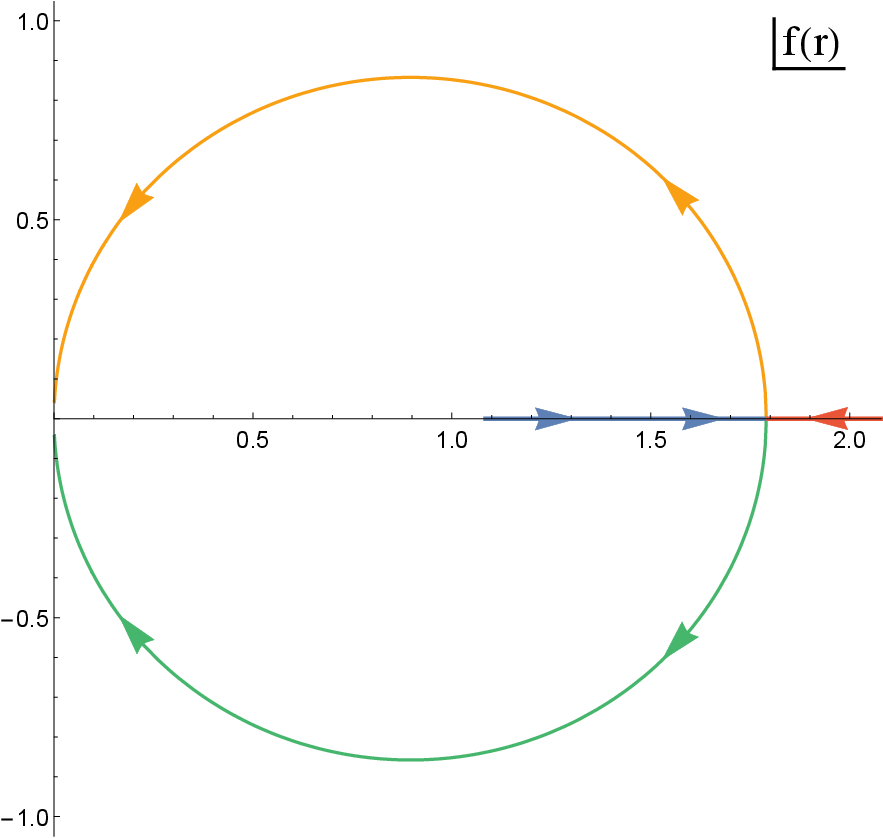}
\subcaption{$\left(\textrm{Re}[f(r)],\textrm{Im}[f(r)]\right)$:
Arrows denote an increase of $r$. Yellow and green lines are complex conjugate to each other.}
\end{subfigure}
\caption{Various solutions $f(r)$ of \eqref{treq}}
\label{fig: f_r}
\end{figure}

Now, for simplicity, we consider the case with equal charge: $Q_1=Q_2$.
Below, we will frequently use (\ref{five-Li2}) at $x=y$ and
the Euler's reflection formula:
\begin{equation}
\begin{aligned}
&\textrm{Li}_2 \left(x^2\right)   - 2\textrm{Li}_2 \left(x\right) +\textrm{Li}_2 \left(1 \right)  = \textrm{Li}_2 \left(\frac{1}{1+x} \right) -\textrm{Li}_2 \left(\frac{x}{1+x} \right) -  \log (x) \log\left (1+x\right),  \\
&\textrm{Li}_2(x)+\textrm{Li}_2(1-x)=\textrm{Li}_2(1)-\log(x)\log(1-x).
\end{aligned}
\end{equation}
Then, setting $\Delta \equiv \Delta_1=\Delta_2$, one obtains
{\allowdisplaybreaks
\begin{align}\label{NCR}
q \equiv \frac{Q_1+J_2}{N^2}&=\frac{Q_2+J_2}{N^2}=\frac{1}{\omega_1} \log \left(1+e^{-\Delta }\right) , \\
j\equiv \frac{J_1-J_2}{N^2}&=-\frac{1}{\omega_1^2} \left[\textrm{Li}_2 \left(1 \right) - 2\textrm{Li}_2 \left(e^{-\Delta }\right)  +\textrm{Li}_2 \left(e^{-2\Delta}\right)\right]
\nonumber\\
&=-\frac{1}{\omega_1^2} \left[\textrm{Li}_2 \left(1 \right) - 2\textrm{Li}_2 \left(\frac{1}{1+e^{\Delta }}\right)  -\left(\log\left(1+e^{-\Delta}\right)\right)^2\right]
\nonumber\\
&=q^2-\frac{1}{\omega_1^2} \left[\textrm{Li}_2 \left(1 \right) - 2\textrm{Li}_2 \left(1-e^{-q\omega_1}\right)\right],\nonumber\\
s\equiv \frac{S}{N^2}&=2(q\Delta + j \omega_1)=\frac{2}{\omega_1} \left[\textrm{Li}_2\left(\frac{1}{1+e^\Delta}\right)
-\textrm{Li}_2\left(\frac{1}{1+e^{-\Delta}}\right)\right]\nonumber\\
&=\frac{2}{\omega_1} \left[\textrm{Li}_2\left(1-e^{-q\omega_1}\right)-\textrm{Li}_2\left(e^{-q\omega_1}\right)\right] =2\omega_1(j-q^2)-2q \log (1-e^{-q\omega_1}) \nonumber\\
&\equiv 2q \left[r f\left(r\right) - \log \left(1-e^{-f(r)}\right) \right]\ ,\nonumber
\end{align}
}where $r \equiv \frac{j}{q^2}-1, \, f\left(\frac{j}{q^2}-1\right) \equiv q\omega_1$, and $f(r)$ is defined implicitly by the following equation:
\begin{equation}\label{treq}
f(r)^2 r = 2 \, \textrm{Li}_2 (1-e^{-f(r)})-\textrm{Li}_2(1).
\end{equation}
Note that $\textrm{Li}_2 (1) = \frac{\pi^2}{6}$. We expect
macroscopic physical solutions only when $q>0$ and $j>0$.
Indeed, with some efforts, one can check this fact explicitly from the above formulae.

Due to the complexity of these equations, we numerically/graphically solve this problem.
For $r=\frac{j}{q^2}-1>0$, one finds that $f(r)$ is a double-valued, while for
$-1<r<0$, it is single-valued. See Fig. \ref{fig: f_r}.
We find that only when $r>r_0 \equiv 0.2003559478...$, $\textrm{Im}(f(r)) \neq 0$. If $r$ is smaller than this critical value $r_0$, $f(r)$ is strictly real. Then, one finds
that $\omega_1,\Delta$ are also real, from the definition of $f$ and the first equation
of (\ref{NCR}), since $f=q\omega_1>0$.
Namely, only when $j>(1+r_0) q^2$, $\textrm{Im} (\omega_1), \textrm{Im} (\Delta) \neq 0$, and we may expect a solution with macroscopic entropy and positive chemical potentials.
One can see that we have two distinct solutions $f(r)=x(r) \pm iy(r)$ when $r>r_0$. In fact, one can analytically show that if $f(r)=x(r)+i y(r)$ is one solution of its defining equation \eqref{treq} at certain $r$, then $(f(r))^*=x(r)-iy(r)$ becomes another solution. Correspondingly, for given $j,q$, one will find the following form of two distinct solutions for the chemical potentials and entropy: $\omega_1 = \omega_1^R \pm i \omega_1^I, \, \Delta = \Delta^R \pm i \Delta^I$, and $S=S^R \pm i S^I$. So the directly observable physical quantities, given by the real parts $\omega_1^R,\Delta^R,S^R$, are uniquely determined in terms of $j, q$.
As commented below (\ref{comment}), the region $r<r_0$ does not yield sensible
saddle points.

\begin{figure}[!t]
\centering
\begin{subfigure} [b]{0.45\textwidth}
\includegraphics[width=\textwidth]{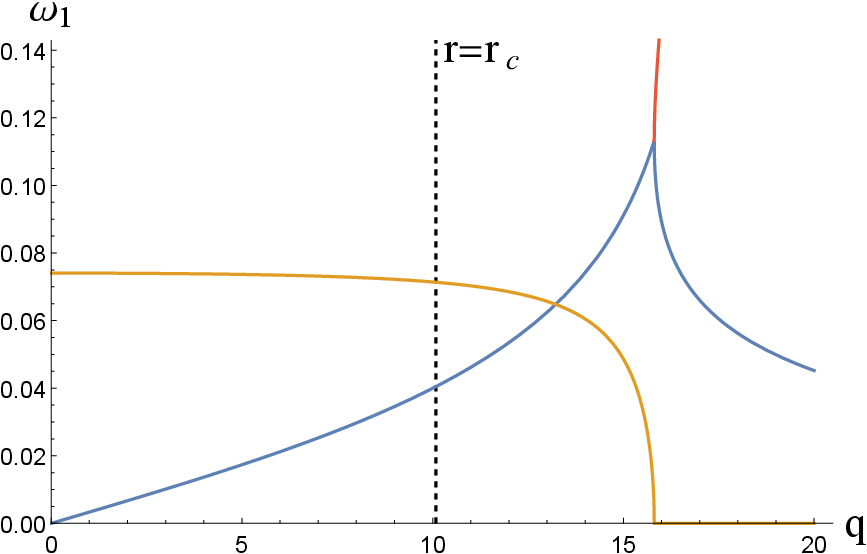}
\subcaption{$\omega_1$ as a function of $q$ at $j=300$}
\end{subfigure}
\quad
\begin{subfigure} [b]{0.45\textwidth}
\includegraphics[width=\textwidth]{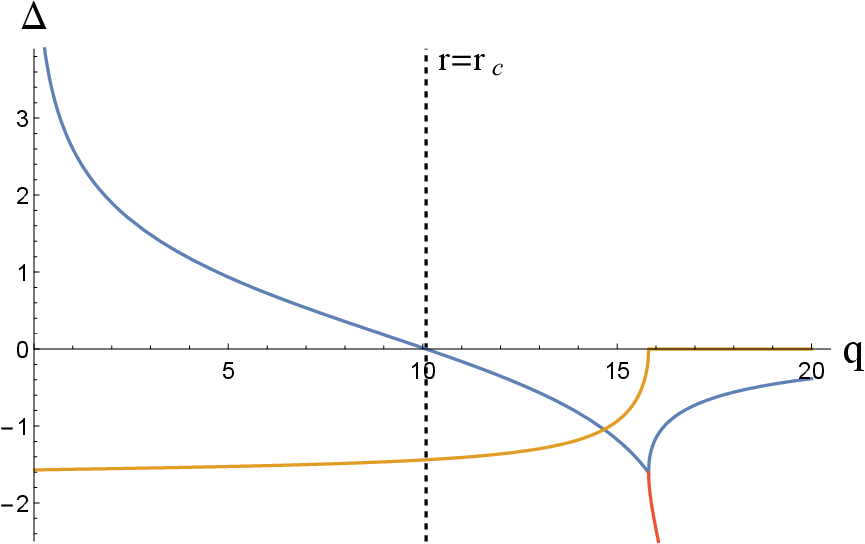}
\subcaption{$\Delta$ as a function of $q$ at $j=300$}
\end{subfigure}
\caption{Blue/yellow line denotes the real/imaginary part of (a) $\omega_1$, (b) $\Delta$. Red line denotes $\omega_1, \Delta$ corresponding to $f(r)$ described as the red line in
Fig. \ref{fig: f_r} (a), which we dismiss.}
\label{fig: omegadelta}
\end{figure}

\begin{figure}[!t]
\centering
\includegraphics[width=0.45\textwidth]{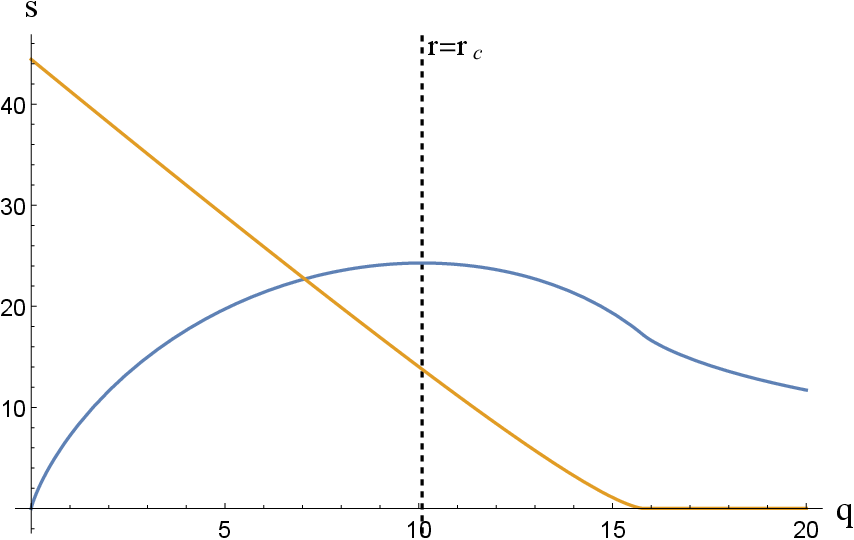}
\caption{$s(q)$ at $j=300$:
blue/yellow line denotes its real/imaginary part.}
\label{fig: s}
\end{figure}

For $r>r_0$, we study whether ${\rm Re}(\Delta),{\rm Re}(\omega_1)$ are actually positive.
In Fig. \ref{fig: omegadelta}, $\omega_1, \Delta$ are plotted with respect to
$q$, at fixed $j$.
Note that among two solutions of $f(r)$, we chose the blue one and the yellow one in Fig. \ref{fig: f_r}.
From Fig. \ref{fig: omegadelta}(b), $\textrm{Re}(\Delta)$ decreases to zero
as $q$ increase to a finite quantity, $q_{\rm max}(j)$. We find that only for
$r>r_{\rm c}\approx1.9488532...$, i.e. $j>(1+r_{\rm c}) q^2\approx 2.9488532q^2$,
$\textrm{Re}(\Delta) >0$. So at given angular momentum $j$,
a sensible saddle point at ${\rm Re}(\Delta)>0$ exists only when the
electric charge $q$ is smaller
than a maximal value $q_{\rm max}(j)=\sqrt{\frac{j}{1+r_{\rm c}}}\approx 0.582336j^{\frac{1}{2}}$.
If $r$ is smaller than this critical value $r_c$, $\textrm{Re}(\Delta)<0$.
Note that in the BPS partition function, ${\rm Re}(\Delta)\rightarrow 0^+$ is analogous
to infinite temperature limit, since its dual charge is positive.
It is curious to find such an `infinite temperature limit' at finite $q_{\rm max}(j)$.
See a related comment below.
In Fig. \ref{fig: s}, $s$ is plotted with respect to $q$ at particular $j$. As before, we chose the blue and yellow solution of $f(r)$. One can see that $\textrm{Re} (s)>0$ for arbitrary $j,
q>0$. Also, when $j>(1+r_c) q^2$, the entropy $S$ increases as the charges $j,q$ increases,
as expected.

One may want to find explicit forms of chemical potentials and entropy, in terms of charges,
at least in certain asymptotic regime. This amounts to knowing the function $f(r)$.
An explicit asymptotic form of $f(r)$ can be deduced at very large $r$.
When $r\gg 1$, $f(r) \to 0$. Hence, we can approximate the equation \eqref{treq} as
\begin{equation}
  (f(r))^2 r \sim 2 \textrm{Li}_2 (f(r))-\textrm{Li}_2 (1) \sim 2f(r)-\textrm{Li}_2 (1)
  \rightarrow f(r) \sim \frac{1}{r}\left(1 \pm i \pi \sqrt{\frac{r}{6}}\right)\ .
\end{equation}
So when $r \gg 1$, i.e. $j \gg q^2$, one obtains the asymptotic formula of the chemical
potentials and the entropy in terms of $j,q$ as follows:
\begin{equation}
\begin{aligned}
\omega_1&= \frac{f(r)}{q} \sim  \frac{1}{qr}\left(1 \pm i \pi \sqrt{\frac{r}{6}}\right)\sim \frac{1}{j}\left(q \pm i \pi \sqrt{\frac{j}{6}}\right), \\
\Delta &= -\log (e^{q\omega_1}-1) = -\log (e^{f(r)}-1) \sim -\log f(r) \sim \log r - \log\left(1 \pm i \pi \sqrt{\frac{r}{6}}\right) \\
&  \sim \frac{1}{2} \log r - \frac{1}{2} \log \frac{\pi^2}{6} \mp \log i \sim \frac{1}{2} \log  \frac{j}{q^2}- \frac{1}{2} \log \frac{\pi^2}{6}  \mp \log i, \\
s& = 2(q\Delta+j\omega_1)  \sim q\log  \frac{j}{q^2} +\left(2-\log \frac{\pi^2}{6} \mp 2 \log i\right)q  \pm  i \pi \sqrt{\frac{2j}{3}}.
\end{aligned}
\end{equation}
One finds that the Cardy-like condition $|\omega_1|\ll 1$ is met in this regime,
since ${\rm Re}(\omega_1)\sim\frac{q}{j}\ll 1$ and ${\rm Im}(\omega_1)\sim
j^{-\frac{1}{2}}\ll 1$.
In fact, just as a side comment, the above approximate entropy formula is very
well-fitted even from $r \gtrsim r_c$. At, $r=r_c$,
$\left|\frac{S-S_{\textrm{approx}}}{S}\right| \sim 0.07$.

\begin{figure}[!h]
\centering
\includegraphics[width=0.6\textwidth]{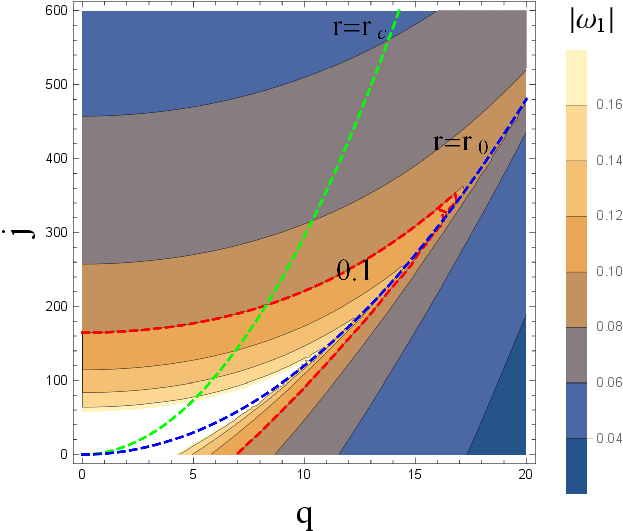}
\caption{$|\omega_1|>0.1$ in the region encircled by the red dashed line.}
\label{fig: contour}
\end{figure}

We study the validity of our Cardy approximation $\omega_1\ll 1$ for more general
$q,j$'s, at $r>r_c$. This can be easily seen in Fig. \ref{fig: contour}, where we showed
the lines with constant $|\omega_1|$ on the $q$-$j$ space.
We can highly trust our approximation when $|\omega_1| \ll 1$. When $r > r_c$,
one can see that if $j \gtrsim 200$, then $|\omega_1|<0.1$.
Therefore, we can say that when $r>r_c$ and $j \gtrsim 200$, our results are
within the Cardy regime.

In summary, only when $j>(1+r_c)q^2$, or
$q<q_{\rm max}(j)=\sqrt{\frac{j}{1+r_c}}\approx 0.582532j^{\frac{1}{2}}$, all
chemical potentials $\omega_1, \Delta$ and the macroscopic entropy $S$ have
positive chemical potentials. Otherwise, we find solutions with ${\rm Re}(\Delta)<0$,
which we disregard.

So far, we presented a semi-analytical analysis, using some identities of
${\rm Li}_2$ functions to simplify the structures. However, to be absolutely sure,
we plugged in our numerical saddle points back to the original extremization conditions
(\ref{MDCh}) without any analytic treatement, to numerically reconfirm the correctness
of our results, at least when ${\rm Re}(\Delta)>0$ in which case ${\rm Li}_2(e^{-\Delta})$,
${\rm Li}_2(e^{-2\Delta})$ are very safely well defined.


We also note that, in the regime $q<q_{\rm max}(j)$, we numerically
analyzed the Hessian
\begin{equation}
  H_{ij}\equiv -\frac{\partial^2 {\rm Re}(S(Q))}{\partial Q_i\partial Q_j}\ ,\ \
  (Q_1=q,\ Q_2=j)
\end{equation}
for $S$ at the saddle point, to study the local thermodynamic stability.
At least for $q<q_{\rm max}(j)$, we find that both eigenvalues of $H_{ij}$
are positive, implying that all susceptibility parameters are positive.
Also, we find that $\log Z$ at the saddle point is always positive in our
Cardy regime with large charges, making it more dominant than the gravitons.

Now we turn to discuss some aspects of our results.
First of all, it is interesting to see where the small black holes
satisfying $Q^2=\frac{N^2}{2}J_1$ are located. Since $J_2=0$ on the known solutions,
this charge condition translates to
$q=\frac{j^{\frac{1}{2}}}{\sqrt{2}}\approx 0.707j^{\frac{1}{2}}$.
This is the charge region where our new predicted saddle points cannot exist, since
its $q$ is larger than $q_{\rm max}(j)$.
So to conclude, our free energy predicted new $\frac{1}{8}$-BPS black hole-like
saddle points with macroscopic entropy,
when $q<q_{\rm max}(j)=\sqrt{\frac{j}{1+r_c}}\approx 0.582532j^{\frac{1}{2}}$,
in the Cardy regime. Since no such black holes are known so far in this sector,
including the small black hole limits of \cite{Gutowski:2004ez},
one may ask where to seek for such objects in the gravity dual.

Here we note that there has been some endeavors to construct black holes
beyond those known in the literature, based on allowing condensations of matters
outside the event horizon. These black holes are called hairy black holes.
In the context of global $AdS_5\times S^5$, \cite{Bhattacharyya:2010yg,Markeviciute:2016ivy}
made studies of hairy black holes with one electric charge $Q\equiv Q_1=Q_2=Q_3$
at $J_1=J_2=0$. At zero angular momentum, one finds that the hairy black hole horizon
disappears as one reduces the energy to its BPS bound $E\searrow 3Q$,
with fixed $Q$. The end point is either a smooth AdS soliton
when $Q$ is smaller than a critical value $Q_c$, or a singular horizonless solution
if $Q>Q_c$. Studying the temperature as $E\searrow 3Q$, the subcritical solutions
have zero temperature $T=0$, while the supercritical solutions have $T=\infty$.
As for hairy black holes with nonzero angular momenta,
\cite{Markeviciute:2018cqs,Markeviciute:2018yal} studied those at nonzero
$Q\equiv Q_1=Q_2=Q_3$ and $J\equiv J_1=J_2$. In this case, as $E$ is reduced
to its BPS bound $M\searrow 3Q+2J$ at fixed $Q,J$, one still finds black holes
with nonzero entropy. Again here, one finds a signal of
two different types of endpoints. In the subcritical region $Q< Q_{\rm max}(J)$,
the temperature of the limiting hairy black hole goes to $0$. In the supercritical
region, $Q>Q_{\rm max}(J)$, the temperature blows up to $\infty$. The critical
charge depends on $J$. It seems that due to numerical limitations, the
precise value of $Q_{\rm max}(J)$ could not be determined \cite{Markeviciute:2018cqs}.

Even if the hairy black holes explained above are in a different charge sector,
we find some qualitative similarities with the new saddle points that we find in
the Macdonald-Cardy limit. This is because our new saddle points also exist only
in a subcritical region $q<q_{\rm max}(j)\approx 0.582532 j^{\frac{1}{2}}$.
The reason why this gets spoiled at $q=q_{\rm max}(j)$ is because the chemical
potential ${\rm Re}(\Delta)$ approaches zero, which is analogous to the high temperature
limit in the BPS sector. It will be interesting to see if this more than
just an analogy.

\section{Large supersymmetric AdS$_7$ black holes}

In this section, we apply the method of section 2.2 to the 6d
$\mathcal{N}=(2,0)$ SCFT living on $N$ M5-branes. We shall again rely on a background
field method on $S^5$, reducing the system on small temporal $S^1$ in a Cardy-like limit.
The results in this section are by no means a `derivation' or `full microscopic account'
of AdS$_7$ black holes, even in our highly progressive standard.
Technically, in the setup of section 2.2, this is mainly due to the
fact that we do not have arguments on why we can ignore finite number of
gauge invariant Chern-Simons terms of background fields. We shall assume this, probably
appealing to a $\frac{1}{N}$ suppression. Other than this drawback, we show that gauge
non-invariant Chern-Simons terms determined by 't Hooft anomalies derive the free energy
suggested in \cite{Hosseini:2018dob} in the Cardy limit, which completely captures the large
supersymmetric AdS$_7$ black holes. And then we explain that other higher derivative terms
are suppressed in
our BPS Cardy limit. So in a sense, our studies reduce the problem of
large BPS black holes to studies of finitely many gauge-invariant CS terms on $S^5$.
Note that the absence or $\frac{1}{N}$ suppression of some terms are already partly
addressed in the literature \cite{DiPietro:2014bca,Kim:2012ava}, as we shall explain below.

The SCFT is put on $S^5\times\mathbb{R}$.
The 6d partition function is given by
\begin{equation}
  Z={\rm Tr}\left[e^{-\beta E}e^{-\Delta_1Q_1-\Delta_2Q_2}e^{-\sum_{i=1}^3\omega_iJ_i}\right]\ ,
\end{equation}
where $Q_1,Q_2$ are two charges for $U(1)^2\subset SO(5)_5$, and $J_{1,2,3}$ are
three $U(1)^3\subset SO(6)$ angular momenta on $S^5$. The 6d theory has $16$
Poincare supercharges $\mathcal{Q}^{Q_1,Q_2}_{J_1,J_2,J_3}$ where
$(Q_1,Q_2)=(\pm\frac{1}{2},\pm\frac{1}{2})$, and $J_i=\pm\frac{1}{2}$ with the
product of three $\pm$ signs of $J_i$'s being $-1$. We choose
$\mathcal{Q}\equiv\mathcal{Q}^{++}_{---}$ and its conjugate $\mathcal{S}$, and
constrain $\Delta_I,\omega_i,\beta$ to make $Z$ an index. One should constrain
\begin{equation}
  \Delta_1+\Delta_2-\omega_1-\omega_2-\omega_3=2\pi i\ \ (\textrm{mod }4\pi i)
\end{equation}
and take $\beta\rightarrow 0^+$. We will study $\log Z$ at $|\omega_i|\ll 1$, again
keeping finite imaginary parts of $\Delta_I$ to admit saddle points in which
boson/fermion cancelations are obstructed.

We consider the 6d QFT on $S^5\times S^1$ coupled to the following background fields:
\begin{eqnarray}
  ds^2&=&r^2\sum_{i=1}^3\left[dn_i^2
  +n_i^2\left(d\phi_i-\frac{i\omega_i}{\beta}d\tau\right)^2\right]+d\tau^2
\end{eqnarray}
where $n_i$ label two of the coordinates of $S^5$, constrained as
$n_1^2+n_2^2+n_3^2=1$. The other angles satisfy $\phi_i\sim\phi_i+2\pi$. $\tau$
has period $\beta$. The $U(1)^2\subset SO(5)_R$ gauge fields are given by
\begin{equation}
  A^I=-\frac{i\Delta_I}{\beta}d\tau\ .
\end{equation}
In the absence of any 6d Lagrangian description, we find it awkward to concretely
discuss the KK modes and follow all the discussions presented in section 2.2.
However, the structure of zero modes are well known, given by 5d maximal SYM (deformed by
various parameters) on $S^5$. If the $S^1$ radius for KK reduction is small,
the 5d zero modes are weakly coupled. Also, we simply assume here that nontrivial holonomy
issues
of \cite{Ardehali:2015bla} are absent, at least for the $A_{N-1}$ type theory which is of
our main concern.\footnote{It will be interesting if one can address whether there are
nontrivial issues with outer automorphism twists \cite{Tachikawa:2011ch}, whose zero modes
are 5d Yang-Mills theories with non-ADE gauge groups. \cite{Hwang:2016gfw} studied such
partition functions on $\mathbb{R}^4\times T^2$ from 5d instanton calculus,
which may provide microscopic clues to this question.} The contribution from 5d zero modes'
perturbative partition function on $S^5$ can surely be ignored. This can be seen
either by relying on arguments similar to section 2.2, or simply by a $\frac{1}{N}$
suppression since this part will be proportional to $N^2$.

So we study the structure of the effective action of our background fields, which
encodes the effects of 6d KK modes along $S^1$. We organize the background fields
to the following 5d fields after the KK reduction:
\begin{align}
ds^2_6&=ds_5^2+e^{-2\Phi} (d\tau+ a)^2
\nonumber \\
ds^2_5&=r^2 \Big[ d\theta_1^2+\sin^2\theta_1 d\theta_2^2+n_i^2 d\phi_i^2+{r^2 (\omega_i n_i^2 d\phi_i)^2\over \beta^2 (1-r^2 {n_i^2 \omega_i^2\over \beta^2})} \Big]
\end{align}
where the dilaton field \(\Phi\) and the gravi-photon field \(a\) are given by
\begin{align}
e^{-2\Phi}=1-r^2 {n_i^2 \omega_i^2\over \beta^2}, \quad a=-i{r^2 \omega_i n_i^2 d\phi_i\over \beta(1-r^2 {n_i^2 \omega_i^2\over \beta^2})}
\label{168}
\end{align}
The 6d background fields \(A^I\) are rewritten as 5d gauge fields \(\mathcal{A}^I\) and
scalars \(A_6^I\) as \(A^I=A_6^I(d\tau+a)+\mathcal{A}^I\), where
\begin{align}
A_6^I=-{i\Delta_I\over \beta}, \quad \mathcal{A}^I = -A_6^I a\ .
\label{172}
\end{align}

In our scaling limit  (\(\beta \ll |\omega_i| \ll 1\)), the leading terms will turn
out to come
from Chern-Simons terms, at order $\frac{\beta^0}{\omega_1\omega_2\omega_3}$. So
it is crucial to know all their coefficients to get the free energy in our Cardy limit.
The gauge non-invariant CS terms are again dictated by the 't Hooft anomalies of
$SO(5)_R$, which will be presented below. The gauge invariant Chern-Simons terms of
\(\mathcal{A}^I\) and \(a\) take the forms of \cite{DiPietro:2014bca}
\begin{align}\label{5d-CS-invariant}
&\beta^{-3} a\wedge da\wedge da, &
&\beta^{-2} \mathcal{A}^I \wedge da \wedge da,&
&\beta^{-1}  \mathcal{A}^I \wedge d\mathcal{A}^J \wedge da,&
& \mathcal{A}^I \wedge d\mathcal{A}^J \wedge d\mathcal{A}^K\ .
\end{align}
Here, just like in section 2.2, we do not discuss Chern-Simons terms involving gravitational
fields since they will be absent or subleading in our scaling limit. (See below in this
section.) Now, unlike the 3d CS terms for 4d $\mathcal{N}=4$ theory,
we are not given enough discrete symmetries of 6d $(2,0)$ theory to forbid them all.
In fact, some of them are believed to be nonzero.

Trying to see if one can use abstract symmetry-based arguments to forbid CS terms,
one can only partly achieve the goal. Firstly,
$\mathcal{A}^I \wedge d\mathcal{A}^J \wedge d\mathcal{A}^K$ at
$I,J,K=1,2$, $\mathcal{A}^I \wedge d\mathcal{A}^J \wedge da$ at $I\neq J$ and
$\mathcal{A}^I \wedge da \wedge da$  can be forbidden from the Weyl symmetry of $SO(5)_R$, just like we
excluded $\mathcal{A}^I\wedge da$ or $\mathcal{A}^I\wedge d\mathcal{A}^J$ at $I\neq J$
in section 2.2. In section 2.2, one used parity (suitably blind
to $SO(6)_R$) to forbid other terms. However, in 6d $(2,0)$ theory, the system is
intrinsically chiral, so that we have no simple argument to forbid
\begin{equation}\label{5d-CS-unknown}
  \beta^{-3}a\wedge da\wedge da\ \ ,\ \
  \beta^{-1}\sum_{I=1}^2\mathcal{A}^I\wedge d\mathcal{A}^I\wedge da\ .
\end{equation}
A proposal made in \cite{DiPietro:2014bca} had a consequence that the
coefficient of $a\wedge da\wedge da$ is zero for the  $(2,0)$ theory. This is partly
supported from a SUSY calculus of the index on $S^5\times S^1$ at high temperature
\cite{Kim:2012ava}, by not exhibiting a free energy at order $\beta^{-3}$ (although the calculus
was carried out after turning off many chemical potentials). Also, the $\beta^{-1}$
term of the free energy studied in \cite{Kim:2012ava} was at order $N^1$. This may be
related to an argument that the second term of (\ref{5d-CS-unknown}) is $\frac{1}{N}$
suppressed. Anyway, in the remaining part of this section, we shall assume the vanishing
or suppression of (\ref{5d-CS-unknown}). Perhaps carefully studying the microscopically
computed partition functions of the $(2,0)$ theory at high temperature (e.g. see
\cite{Kim:2017zyo}), one may be able to determine these coefficients.

The gauge non-invariant Chern-Simons terms for $\mathcal{A}^I$, $A^I_6$ can be determined
from the 't Hooft anomaly of $SO(5)_R$. Note that the
anomaly 8-form of 6d (2,0) \(A_{N-1}\) theory is
\begin{align}
I_8={N^3-N\over 24}p_2(N)+{N\over 48}\Bigr[ p_2(N)-p_2(T)+{1\over 4}(p_1(T)-p_1(N))^2 \Bigr]
\end{align}
with
\begin{align}
p_1(N)=-{1\over 2(2\pi)^2}\text{tr} F^2, \quad p_2(N)={1\over (2\pi)^4}\Bigr(-{1\over 4}\text{tr}F^4+{1\over 8}(\text{tr}F^2)^2 \Bigr)\ .
\end{align}
\cite{Kim:2017zyo} discussed the gauge non-invariant Chern-Simons term for
$A^1_6+A^2_6=0$, $\mathcal{A}^1+\mathcal{A}^2=0$, to study certain asymptotic aspects
of the free energy of $(2,0)$ theory on $\mathbb{R}^4\times T^2$.
Generalizing the calculus of \cite{Kim:2017zyo} for $U(1)^2$, one
obtains\footnote{We flipped the overall sign of $S_{\rm CS}$ compared with
\cite{Kim:2017zyo}, due to opposite 6d chirality conventions. E.g.,
in \cite{Kim:2017zyo}, supercharges contain (anti-chiral)$_{\mathbb{R}^4}\times$
(right chiral)$_{T^2}$, which is in $(0,2)$ spinors in our convention here.}
\pagebreak
\begin{align}
S_{CS}&={i(N^3-{N\over 4})\beta\over 192\pi^3} \int_{S^5} \Bigr[ 2\Bigr(A_6^1 \mathcal{A}^1 \wedge d\mathcal{A}^2 \wedge d\mathcal{A}^2+A_6^2 \mathcal{A}^2 \wedge d\mathcal{A}^1 \wedge d\mathcal{A}^1 \Bigr)
\nonumber \\
&+\Bigr(4A_6^1 A_6^2 \mathcal{A}^1 \wedge d\mathcal{A}^2 \wedge da+(A_6^1)^2 \mathcal{A}^2 \wedge d\mathcal{A}^2 \wedge da+(A_6^2)^2 \mathcal{A}^1 \wedge d\mathcal{A}^1 \wedge da \Bigr)
\nonumber \\
&+2\Bigr ((A_6^2)^2  A_6^1\mathcal{A}^1 \wedge da \wedge da+ (A_6^1)^2  A_6^2\mathcal{A}^2 \wedge da \wedge da \Bigr)
+(A_6^1)^2 (A_6^2)^2 a \wedge da \wedge da \Bigr]
\nonumber \\
&+{i N \beta \over 1536 \pi^3}\sum_{I=1}^2\int_{S^5} \Bigr[ 4 A_6^I \mathcal{A}^I \wedge d\mathcal{A}^I  \wedge d\mathcal{A}^I
+6(A_6^I)^2 \mathcal{A}^I \wedge d\mathcal{A}^I \wedge da
\nonumber \\
&+4(A_6^I)^3 \mathcal{A}^I \wedge da \wedge da
+(A_6^I)^4 a\wedge da\wedge da\Bigr]\ .
\label{179}
\end{align}
Inserting (\ref{168}), (\ref{172}) to (\ref{179}), one obtains
\begin{align}
S_{CS}=-{iN^3 \over 192\pi^3} {\Delta_1^2 \Delta_2^2\over \beta^3}\int_{S^5} a\wedge da \wedge da+\mathcal{O}(N^1)\ .
\end{align}
Evaluating $\int a\wedge da\wedge da$
with (\ref{168}), one obtains
\begin{align}
\int_{S^5} a \wedge da \wedge da
&=-{(2\pi)^3 (-i)^3 r^6 \omega_1 \omega_2 \omega_3\over \beta^3}
{1 \over \Bigr(1-{r^2 \omega_1^2\over \beta^2} \Bigr) \Bigr(1-{r^2 \omega_2^2\over \beta^2} \Bigr)  \Bigr(1-{r^2 \omega_3^2\over \beta^2} \Bigr) }\ .
\end{align}
Taking the \(\beta\to 0^+\) limit, one obtains
\begin{align}\label{5d-CS-free}
S_{CS}={N^3 \over 24} {\Delta_1^2 \Delta_2^2\over \omega_1 \omega_2 \omega_3}\ .
\end{align}
Therefore, the asymptotic free energy one obtains from $S_{\rm CS}$ is
\begin{equation}\label{6d-free-final}
  \log Z\sim -S_{\rm CS}=
  -{N^3 \over 24} {\Delta_1^2 \Delta_2^2\over \omega_1 \omega_2 \omega_3}\ ,
\end{equation}
supposing that other higher derivative terms are suppressed.

We now examine other background terms in the $S^5$ effective action, assuming the absences
or large $N$ suppressions of particular low-order terms (\ref{5d-CS-invariant}), as discussed
above. All other terms arranged in an infinite tower of derivative expansion
will turn out to be suppressed in the scaling limit ${\beta}/{r} \ll \omega \ll 1$,
as we shall illustrate with sample terms below.
We shall study the case without $\epsilon^{\mu\nu\rho\sigma\lambda}$ first and then
the other case. The analysis on the $S^5$ background action will be parallel to that on the $S^3$ action done in section~2.2. So we shall keep our discussion more concise, inspecting a few sample terms rather than attempting an exhaustive list of  corrections to certain order, as in \eqref{eq:s3action}. Below we assume $\omega_1 = \omega_2 = \omega_3 \equiv \omega$ for simplification, so that the dilaton $\Phi$ becomes a constant.

\pagebreak
We first consider the background action built from the scalar contraction of tensors without $\epsilon^{\mu\nu\rho\sigma\lambda}$. Evaluating a few terms which involve $0$, $2$, and $4$ derivatives, we find
{\allowdisplaybreaks
\begin{align}
\label{eq:s5action}
   \frac{1}{(2\pi)^3}\int\beta^{-5} e^{5\Phi} \sqrt{g} & = \frac{\beta  r^5}{8 (\beta ^2-r^2 \omega ^2 )^3} = -\frac{\beta }{8 r \omega ^6} + \mathcal{O}\left(\frac{\beta^3}{r^3 \omega^8}\right)\\
   \frac{1}{(2\pi)^3}\int\beta^{-3} e^{3\Phi} \sqrt{g} \,\mathcal{R}^{\mu\nu}{}_{\mu\nu} & = \frac{5 \beta ^3 r^3-6 \beta  r^5 \omega ^2}{2 (\beta ^2-r^2 \omega ^2 )^3} =  \frac{3 \beta}{r \omega ^4} + \mathcal{O}\left(\frac{\beta^3}{r^3 \omega^6}\right) \nonumber \\
   \frac{1}{(2\pi)^3}\int\beta^{-1} e^{\Phi} \sqrt{g} \,\mathcal{F}^I_{ab} \mathcal{F}^{Jab} &= \frac{\beta  r^5 \omega ^2 \Delta^I \Delta^J }{2 (\beta ^2-r^2 \omega ^2)^3 } = -\frac{\beta\Delta^I \Delta^J  }{2 r \omega ^4 } + \mathcal{O}\left(\frac{\beta^5}{r^5 \omega^8}\right)  \nonumber\\
  \frac{1}{(2\pi)^3}\int \beta  e^{-\Phi} \sqrt{g} \,(\nabla_c \mathcal{F}^I_{ab})(\nabla^c\mathcal{F}^{J ab}) &=  \frac{\beta ^3  \Delta^I\Delta^J  r^3 \omega ^2}{(\beta ^2-r^2 \omega ^2)^{3}}  = -\frac{\beta ^3 \Delta^I\Delta^J }{r^3 \omega ^4}  + \mathcal{O}\left(\frac{\beta^5}{r^5 \omega^6}\right)  \nonumber\\
\frac{1}{(2\pi)^3}\int\beta^{-1} e^{\Phi} \sqrt{g} \,\mathcal{R}_{\mu\nu\rho\sigma}\mathcal{R}^{\mu\nu\rho\sigma}& = \frac{24 \beta  r^5 \omega ^4-12 \beta ^3 r^3 \omega ^2+5 \beta ^5 r}{(\beta ^2-r^2 \omega ^2)^3} = -\frac{24 \beta }{r \omega ^2} + \mathcal{O}\left(\frac{\beta^3}{r^3 \omega^4}\right)
\nonumber\\
\frac{1}{(2\pi)^3}\int \beta  e^{-\Phi} \sqrt{g} \,\mathcal{F}^I_{ab}\mathcal{F}^J_{cd}\mathcal{R}^{abcd}&= -\frac{ \Delta^I \Delta^J  (6 \beta  r^5 \omega ^4-\beta ^3 r^3 \omega ^2)}{  (\beta ^2-r^2 \omega ^2)^{3}} = \frac{6 \beta \Delta^I \Delta^J  }{r  \omega ^2 } + \mathcal{O}\left(\frac{\beta^3}{r^3 \omega^4}\right)  \nonumber\\
\frac{1}{(2\pi)^3}\int\beta^5 e^{-5\Phi} \sqrt{g} \,\mathcal{F}^{Iab} \mathcal{F}^J_{ab} \mathcal{F}^{Kcd} \mathcal{F}^L_{cd} &=  \frac{2 \beta   r^5 \omega ^4 \cdot \Delta^I \Delta^J \Delta^K \Delta^L }{(\beta ^2-r^2 \omega ^2)^3} = -\frac{2 \beta  \cdot \Delta^I \Delta^J \Delta^K \Delta^L}{r \omega ^2} + \mathcal{O}\left(\frac{\beta^3}{r^3 \omega^4}\right) .\nonumber
\end{align}
where the indices $I,J,K,L$ run over $0,1,2,3$ and $\Delta^0 \equiv -i$.
These terms are all much smaller than (\ref{5d-CS-free}) in the scaling limit ${\beta}/{r} \ll \omega \ll 1$. Moreover, their leading behavior is consistent with the following speculation:} An action made of $n_1$ curvature tensors, $n_2$ graviphoton field strengths, $n_3$ background $U(1)^2 \subset SO(5)_R$ field strengths, $n_4$ derivatives scales as
\begin{align}
\label{eq:scale2}
\frac{\beta^{1 + n_4} \Delta^{n_3}}{r^{1  + n_4}\omega^{6-2n_1 -n_2 - n_3}}  + \mathcal{O}\left( \frac{\beta^{3  + n_4}}{r^{3  + n_4} \omega^{8-2n_1 -n_2 -n_3}}\right),
\end{align}
Notice that it differs from \eqref{eq:scale1} due to the additional factor
$r^2 \cdot (\beta e^{-\Phi})^{-2} \sim \omega^{-2}$. All these terms would be suppressed by taking
the scaling limit ${\beta}/{r} \ll  \omega \ll 1$.

Now we turn to the background action associated to a pseudo-scalar Lagrangian density
which has $\epsilon^{\mu\nu\rho\sigma\lambda}$. It can be either a Chern-Simons action or the
action coming from a gauge invariant Lagrangian density. Gauge non-invariant CS terms have
been determined to be (\ref{179}) from 6d 't Hooft anomaly.
The analogue of the gravitational CS term (2.60) that involves the spin connection $\omega_{\mu}^{ab}$ cannot exist in $5$ dimensions, but only in $3, 7, 11$ dimensions \cite{Zanelli:2012px}.
 The Weyl symmetry of $SO(5)_R$  restricts the other gauge invariant CS terms to be invariant under the simultaneous sign flip of $\mathcal{A}^{I=1}$ and $\mathcal{A}^{I=2}$. Displaying all possible CS terms,
\begin{align}
\frac{\beta^{-3}}{5!(2\pi)^3}\int \epsilon^{\mu\nu\rho\sigma\lambda}a_{\mu}(da)_{\nu\rho}(da)_{\sigma\lambda} &= \frac{i r^6 \omega ^3}{120 \left(\beta ^2-r^2 \omega ^2\right)^3} = -\frac{i}{120 \omega ^3} + \mathcal{O}\left(\frac{\beta^2}{r^2\omega^5}\right)\\
\frac{\beta^{-1}}{5!(2\pi)^3}\int \epsilon^{\mu\nu\rho\sigma\lambda}a_{\mu}
\mathcal{R}_{\nu\rho}{}^{\alpha\beta}\mathcal{R}_{\sigma\lambda\alpha\beta} &= -\frac{ i r^6 \omega ^5}{5\left(\beta ^2-r^2 \omega ^2\right)^3} = \frac{ i}{5\omega } + \mathcal{O}\left(\frac{\beta^2}{r^2\omega^3}\right)\\
\frac{\beta^{-1}}{5!(2\pi)^3}\int \epsilon^{\mu\nu\rho\sigma\lambda}\mathcal{A}^I_{\mu}
\mathcal{F}_{\nu\rho}^J (da)_{\sigma\lambda} &=  -\frac{i \Delta ^I\Delta ^J r^6 \omega ^3}{120 \left(\beta ^2-r^2 \omega ^2\right)^3} = \frac{i \Delta ^I\Delta ^J}{120 \omega ^3} + \mathcal{O}\left(\frac{\beta^2}{r^2\omega^3}\right)\ .
\end{align}
In fact, as asserted earlier, CS terms containing gravitational terms are suppressed,
while other gauge invariant CS terms are not. As noted above, we assume (partly relying on
assertions/observations made in the literature) that their coefficients are either
exactly zero or $\frac{1}{N}$ suppressed. Then we move to study the action associated to
the gauge invariant Lagrangian density containing $\epsilon^{\mu\nu\rho\sigma\lambda}$. We compute
some non-vanishing terms of this kind, e.g.,
\begin{eqnarray}
	&&\hspace*{-.5cm}
\frac{1}{5!(2\pi)^3} \int \beta^6e^{-6\Phi}\, \epsilon^{\mu\nu\rho\sigma\lambda} \mathcal{F}^I_{\mu\nu} \mathcal{F}^I_{\rho\sigma}(\nabla_{\alpha}\mathcal{F}_{\lambda\beta}^J)\mathcal{F}^{J\alpha\delta}\mathcal{F}^0_{\delta}{}^\beta =
 \frac{i \beta ^2  r^4 \omega ^5 (\Delta^I \Delta^J)^2}{30 \left(\beta ^2-r^2 \omega ^2\right)^{3}} = -\frac{i \beta ^2  (\Delta^I \Delta^J)^2}{30 r^2 \omega} + \mathcal{O}\left(\frac{\beta^4}{r^4\omega^3}\right) \nonumber
\\
	&&\hspace*{-.5cm}
\frac{1}{5!(2\pi)^3} \int \beta^{14}e^{-14\Phi}\, \epsilon^{\mu\nu\rho\sigma\lambda}  \mathcal{F}^I_{\mu\alpha} \mathcal{F}^I_{\nu\beta}\mathcal{F}^{I\alpha\beta} \mathcal{F}^I_{\rho\kappa} \mathcal{F}^J_{\sigma\iota}\mathcal{F}^{J\kappa\iota} (\nabla_{\psi}\mathcal{F}_{\lambda\gamma}^J)\mathcal{F}^{J\psi\tau}\mathcal{F}^0_{\tau}{}^\gamma   =\frac{i \beta ^2 r^4 \omega ^9 (\Delta^I \Delta^J)^4}{30 \left(\beta ^2-r^2 \omega ^2\right)^{3}}
\nonumber\\
	&&\hspace{.5cm}= -\frac{i \beta ^2 \omega ^3 (\Delta^I \Delta^J)^4}{30 r^2}  + \mathcal{O}\left(\frac{\beta^4 \omega^1}{r^4 }\right)\ .\nonumber
\end{eqnarray}
We observe that their scaling behavior in the limit ${\beta}/{r} \ll  \omega \ll 1$ follows \eqref{eq:scale2}. All these terms would be subleading corrections to the free energy.

Now, we perform Legendre transformation of (\ref{6d-free-final})
to the microcanonical ensemble. One should extremize the following entropy function:
\begin{align}
S(\Delta_I, \omega_i; Q_I, J_i)&=-{N^3\over 24}{\Delta_1^2 \Delta_2^2\over \omega_1 \omega_2 \omega_3}+\sum_{I=1}^2 Q_I \Delta_I+\sum_{i=1}^3 J_i \omega_i\ .
\label{36}
\end{align}
This problem was studied in \cite{Hosseini:2018dob}, reproducing the entropy of known
BPS AdS$_7$ black holes of \cite{Chong:2004dy,Chow:2007ts}.
We will review the calculation in \cite{Hosseini:2018dob}.
As in section 2.3, we also
extend the studies of \cite{Hosseini:2018dob} by checking the agreements of chemical
potentials. This allows us to regard the real part of (\ref{6d-free-final}) as the
free energy of known BPS black holes, even away from the Cardy limit.

Since we consider the index, the extremization should be performed on the specific
surface of the chemical potential space where
\begin{align}
\Delta_1+\Delta_2-\omega_1-\omega_2-\omega_3=2\pi i\ .
\label{40}
\end{align}
This also reflects the ignorance of the index on one of the five charges.
The relevant BPS states saturate the bound \(E\geq 2Q_1+2Q_2+J_1+J_2+J_3\).
On the surface (\ref{40}), one can reparameterize the chemical potentials with four unconstrained complex variables \(z_{1,2,3,4}\).
\begin{align}
\Delta_I&={2\pi i z_I\over 1+z_1+z_2+z_3+z_4}, & \quad I&=1,2
\nonumber \\
\omega_1&={-2\pi i z_3\over 1+z_1+z_2+z_3+z_4}, & \omega_2&={-2\pi i z_4\over 1+z_1+z_2+z_3+z_4}, & \omega_3&={-2\pi i \over 1+z_1+z_2+z_3+z_4}\ .
\label{214}
\end{align}
With this reparametrization, the entropy function (\ref{36}) becomes
\begin{align}
S={2\pi i \over 1+z_1+z_2+z_3+z_4}\Bigr( {N^3\over 24}{z_1^2 z_2^2\over z_3 z_4}+Q_1 z_1+Q_2 z_2-J_1 z_3 -J_2 z_4-J_3 \Bigr)
\end{align}
Extremization in \(z_i\) yields four saddle point equations,
which can be reorganized as follows:
\begin{align}
Q_I+J_3&=-{N^3\over 24}{(z_1 z_2)^2 \over z_3 z_4}
\Bigr(1+{2\over z_I} \Bigr), \nonumber \\
J_1-J_3&=-{N^3 \over 24}{(z_1 z_2)^2 \over z_3 z_4}\Bigr(-1+{1\over z_3} \Bigr), \quad
J_2-J_3=-{N^3\over 24}{(z_1 z_2)^2 \over z_3 z_4}\Bigr(-1+{1\over z_4} \Bigr)\ .
\label{59}
\end{align}
At the saddle point, the black hole entropy becomes
\begin{align}
S=2\pi i \Bigr( -{N^3\over 24} {(z_1 z_2)^2\over z_3 z_4} -J_3\Bigr)\ .
\label{226}
\end{align}
Using the last expression, one can replace the common factor
\(-{N^3\over 24}{(z_1 z_2)^2\over z_3 z_4}\) in (\ref{59}) into \({S\over 2\pi i }+J_3\).
Then the saddle point values of \(z_i\) can be expressed in terms of the charges and the
entropy as follows:
\begin{align}
z_I=-2{S+2\pi i J_3\over S-2\pi i Q_I}, \quad z_3={S+2\pi i J_3\over S+2\pi i J_1}, \quad z_4={S+2\pi i J_3\over S+2\pi i J_2}
\label{234}
\end{align}
Plugging in these values for $z_{1,2,3,4}$ to (\ref{226}), one obtains
a simple quartic equation for $S$ in terms of charges:
\begin{align}
\Bigr(S-2\pi  i Q_1 \Bigr)^2 \Bigr(S-2\pi  i Q_2 \Bigr)^2 + {4\pi i N^3\over 3} \Bigr( S+2\pi i J_1\Bigr)\Bigr( S+2\pi i J_2\Bigr)\Bigr( S+2\pi i J_3\Bigr)=0\ .
\label{71}
\end{align}

The equation (\ref{71}) has four complex solutions $S$, at given five real charges.
Again, our general attitude on ${\rm Im}(S)$ is that it is the phase factors that one may
end up with, by allowing imaginary parts of chemical potentials to ideally obstruct
boson/fermion cancelations. However, just as in the case of section 2.2, special solutions
are somehow known at the surface ${\rm Im}(S)=0$. So among the four solutions of (\ref{71}),
we study the special sets of charges which allow a real and positive solution for $S$.
Note that (\ref{71}) has the form of \((a_4 S^4+a_2 S^2+a_0)+i(a_3 S^3+a_1 S^1)=0\) with real coefficients \(a_i\). Demanding a real solution requires
\(a_4 S^4+a_2 S^2+a_0\) and \(a_3 S^3+a_1 S^1\) to separately vanish.
This leads to the two alternative expressions for the entropy:
\begin{eqnarray}
  \left(\frac{S}{2\pi}\right)^2&=&
  {3(Q_1^2 Q_2+Q_1 Q_2^2)-N^3(J_1 J_2+J_2 J_3+J_3 J_1) \over 3(Q_1+Q_2) -N^3}
  \nonumber \\
  \left(\frac{S}{2\pi}\right)^2&=&\left( {N^3\over 3}(J_1 +J_2+J_3)+{Q_1^2+Q_2^2\over 2}
  +2 Q_1 Q_2\right)
\nonumber \\
  &&\times \left( 1-{\sqrt{1-{{2\over 3}N^3 J_1 J_2 J_3+Q_1^2 Q_2^2\over
  [ {N^3\over 3}(J_1 +J_2+J_3)+{Q_1^2+Q_2^2\over 2}+2 Q_1 Q_2]^2}  } } \right)\ .
\label{199}
\end{eqnarray}
The compatibility of two expressions require a charge relation for ${\rm Im}(S)=0$.

Here, note that the known BPS black hole solutions also satisfy a charge relation.
Unfortunately, black hole solutions with all unequal $Q_1,Q_2,J_1,J_2,J_3$ are yet
unknown. This is most probably just a technical limitation. A class of non-extremal solutions
studied in \cite{Chong:2004dy} has unequal $Q_1,Q_2$, but equal angular momentum
$J_1=J_2=J_3\equiv J$. Together with energy $E$, there are $4$ parameter solutions for
independent $E,Q_1,Q_2,J$. However, imposing a BPS limit for
$E=2Q_1+2Q_2+3J$, one also has to impose a separate
condition that the smooth horizon is not spoiled. So one ends up with a $2$ parameter
solution with nonzero $Q_1,Q_2,J$, where the last three charges meet a relation.
A different slice of black hole solutions was found in \cite{Chow:2007ts}. The solutions
here satisfy $Q_1=Q_2\equiv Q$, with independent $J_1,J_2,J_3$ and energy $E$. Again imposing
the BPS condition for $E=4Q+J_1+J_2+J_3$ and smooth horizon condition, one obtains
a $3$ parameter solution with $Q,J_1,J_2,J_3$, so that the charges again meet a relation.
In both cases, one finds that the charge relation is precisely the two right hand sides
of (\ref{199}) being equal. So the known BPS black holes happen to live on the
surface ${\rm Im}(S)$, for which we again do not have a good physical insight.
On this surface, one can again show that the Bekenstein-Hawking entropy of these black
holes precisely agree with our (\ref{199}). The results summarized in this paragraph have
all been reported in \cite{Hosseini:2018dob} already. Now we have a sort of
`derivation' of (\ref{36}) in the Cardy regime $|\omega_i|\ll 1$, with certain assumptions
stated earlier in this section. We hope the discussions presented so far in this section
to shed good lights on AdS$_7$ black holes, and also to 6d $(2,0)$ theory (especially
about the CS coefficients in the high temperature expansion).

In the remaining part of this section, we supplement
\cite{Hosseini:2018dob} by showing that the chemical potentials of black holes agree with
the real parts of $\Delta_I,\omega_i$. We only do so for the case with
general $Q_1,Q_2$ and equal angular momenta $J_1=J_2=J_3\equiv J$. As in section 2.3,
we should start from non-BPS solutions and take $T\rightarrow 0$ BPS limit to read off
BPS chemical potentials.

The energy, charges and entropy for non-extremal black holes
of \cite{Chong:2004dy,Cvetic:2005zi} are given in terms of four parameters
\(\delta_{1,2}, \ m\) and \(a\)\footnote{We change the normalization of (4.7) and (4.9)
in \cite{Cvetic:2005zi}, by multiplying the first factors put before $\cdot$ on all
right hand sides of our (\ref{263}). This is mostly to convert to our convention.}:
{\allowdisplaybreaks
\begin{align}
E&= {1\over g G_N} \cdot{m\pi^2\over 32 \Xi^4}\Bigr[12 \Xi_+^2 (\Xi_+^2 -2)-2c_1 c_2 a^2 g^2 (21 \Xi_+^4-20\Xi_+^3-15\Xi_+^2-10\Xi_+-6) \nonumber \\
&\hspace{3cm}+(c_1^2+c_2^2)(21\Xi_+^6-62\Xi_+^5+40 \Xi_+^4+13\Xi_+^2-2\Xi_++ 6)\Bigr]
\nonumber \\
J&= -{1\over G_N} \cdot{ma\pi^2\over 16\Xi^4}\Bigr[4ag\Xi_+^2-2c_1 c_2 (2\Xi_+^5-3\Xi_+^4-1)+ag (c_1^2+c_2^2)(\Xi_++1)(2\Xi_+^3-3\Xi_+^2-1) \Bigr]
\nonumber \\
Q_1&={1\over 2g G_N} \cdot
{m\pi^2 s_1\over 4\Xi^3}\Bigr[a^2 g^2 c_2 (2\Xi_++1)-c_1(2\Xi_+^3-3\Xi_+-1) \Bigr]
\nonumber \\
Q_2&= {1\over 2g G_N} \cdot
{m\pi^2 s_2\over 4\Xi^3}\Bigr[a^2 g^2 c_1 (2\Xi_++1)-c_2(2\Xi_+^3-3\Xi_+-1) \Bigr]
\nonumber \\
S&={1\over 4G_N}\cdot{\pi^3 (r^2+a^2) \over  \Xi^3}
\sqrt{  f_1(r_+)}\ .
\label{263}
\end{align}
\pagebreak
Here, the parameters and functions are defined by\footnote{We corrected a typo in
(4.5) of \cite{Cvetic:2005zi}, where we correct \(\rho_{\rm theirs}=\sqrt{r^2+a^2}\) by
$\rho_{\rm ours}=\sqrt{\Xi}r$.}
\begin{align}
s_i&=\sinh{\delta_i}, \quad c_i=\cosh{\delta_i }, \quad \Xi_\pm=1\pm ag, \quad \Xi=1-a^2 g^2, \quad \rho=\sqrt{\Xi}r , \quad H_i=1+{2ms_i^2\over \rho^4}
\nonumber \\
\alpha_1&=c_1-{1\over 2}(1-\Xi_+^2)(c_1-c_2), \quad \alpha_2=c_2+{1\over 2}(1-\Xi_+^2)(c_1-c_2), \quad \beta_1=-a\alpha_2, \quad \beta_2=-a\alpha_1
\nonumber \\
f_1(r)&=\Xi \rho^6 H_1 H_2-{4\Xi_+^2 m^2 a^2 s_1^2 s_2^2 \over \rho^4}+{1\over 2}m a^2 \Bigr(4\Xi_+^2 +2 c_1 c_2(1-\Xi_+^4)+(1-\Xi_+^2)^2(c_1^2+c_2^2) \Bigr)
\nonumber \\
f_2(r)&=-{1\over 2}g\Xi_+ \rho^6 H_1 H_2 +{1\over 4}ma \Bigr(2(1+\Xi_+^4)c_1 c_2+(1-\Xi_+^4)(c_1^2+c_2^2) \Bigr) \Bigr)
\nonumber \\
Y(r)&=g^2 \rho^8 H_1 H_2 +\Xi \rho^6 +{1\over 2}ma^2 \Bigr(4\Xi_+^2+2(1-\Xi_+^4)c_1 c_2+(1-\Xi_+^2)^2(c_1^2+c_2^2) \Bigr)
\nonumber \\
&-{1\over 2}m \rho^2 \Bigr(4\Xi+2a^2 g^2 (6+8ag+3a^2 g^2)c_1 c_2  -a^2 g^2(2+ag)(2+3ag)(c_1^2+c_2^2) \Bigr)\ .
\end{align}
\(r=r_+\) is the largest positive root of \(Y(r)=0\). The BPS limit is achieved
by setting\footnote{We corrected a typo in (4.46) of \cite{Cvetic:2005zi}:
$(3e^{\delta_1+\delta_2}-1)^2\rightarrow (3e^{\delta_1+\delta_2}-1)^3$
in the numerator of $m$.}
\begin{align}
a&={2\over 3g}{1\over 1-e^{\delta_1+\delta_2}}, \quad
m={128 e^{\delta_1+\delta_2} (3e^{\delta_1+\delta_2}-1)^{3}\over 729g^4 (e^{2\delta_1}-1) (e^{2\delta_2}-1)(e^{\delta_1+\delta_2}+1)^2 (e^{\delta_1+\delta_2}-1)^4 }\ .
\label{274}
\end{align}
Then the outer horizon is located at
\begin{align}
r_+&=\sqrt{16\over 3g^2 (e^{\delta_1+\delta_2}+1) (3e^{\delta_1+\delta_2}-5) }\ .
\label{279}
\end{align}
Inserting (\ref{274}) and (\ref{279}) to (\ref{263}), one can obtain BPS relation \(E=3J+2Q_1+2Q_2\). Here, the seven dimensional Newton's constant is given by
\(G_N={3\pi^2\over 16 g^5 N^3}\) for AdS$_7\times S^4$ for $N$ M5-branes.
$g$ is the inverse-radius of AdS$_7$.
}

The first law of black hole thermodynamics is given by
\begin{align}
dE=T dS+3\Omega dJ+\Phi_1 dQ_1+\Phi_2 dQ_2\ ,
\label{287-2}
\end{align}
with the chemical potentials are\footnote{We changed normalization and corrected typo in (4.7)
of \cite{Cvetic:2005zi}, by all the factors shown with red colors.
The correct temperature and chemical potentials can be derived from the metric (2.5) of \cite{Hosseini:2018dob}. }
\begin{align}
T&= {\color{red}{1\over 4\pi g \rho^3 \sqrt{ \Xi f_1}} } {\partial Y\over \partial r} , &
\Omega&={\color{red}-{1\over g}}\Bigr(g+{2f_2\over  f_1} \Xi_-\Bigr), &
\Phi_i&={{\color{red}4}ms_i\over \rho^4 \Xi H_i} \Bigr(\alpha_i {\color{red}\Xi_-} +\beta_i {2f_2 \Xi_-\over f_1} \Bigr)\ .
\end{align}
All functions are evaluated at \(r=r_+\).  The free energy \(F\) in the canonical ensemble is given by
\begin{align}
F=E-TS-3\Omega J-\Phi_I Q_I
\end{align}
Defining \(\Delta E=E-2\sum_I Q_I-3J\), one finds
\begin{align}
{F\over T}&={\Delta E\over T}-S+\sum_I {2-\Phi_I\over T}Q_I+3{1-\Omega\over T}J
\end{align}
Taking the BPS limit (\ref{274}), the black hole chemical potentials approach \(\Phi_I\to 2\) and \(\Omega \to 1\). Therefore, we can define BPS chemical potentials as
\begin{align}\label{AdS7-chemical}
\xi_I=\lim_{T\to 0}{2-\Phi_i\over T}, \quad \zeta=\lim_{T\to 0}{1-\Omega\over T}\ .
\end{align}
Since the entropy \(S\) is finite in BPS limit,
\(F_\text{BPS}\equiv{F-\Delta E\over T}\) should remain finite. Therefore,
\begin{align}
S=-F_\text{BPS}+\sum_I \xi_I Q_I +3\zeta J\ .
\end{align}
We checked that $\xi_I$, $\zeta$ computed from (\ref{AdS7-chemical})
agree with ${\rm Re}(\Delta_I)$, ${\rm Re}(\omega)$, computed from (\ref{36}).

\section{Discussions and future directions}

We first discuss possible subtleties of our results.
We also try to suggest conservative interpretations of our results, in
case some readers might be worrying about subtleties.
\begin{itemize}

\item Throughout this paper, we mostly took (with one exception) Cardy-like
limits which suppress the fluctuations relying on large $J$.
However, general black holes are semi-classical saddle points at large $N$, rather than
large charges. So we are assuming an interpolation, which connects large $N$ saddle points
given by black holes and large $J$ saddle points of our QFT. This often turned out to provide
the correct quantitative results, starting from the seminal work \cite{Strominger:1996sh}.
The fact that our Cardy free energy successfully captures known black holes of
\cite{Gutowski:2004ez,Chong:2005da} makes us to hope that a similar situation is
happening here.

\item In our Cardy limit, we took the $U(N)$ gauge holonomies
$\alpha_a$ to be at the maximally deconfining point. One cannot imagine such saddle points
at finite charges (or finite $\omega$), because the Haar measure repulsion forbids
$\alpha_a$'s to be on top of another \cite{Aharony:2003sx,Kinney:2005ej}. We expect
our maximally deconfining saddle point to actually mean that the distances of
$\alpha_a$'s are suppressed by small $\omega$. It is easy to check that
this is the local saddle point in the Cardy limit, but one may ask if this is the global
minimum of free energy. There are examples of 4d $\mathcal{N}=1$ QFTs in which this
fails to be true \cite{Ardehali:2015bla}. Considering the empirical relation between
more nontrivial saddle points and the behaviors of $Z[S^3]$ \cite{Ardehali:2015bla},
it seems that our model should be safe of this issue. We checked
that our saddle point is the global minimum, but only in a self-consistent way
at the specific value of $\Delta_I,\omega_i$ for equal charge black holes.
Studies at more general values of $\Delta,\omega$ appear to be
cumbersome. However, at the very least, we have identified their dual black hole
saddle points, no matter stable or metastable. So our maximally deconfining saddle points
should have substantial physical implications to the large $N$ gravity dual.

\item The fact that BPS black holes exist only with a charge relation might be somewhat
puzzling from the QFT dual side, especially after we claimed that we have counted them
(at large charges). We have little to comment on it, especially in our Cardy regime
in which other solutions seem to be unknown so far
\cite{Markeviciute:2018cqs,Markeviciute:2018yal}. Especially, intertwined with the ignorance
of the index on one of the $5$ charges, the possibility of more general black holes seems
not easy to address within our results. However, technically from the gravity
side, such charge relations of BPS black holes are ubiquitous. Familiar examples are
single-centered 4d black holes \cite{Ferrara:1995ih} at zero angular momentum,
or 5d BMPV black holes \cite{Breckenridge:1996is} with self-dual angular momenta.
By now we know much richer families of BPS
black solutions, such as 4d multi-centered black holes
\cite{Denef:2000nb} or 5d black rings \cite{Elvang:2004rt}, which
violate such charge relations. In AdS, one can naturally seek for
hairy black holes. The BPS version of such black holes were recently reported
\cite{Markeviciute:2018cqs,Markeviciute:2018yal}, even though it appears
not in our large rotation regime (at least from the data presented there).

\item We studied Cardy-like and non-Cardy-like scaling limits of the
$\frac{1}{8}$-BPS Macdonald index. In the latter, we have identified the small black hole
limit of the known BPS solutions (third reference of \cite{Gutowski:2004ez}).
In the former, our Cardy free energy is quite nontrivial, and exhibits rich saddle points.
These saddle points exhibit properties very reminiscent of hairy black holes
\cite{Bhattacharyya:2010yg,Markeviciute:2018cqs}. If one can again trust the smooth
interpolation between our Cardy saddle point and the large $N$ saddle point, we can claim
that we have predicted new (hairy) black holes in the Macdonald sector. Since no solutions
are actually constructed yet, we are much less confident about the issues raised above
in this section. Perhaps actual constructions of such gravity solutions can clear the
uncertainty. Recently, it has been pointed out that the Macdonald-Cardy saddle point responsible for \eqref{macdonald-free} may be invalidated due to some subtleties related to the Picard-Lefschetz theory \cite{Chang:2023ywj}. We leave detailed studies to the future.

\item There were extra assumptions in our discussions of large AdS$_7$ black holes.
One issue is the unknown coefficients of gauge invariant CS terms on $S^5$, in the
high temperature expansion.
This issue has been resolved in \cite{Nahmgoong:2019hko}.

\end{itemize}

We think there are many interesting future directions to pursue.
We finish this paper by briefly mentioning some of them.
\begin{itemize}

\item Having seen macroscopic entropies from the index, one should expect
an explicit construction of such operators at weak-coupling. At 1-loop level,
the BPS states are mapped to cohomologies of the supercharge $\mathcal{Q}$.
\cite{Kinney:2005ej,Berkooz:2006wc,Berkooz:2008gc,Grant:2008sk,Chang:2013fba}.
Considering the free QFT analysis of section 2.1, (\ref{fermion-letter-limit}) and
comments above it, fermionic fields may be responsible for our
asymptotic free energy. \cite{Berkooz:2006wc,Berkooz:2008gc} considered a class
of such operators called `Fermi liquid operators.'
Unfortunately, the operators discussed there
were shown to be (weakly) renormalized, even at weak coupling. As already
mentioned in \cite{Berkooz:2008gc} as a possible scenario, dressing these operators
with other fermion fields might yield large number of new BPS states.
Perhaps a clever `ansatz' for such operators using all four fermions should be
discovered, generalizing \cite{Berkooz:2006wc}. \cite{Chang:2013fba} performed a
systematic analysis of this cohomology at $N=2,3$, up to certain
energy order, without using an ansatz. However, it is not completely clear to us whether
the energy orders covered in \cite{Chang:2013fba} are definitely well above $N^2$.
For instance, our Cardy limit demands $\omega$ to be small. Its conjugate $J$ is given by
$J\sim\frac{1}{\omega^3}$. So even if one generously accepts $\omega\sim 0.1$ to be small,
the associated charge will be $J\sim 10^3$,  definitely out of reach in
\cite{Chang:2013fba}. For recent progress on this subject, see \cite{Chang:2022mjp,Choi:2022caq,Choi:2023znd,Chang:2023zqk,Budzik:2023vtr,Choi:2023vdm,Chang:2024zqi} and references thereof.

\item On the other hand, the roles of fermions seen around (\ref{fermion-letter-limit})
might be an `emergent' one. This is because, if we study the Cardy limit honestly from
the index, (\ref{Li3-index}) is obtained by both bosons and fermions. Here, note that
there is a known toy model in which a fermion picture emerges. This is the half-BPS sector
of 4d $\mathcal{N}=4$ Yang-Mills theory, exhibiting a Fermi droplet picture
\cite{Berenstein:2004kk,Lin:2004nb}. It may be interesting to clarify the true nature
of the `fermion picture' we think we see around (\ref{fermion-letter-limit}).

\item As also commented at various places earlier, it will be interesting to
see what one obtains by going beyond the Cardy limit, seeking for large $N$ saddle points
of $N$ integral variables, again carefully tuning the imaginary parts of the
chemical potentials. The analysis of \cite{Kinney:2005ej} already seems to set some
limitation of this approach, but it would be interesting (if possible) to see how
their results at order $1$ chemical potentials get connected to our results in
the Cardy-like limit. However, at least at the moment, this appears to be a
very challenging calculus.

\item In the $\frac{1}{8}$-BPS Macdonald sector, our studies `predict' that
there should be black holes, in case one believes that our Cardy saddle points will
transmute to large $N$ saddle points. Known black holes
reduce to small black holes with vanishing entropy in this limit. Considering
some qualitative aspects similar to the recently explored hairy black holes,
we speculate that they might be hairy $\frac{1}{8}$-BPS black holes.
Since one is now equipped with $4$ real Killing spinors, perhaps combining
the general SUSY analysis with a clever ansatz may shed lights on such solutions.

\item It may be straightforward to generalize the background field methods
of sections 2.2 and 4 to 4d $\mathcal{N}=1$ or 6d $\mathcal{N}=(1,0)$ SCFTs, with
or without gravity duals. For those with gravity duals,
\cite{Hosseini:2017mds,Hosseini:2018dob} already suggest expressions in terms
of the anomaly polynomials of the SCFTs. It will also be interesting to find
possible caveats of our discussions in various models, coming from zero mode
structures, as explored in \cite{Ardehali:2015bla}.

\item It may be useful to employ the background field approach at small $S^1$,
to explore large non-BPS AdS black holes. Of course in this
case, we expect that additional dynamical information has to be put in, unlike
BPS black holes. Maybe not too surprisingly, we find
similar structures as the hydrodynamic approach to the large AdS black holes
\cite{Bhattacharyya:2007vs}.

\item One is naturally led to the question of BPS black holes in AdS$_4$
\cite{Kostelecky:1995ei,Cvetic:2005zi} and
AdS$_6$ \cite{Chow:2008ip}. Some macroscopic/microscopic studies have been appeared recently, e.g. \cite{Choi:2018fdc,Choi:2019zpz,Nian:2019pxj,Choi:2019dfu,Bobev:2022wem,Bobev:2024mqw,Choi:2019miv,Crichigno:2020ouj}.

\end{itemize}

\vskip 0.5cm

\hspace*{-0.8cm} {\bf\large Acknowledgements}
\vskip 0.2cm

\hspace*{-0.75cm} We thank Chiung Hwang, Hee-Cheol Kim, Kimyeong Lee,
Jaemo Park and Jaewon Song for helpful discussions. This work is supported in part
by the National Research Foundation of Korea (NRF) Grant 2018R1A2B6004914 (SC, SK, JN),
NRF-2017-Global Ph.D. Fellowship Program (SC), a KIAS Individual Grant PG081602 at Korea Institute for Advanced Study (SC), World Premier International Research Center Initiative (WPI), MEXT, Japan (SC), the NRF grant 2021R1A2C2012350 (SK),
and Hyundai Motor Chung Mong-Koo Foundation (JN).

\end{document}